\documentclass[%
 aip,
 apl,
 reprint,
 amsmath,amssymb
]{revtex4-1}  

\usepackage{graphicx}
\usepackage[alsoload=synchem]{siunitx}
\usepackage{braket}
\usepackage{xspace}

 \usepackage{float}

\usepackage{changepage} 
\usepackage{enumitem}
\usepackage{tabularx} 

\newcolumntype{Y}{>{\centering\arraybackslash}X} 

\begin{document}

  \title[Spin-dependent fluorescence in h-BN]
  {Spin-Dependent Quantum Emission in Hexagonal Boron Nitride at Room Temperature}

\author{Annemarie L. Exarhos} 
\altaffiliation[Present address: ]{Department of Physics, Lawrence University, Appleton, WI 54911, USA}
\affiliation{Quantum Engineering Laboratory, Department of Electrical and Systems Engineering, University of Pennsylvania, Philadelphia, Pennsylvania 19104, United States}

\author{David A. Hopper}
\affiliation{Quantum Engineering Laboratory, Department of Electrical and Systems Engineering, University of Pennsylvania, Philadelphia, Pennsylvania 19104, United States}
\affiliation{Department of Physics and Astronomy, University of Pennsylvania, Philadelphia, Pennsylvania 19104, United States}

\author{Raj N. Patel}
\affiliation{Quantum Engineering Laboratory, Department of Electrical and Systems Engineering, University of Pennsylvania, Philadelphia, Pennsylvania 19104, United States}

\author{Marcus W. Doherty}
\affiliation{Laser Physics Centre, Research School of Physics and Engineering, Australian National University, Canberra ACT 2601, Australia}

\author{Lee C. Bassett}
\email{lbassett@seas.upenn.edu}
\affiliation{Quantum Engineering Laboratory, Department of Electrical and Systems Engineering, University of Pennsylvania, Philadelphia, Pennsylvania 19104, United States}

\date{\today}

\begin{abstract} 
\textbf{Optically addressable spins associated with defects in wide-bandgap semiconductors are versatile platforms for quantum information processing and nanoscale sensing, where spin-dependent inter-system crossing (ISC) transitions facilitate optical spin initialization and readout.  Recently, the van der Waals material hexagonal boron nitride (h-BN) has emerged as a robust host for quantum emitters (QEs), but spin-related effects have yet to be observed.  Here, we report room-temperature observations of strongly anisotropic photoluminescence (PL) patterns as a function of applied magnetic field for select QEs in h-BN. Field-dependent variations in the steady-state PL and photon emission statistics are consistent with an electronic model featuring a spin-dependent ISC between triplet and singlet manifolds, indicating that optically-addressable spin defects are present in h-BN\,---\,a versatile two-dimensional material promising efficient photon extraction, atom-scale engineering, and the realization of spin-based quantum technologies using van der Waals heterostructures.}
\end{abstract}

\maketitle



\def\PLimages{\begin{figure*}[!t]
  \includegraphics[width=7in]{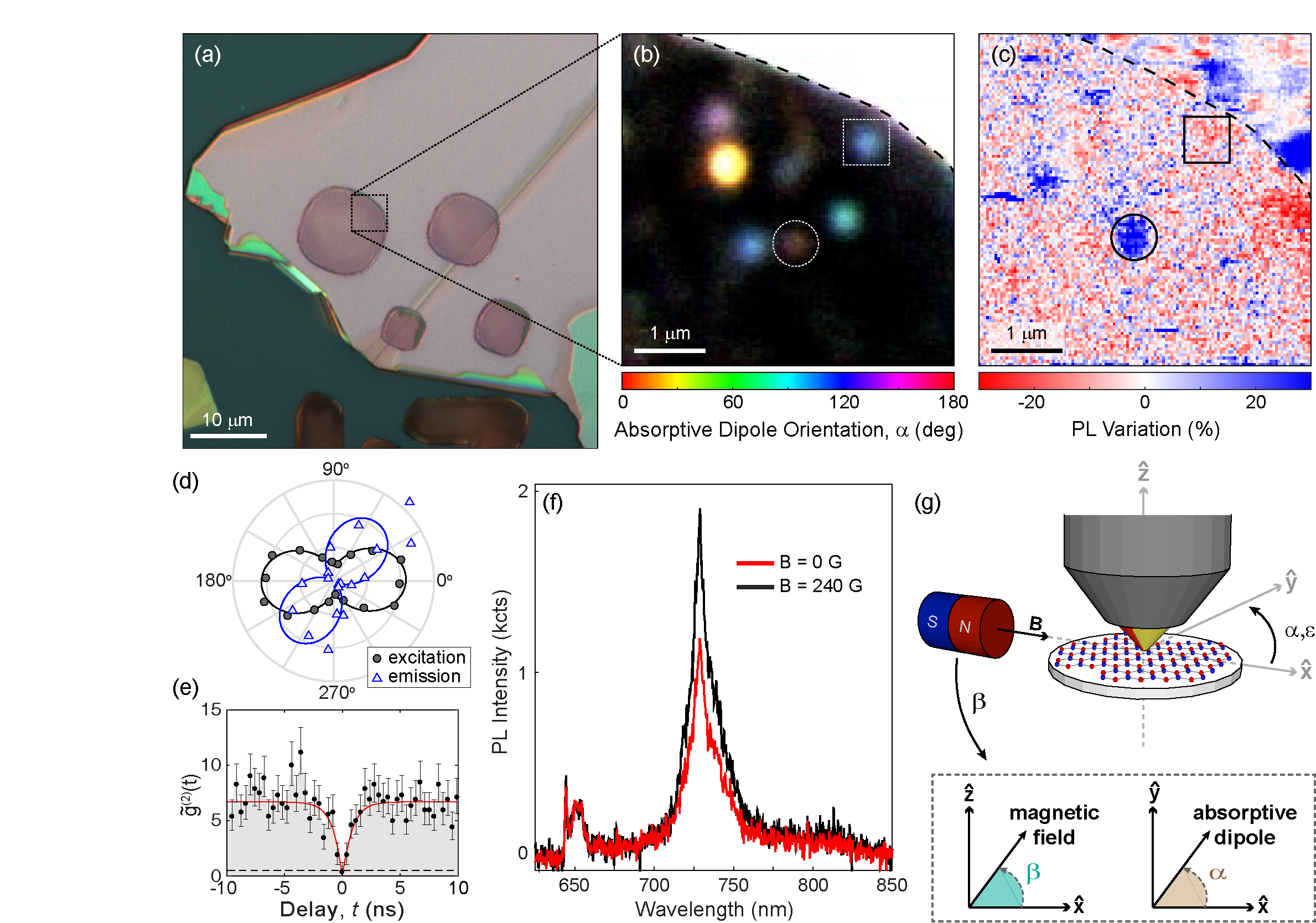}
  \caption{\textbf{Identification of field-dependent quantum emitters.} 
  \textbf{(a)}  Optical microscope image of an exfoliated h-BN flake on a patterned substrate. \textbf{(b)} Polarization-resolved PL image of suspended h-BN (denoted by the dashed box in (a)) at $B=0$ G.  The dashed curve indicates the edge of the suspended region. Color and brightness denote the absorptive dipole orientation and PL intensity, respectively \cite{Exarhos2017}.
  \textbf{(c)}  Background-subtracted differential PL variation image from (b) identifying changes due to an in-plane magnetic field ($B$=240 G).  Blue (red) denotes increased (decreased) PL when $B\neq 0$.
Panels (d)-(f) correspond to the QE circled in (b) and (c): \textbf{(d)} Background-subtracted PL excitation (circles) and emission (triangles) polarization dependences.  Curves denote fits to the data.
\textbf{(e)}  Photon autocorrelation function (points) with a fit to a three-level emission model (curve). Data are corrected for a Poissonian background.  The dashed line shows the single photon emission criterion.
\textbf{(f)} PL spectra with and without an in-plane magnetic field parallel to the QE's absorptive dipole.
\textbf{(g)} Illustration of the coordinate system for magnetic fields with respect to the microscope objective and sample.  $\beta$, in the $\hat{x}$-$\hat{z}$ plane, defines the angle of the magnetic field with respect to the sample plane and $\alpha$ ($\varepsilon$), in the $\hat{x}$-$\hat{y}$ plane, denotes the absorptive (emissive) dipole angle.}
  \label{PLimages}
\end{figure*}}

\def\PLvsB{\begin{figure}[]
\includegraphics[width=3.33in] {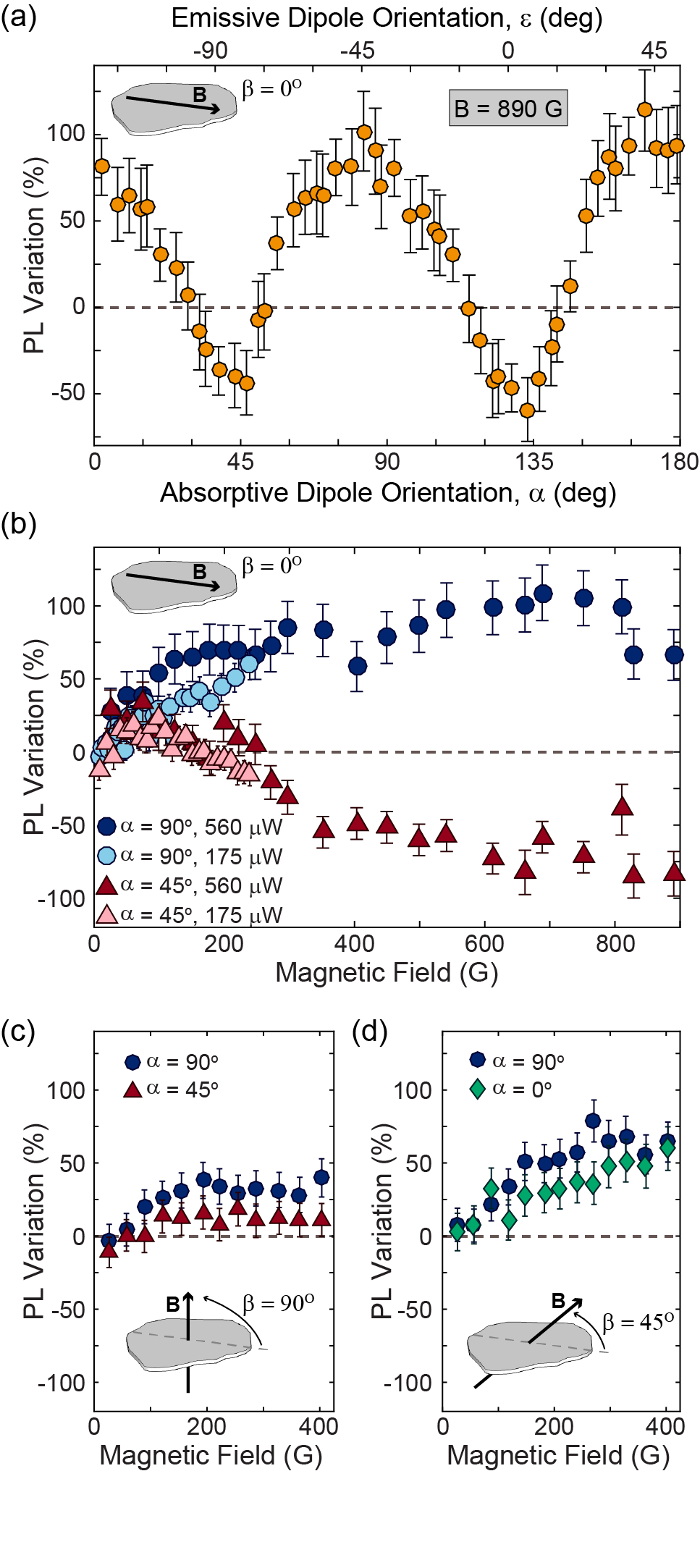}
 \caption{\textbf{Anisotropic magnetic-field dependent PL.} \textbf{(a)}  PL variation as a function of the relative orientation between the emitter's optical dipoles and an in-plane $B$=890~G.
 \textbf{(b-d)} PL variation for magnetic fields \textbf{(b)} parallel ($\beta = 0^{\circ}$), \textbf{(c}) perpendicular ($\beta = 90^{\circ}$), and \textbf{(d)} at $45^{\circ}$ ($\beta = 45^{\circ}$) to the sample plane. All data are taken at 560 $\mu$W (measured before the objective) unless otherwise specified.}
  \label{PLvsB}
\end{figure}}

\def\autocorrelation{\begin{figure}[]
\includegraphics[width=3.33in]{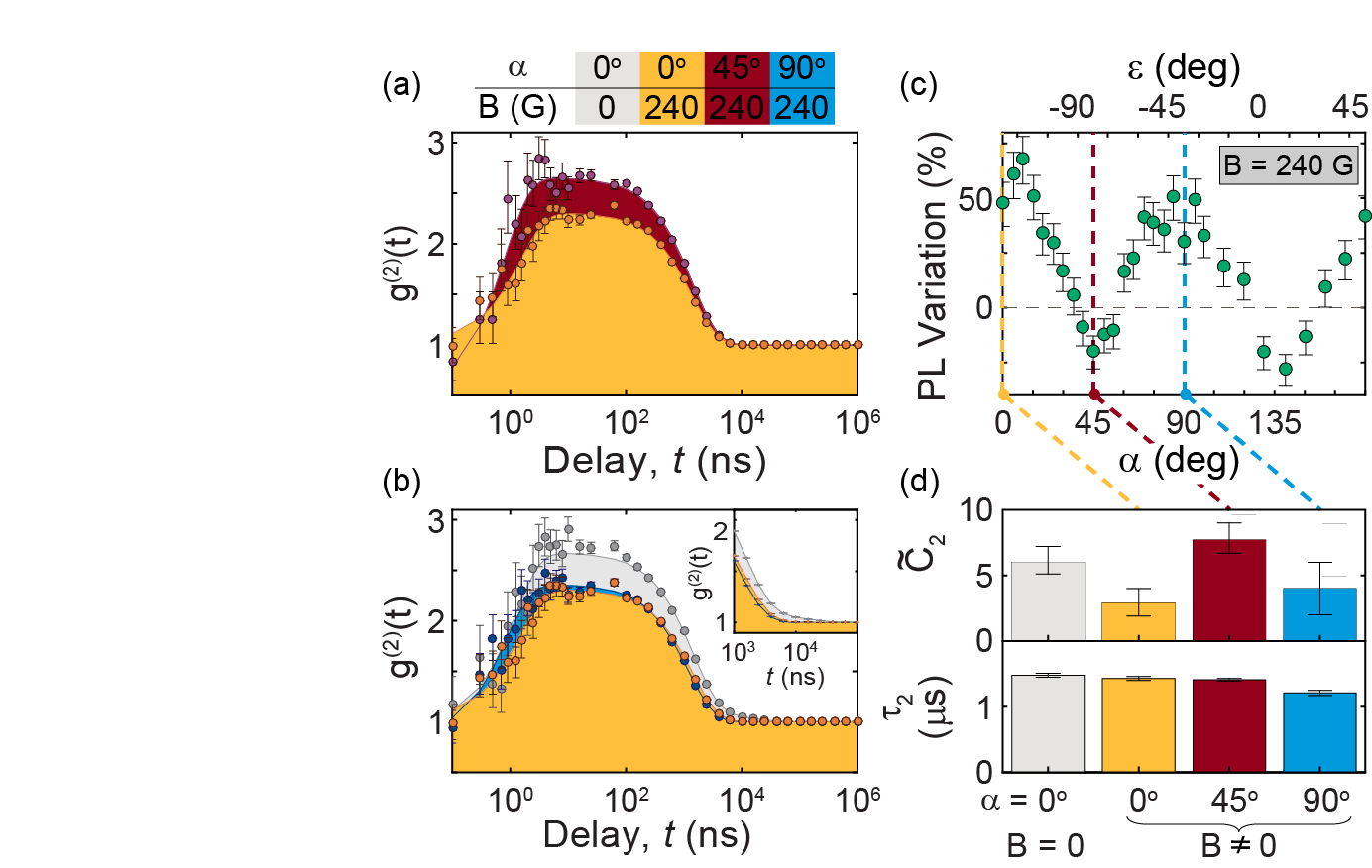}
\caption{\textbf{Photon emission dynamics.} 
\textbf{(a,b)} Measurements of the photon autocorrelation function for different orientations of the defect with respect to an in-plane magnetic field ($\beta = 0^{\circ}$) as indicated by the color-coded caption.  No background correction is applied to the data. \textbf{(b, inset)} Detailed view of the long timescale component ($\tau_3$) visible only when $B=0$ G.
\textbf{(c)} PL variation as a function of sample orientation under an in-plane magnetic field at 240~G. 
\textbf{(d)} Best-fit values of the background-corrected bunching amplitude, $\tilde{C}_2$, and corresponding timescale, $\tau_2$, for the data in (a,b).}
  \label{autocorrelation}
\end{figure}}

\def\model{\begin{figure*}[]
\makebox[\textwidth][c]{\includegraphics[]{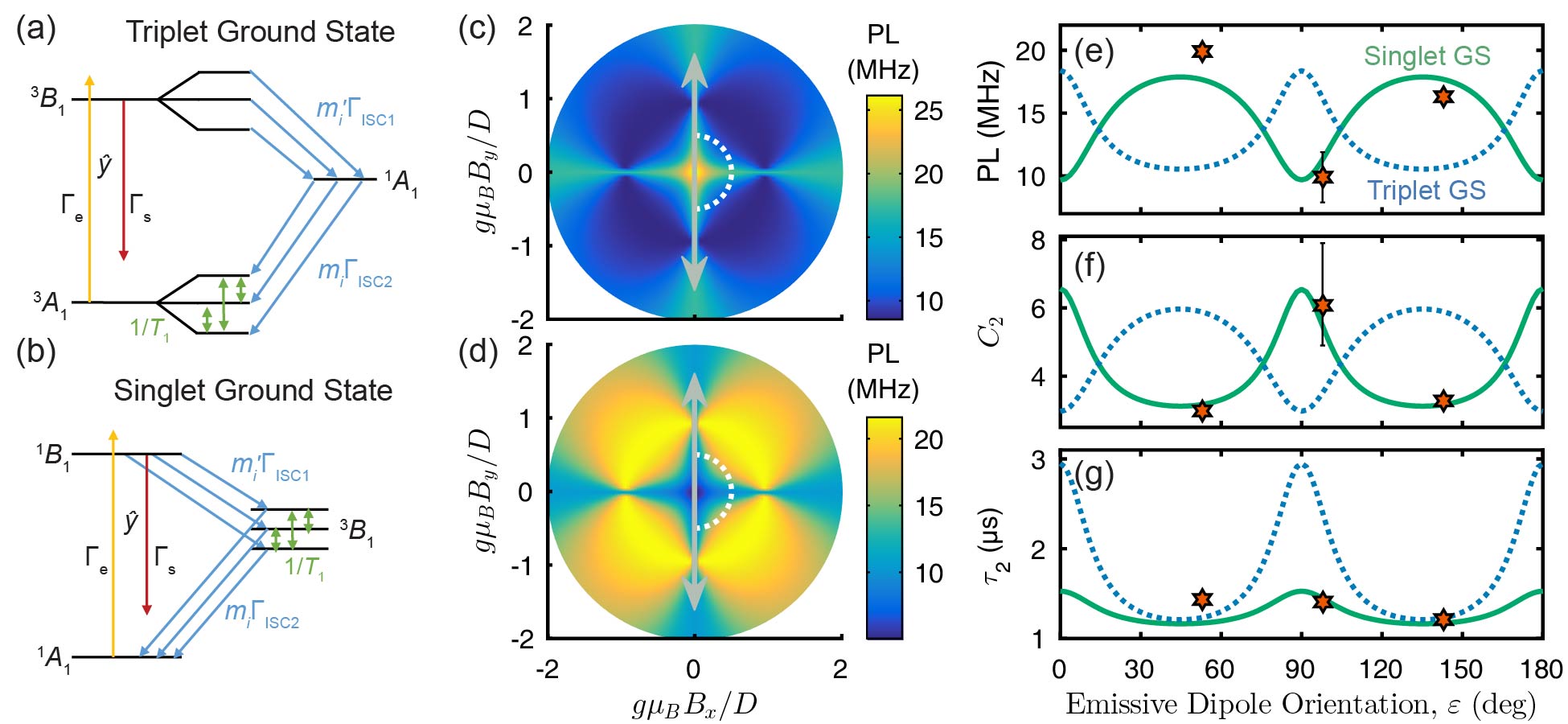}}
\caption{\textbf{Inter-system crossing models.} \textbf{(a,b)} Energy level diagrams showing the symmetry-allowed transitions for a triplet-singlet ISC (a) and a singlet-triplet ISC (b).  \textbf{(c,d)} Steady-state PL simulations for level diagrams in (a,b) respectively, as a function of in-plane magnetic field ($\beta = 0^{\circ}$). \textbf{(e,f)} Simulated PL amplitude (e), bunching amplitude (e), and bunching time (f) as a function of field orientation, $\varepsilon$, shown as dashed blue and solid green curves for calculations along the dashed curves in (c,d), respectively [$g\mu_B B/D=0.5$]. Star symbols indicate background-corrected measurements. PL measurements in (e) are multiplied by 1000. }
  \label{model}
\end{figure*}}

Spins in semiconductors are the elementary units for quantum spintronics \cite{Awschalom2013}, enabling an array of technologies including quantum communication \cite{Sipahigil2016,Kalb2017}, spin-based nanophotonics \cite{Gao2015}, nanoscale nuclear magnetic resonance \cite{Lovchinsky2016,Aslam2017}, and \textit{in vivo} transduction of intracellular magnetic, thermal, and chemical fields \cite{LeSage2013}. 
Leading candidates in diamond \cite{Doherty2013,Sipahigil2017,Rose2017arxiv} and silicon carbide \cite{Christle2015,Widmann2015} exhibit room-temperature, spin-dependent photoluminescence (PL) that facilitates initialization and readout of individual electron spins, along with their proximal nuclear spins \cite{Abobeih2018}.
Substantial progress notwithstanding, synthesis and device fabrication with these three-dimensional semiconductors remains challenging, especially for sensing applications, which demand the use of near-surface spins whose quantum properties are degraded as compared to the bulk.  Intrinsically low-dimensional materials, such as the van der Waals material hexagonal boron nitride (h-BN) offer an appealing alternative\,---\,spins confined to the same two-dimensional (2D) atomic plane, and all at the surface, offer enormous potential to engineer novel quantum functionality.

Magneto-optical effects are the principal means by which individual spins are addressed\cite{Heremans2016} and coupled to light \cite{Buckley2010}.
Quantum emission in h-BN\cite{Tran2016,Tran2016a,Martinez2016,Jungwirth2016,Chejanovsky2016,Exarhos2017,Jungwirth2017} is believed to originate from defects with localized electronic states deep within its bandgap, similarly to 
other wide-bandgap materials exhibiting defect-related single-photon emission \cite{Aharonovich2016}. However, even in this expanding catalog of materials and their numerous fluorescent defects\cite{Zaitsev2001}, room-temperature, spin-dependent PL remains a rare phenomenon due to the necessary alignment of energy levels and symmetry-protected selection rules.  Despite well-established electron paramagnetic resonance signatures for bulk h-BN \cite{Geist1964,Fanciulli1997} and recent theoretical predictions \cite{Abdi2017}, experimental evidence for magneto-optical effects has been elusive to date \cite{Li2017, Koperski2017}. Here, we demonstrate that select QEs in h-BN do exhibit room-temperature, magnetic-field-dependent PL consistent with a spin-dependent ISC, paving the way to the development of 2D quantum spintronics.

\subsection*{Identification of Magnetic-Field-Dependent QEs}

Present understanding of the chemical and electronic structure of h-BN's QEs is impeded by the heterogeneity of their optical properties\cite{Exarhos2017}.  Contending models aim to account for disparate observations; multiple QE species likely play a role\cite{Tran2016,Tawfik2017,Wu2017,Abdi2017_arXiv}. Nonetheless, h-BN's QEs universally exhibit linearly polarized optical absorption and emission consistent with optical dipole transitions from a defect with broken in-plane symmetry \cite{Tran2016,Exarhos2017,Jungwirth2017}. Based on symmetry considerations, any spin-dependent ISC transitions likely produce an anisotropic PL response to in-plane magnetic fields\,---\,a fact we exploit to identify and characterize individual QEs with spin-dependent optical properties.  Unfortunately, the absence of well-characterized defect-specific emission signatures prevents the selective addressing of defect sub-ensembles, as has been essential for statistical studies and the identification of spin qubits in other materials \cite{Oort1988,Heremans2016}.  Consequently, we study QEs in h-BN at the single-defect level. 

\PLimages

We study a 400-nm-thick exfoliated h-BN flake suspended across a set of holes etched in a silicon substrate at room temperature in ambient conditions [Fig.~\ref{PLimages}(a)].  Excitation-polarization-resolved PL images [Fig.~\ref{PLimages}(b)] reveal a number of strongly linearly-polarized emitters in the suspended region. To identify magnetic-field-dependent emitters, we construct differential images [Fig.~1(c)] of the PL variation, $(I_B-I_0)/I_0$, where $I_B$ ($I_0$) is the brightness extracted from composite PL maps with (without) a magnetic field, $\mathbf{B}$, applied along the horizontal in-plane direction (see Methods).

Most of the emitters in the suspended region (below the dashed curve in Fig.~\ref{PLimages}(c)) show no change with the magnet, whereas a few red and blue features highlight potentially interesting spots.  Upon further study, some features are not reproducible, but two emitters in particular, denoted by circles and squares in Figs.~\ref{PLimages}(b-c), exhibit systematic changes in brightness due to the applied field.  Strikingly, the circled emitter brightens whereas the boxed emitter dims in response to $\mathbf{B}$ at this orientation.  The following discussion focuses on the circled defect, which remained stable over several months. Data for the boxed emitter and other field-dependent spots are also available\cite{SI}.

Figures~\ref{PLimages}(d-f) summarize the spatial, temporal, and spectral emission characteristics of the QE circled in Figs.~\ref{PLimages}(b-c).  Like many QEs in h-BN \cite{Exarhos2017,Jungwirth2016,Jungwirth2017}, the PL exhibits incomplete visibility as a function of linear excitation polarization angle, with an optimum excitation axis (hereafter called the absorptive dipole) offset from the fully polarized emission-dipole axis by an angle $\Delta = 53^{\circ}\pm 4^{\circ}$.  The absorptive dipole orientation is independent of $\mathbf{B}$ \cite{SI}.  The background-corrected second-order autocorrelation function, $\tilde{g}^{(2)}(t)$, [Fig.~\ref{PLimages}(e)] exhibits an antibunching dip near zero delay that drops below the threshold, $\tilde{g}^{(2)}(0)<0.5$, indicating the PL is dominated by a single emitter. Figure \ref{PLimages}(f) shows the QE's room-temperature PL spectrum with and without an applied magnetic field.

In Fig.~\ref{PLimages}, the absorptive dipole of the circled QE is horizontal, $\parallel\mathbf{B}$. As illustrated in Fig.~\ref{PLimages}(g), we explore arbitrary field orientations by rotating the sample about the optical axis, where $\alpha$ ($\varepsilon$) denotes the orientation of the absorptive (emissive) dipole, relative to $\hat{x}$, and by adjusting a magnet goniometer in the $\hat{x}$-$\hat{z}$ plane, where $\beta$ is the angle of the field relative to $\hat{x}$.

\PLvsB
\subsection*{Variations in steady-state PL}

Figure \ref{PLvsB}(a) shows the PL variation as a function of sample orientation when an 890~G magnetic field is applied along $\hat{x}$.  The dashed line denotes the zero-field emission rate.  The QE exhibits both increased and decreased emission as a function of the in-plane field direction, with $>$50\% variation in both directions.  Furthermore, the $90^{\circ}$ modulation period is approximately aligned to the optical dipole orientations, such that the PL is brighter (dimmer) when the field is either \textit{aligned or perpendicular} to the  absorptive (emissive) dipole.  While this anisotropic PL modulation is reminiscent of other quantum emitters with spin-dependent ISC transitions \cite{Epstein2005}, the 90$^\circ$ symmetry and bipolar response (\textit{i.e.}, both brightening and dimming) are unique.

The disparate behavior as a function of $B$ for different sample orientations is illustrated in Fig.~\ref{PLvsB}(b). The PL increases monotonically with $B$ at $\alpha = 90^{\circ}$ whereas the response at $\alpha = 45^{\circ}$ is non-monotonic; an initial increase out to $\approx$70~G is followed by decreasing PL which eventually falls below the zero-field emission rate. In both orientations, the variation appears to saturate by $\approx$600~G and is independent of optical excitation power \cite{SI}.  

For out-of-plane fields ($\beta=90^\circ$), the PL increases monotonically [Fig.~\ref{PLvsB}(c)], although the variation saturates by $\approx$200~G and is noticeably smaller than for $\beta=0^\circ$. The offset between data sets at different sample orientations likely reflects uncertainty in estimating the zero-field emission rate.  A similar monotonic increase observed for $\beta=45^\circ$ [Fig.~\ref{PLvsB}(d)] suggests an underlying 180$^\circ$ symmetry for rotations about $\hat{x}$ or $\hat{y}$, contrasting with the 90$^\circ$ periodicity observed for rotations about $\hat{z}$. 

\subsection*{Photo-Dynamic Response}

The QE's photon emission statistics provide insight into the field-dependent optical dynamics that modulate the steady-state PL.  Figs.~\ref{autocorrelation}(a,b) show the observed photon autocorrelation function for several settings of the sample orientation and $B$.  Universally, the QE exhibits antibunching at short ($t\lesssim1$~ns) delay times and bunching over longer ($t\approx1$~$\mu$s) times, qualitatively similar to previous observations of h-BN's QEs \cite{Tran2016,Tran2016a,Exarhos2017,Chejanovsky2016}.  We fit the data using an empirical model: $g^{(2)}(t) = 1-C_1 e^{-t/\tau_1}+\sum_{i=2}^n C_i e^{-t/\tau_i}$, where $n$=2 or 3 depending on the shape of the data\cite{SI}. For quantitative comparisons with simulations, we also calculate background-corrected values, $\tilde{C}_i$ (see Methods).  

\autocorrelation

\model 

The dominant field effect appears in the amplitude of the leading bunching component, $\tilde{C}_2$, which decreases (increases) when the steady-state PL becomes brighter (dimmer) [Figs.~\ref{autocorrelation}(c,d)].  Meanwhile, the bunching timescale remains nearly constant at $\tau_2\approx1.4$~$\mu$s. This behavior is consistent with a QE model including one or more metastable dark states and an ISC modulated by $\mathbf{B}$.  In this model, a larger bunching amplitude reflects an increase in the steady-state population trapped in the dark state, and correspondingly lower PL.  Interestingly, a third lifetime component with $\tau_3\approx 16$~$\mu$s is required to capture the autocorrelation shape when $B=0$ G, but this component vanishes when $\mathbf{B}$ is in plane [Fig.~\ref{autocorrelation}(b, inset)].

\subsection*{Modeling Spin-Dependent Optical Dynamics}

We use a semiclassical master equation to simulate QE optical dynamics, where the relative transition rates between spin and orbital sublevels are determined by the symmetry-defined Hamiltonian and a set of empirical parameters \cite{SI}. We consider systems characterized by the point group $C_{2v}$, encompassing many simple defects in multilayer h-BN including vacancy-impurity complexes such as C$_\mathrm{B}$V$_\mathrm{N}$ and distorted vacancies such as  N$_\mathrm{B}$V$_\mathrm{N}$. Using molecular-orbital theory, we consider all possible combinations of three mid-gap, single-particle orbital levels that can encompass an optical ground and excited state with in-plane optical dipole transitions as well as at least one intermediate state from a different spin manifold [Figs.~\ref{model}(a,b)]. 

We further consider configurations with total spin $S\in \{0,\frac{1}{2},1,\frac{3}{2}\}$. The lack of symmetry-protected orbital multiplets in $C_{2v}$ makes configurations with $S>\frac{3}{2}$ energetically unfavorable. An applied magnetic field mixes the spin sublevels of configurations with $S\geq1$ with a pattern determined by the zero-field splitting terms.  Crucially, although the spin eigenstates for an $S=1$ Hamiltonian vary with 180$^\circ$ periodicity as a function of in-plane field orientation, the mixing and ISC spin-selection rules can lead to 90$^\circ$ periodicity in the steady-state PL and autocorrelation parameters (Fig.~\ref{model}). On the other hand, for $S=\frac{3}{2}$, the spin eigenstates are 360$^\circ$-periodic\cite{SI}.  Furthermore, spin-dependent selection rules do not naturally arise for doublet-quartet transitions in $C_{2v}$, since there is only one double-group representation that must characterize all eigenstates with half-integer spin. We therefore argue that singlet-triplet configurations are most likely to explain the observed behavior.

Of all the configurations we considered \cite{SI}, the level diagrams in Figs.~\ref{model}(a,b) most closely match the observations. Both models exhibit 90$^\circ$-periodic PL variations as a function of in-plane field angle [Figs.~\ref{model}(c,d)], with corresponding changes in the intermediate bunching parameters $\tilde{C}_2$ and $\tau_2$ [Figs.~\ref{model}(f,g)].  However, simulations of the triplet-ground-state model predict larger variations in $\tau_2$ than we experimentally observe.  Moreover, the simulated PL is at a maximum when the triplet optical dipole [grey arrow in Fig.~\ref{model}(c)] is aligned or perpendicular to $\textbf{B}$, whereas experimentally we observe a minimum when $\varepsilon=0^\circ$ or 90$^\circ$ (we assume the emission dipole axis reflects the QE's underlying symmetry).  The singlet-ground-state configuration of Fig.~\ref{model}(b) matches the experiments on these points, hence we tentatively identify it as a potential model for the physical system \cite{SI}.

So far we have not considered possible chemical structures that could produce the proposed level diagram.  Recent calculations \cite{Tran2016,Tawfik2017,Wu2017,Abdi2017_arXiv} focus on simple configurations with light elements such as  C$_\mathrm{B}$V$_\mathrm{N}$ or N$_\mathrm{B}$V$_\mathrm{N}$.  These defects have $C_{2v}$ symmetry and share some features with our models. However, electronic structure calculations in 2D materials remain challenging \cite{Wu2017}, and uncertainty persists regarding the energy-level ordering even for these simple candidates. Exploration of structures involving heavier elements or larger complexes remains an important goal, ideally guided by atom-scale structural imaging\cite{Hong2017_defectsin2dreview} correlated with optical experiments.

While the models in Fig.~\ref{model} capture many experimental features, they do not account for all observations.  In particular, the simulations in Fig.~\ref{model}(d) predict a minimum for the PL at $B=0$ G, with no change when $\mathbf{B} \parallel \hat{z}$. Also, whereas longer-lifetime bunching components do emerge from the simulations in certain circumstances, we have been unable to quantitatively reproduce the observations in Fig.~3(c) using a single set of field-independent parameters. These discrepancies could be related to hyperfine coupling, which is not included in our model, but likely becomes important near $B=0$ G \cite{SI}.

Future experiments are required to answer these important questions.  Field-dependent emission appears to be relatively rare for h-BN's visible emitters, occurring for only a few percent of spots in our samples, but the underlying difference between field-dependent and field-independent emitters remains unknown.  Other measurement modalities, especially optically detected magnetic resonance (ODMR), will be crucial to confirm the predictions of our model and to determine the underlying spin Hamiltonian parameters. Calculations suggest a large ODMR contrast will be observed under the right conditions\cite{SI}.  From a materials perspective, significant further work is needed to reproducibly create these spin defects and incorporate them in devices.

\subsection*{Conclusions and Outlook}

The observation of room-temperature, spin-dependent PL from select QEs in h-BN expands the role of h-BN for use in quantum technologies.  Nanophotonic and nanomechanical devices will exploit optically-addressable spins in h-BN for quantum optics\cite{Buckley2010,Bassett2014,Sipahigil2016} and optomechanics\cite{Abdi2017,Chen2018}.  QE electron spins can be used as actuators to address nearby nuclear spins\cite{Lovchinsky2017,Abobeih2018}, offering a platform to study strongly-interacting spin lattices \cite{Britton2012} and perform quantum simulations \cite{Cai2013}. As sensors, the striking PL variation in response to relatively weak magnetic fields bodes well for ultrasensitive detection of nanomagnetism\cite{Pelliccione2016,Gross2017} and chemical characterization \cite{Lovchinsky2016,Aslam2017}.
A spinless singlet ground-state, as proposed in our electronic model, benefits these applications by removing electron-induced nuclear decoherence and unwanted sensor backaction \cite{Lee2013}. 

Additionally, van der Waals heterostructures offer unprecedented opportunities to engineer the QEs' local environment and control their functionality, enabling alternative mechanisms for electro-optical addressing.  For example, QE spins in h-BN could couple to free carriers or excitons in graphene or transition-metal dichalcogenides, where spin-dependent quantum emission in h-BN could be used to initialize or read out spin-valley qubits for cascaded information transfer between layers \cite{Doherty2016,Luo2017_mos2graphenespintransfer}.


\section*{Methods}
\subsection*{Sample Preparation and Mounting}
Earlier work highlighted the strong influence of substrate interactions during irradiation and annealing treatments on h-BN's visible fluorescence \cite{Exarhos2017}.  To eliminate these effects, we study emitters present in freely-suspended h-BN membranes that have been exfoliated from commercially available bulk single crystals and treated as described.  All measurements are performed in ambient conditions.  

H-BN samples are prepared by exfoliating single-crystal h-BN purchased from HQ Graphene onto patterned 90 nm-thick thermal SiO$_2$ on Si according to Ref.~\onlinecite{Exarhos2017}.  Following exfoliation, samples undergo an O$_2$ plasma clean in an oxygen barrel asher (Anatech SCE 108) and are  annealed in Ar at 850$^{\circ}$~C for 30 minutes.  They are imaged using a scanning electron microscope operating at 3 kV (FEI Strata DB235 FIB SEM), after which the samples are again annealed in Ar at 850$^{\circ}$~C for 30 minutes.  

An exfoliated and prepared sample is mounted on a rotation stage enabling in-plane rotation of the sample, and thus the optical dipoles of individual defects, with respect to the rest of the setup in a home-built confocal fluorescence microscope with 592 nm continuous wave excitation.\cite{SI}  Excitation powers used for both bleaching and imaging range from 175-550 $\mu W$ and the PL variation is roughly constant over this range.\cite{SI}  Additional control over the direction of excitation linear polarization is facilitated with a half-waveplate.  An external magnetic field is applied using neodymium magnets mounted on a home-built goniometer that enables variations between the direction of the applied field and the sample plane, as shown in Figure \ref{PLimages}(g).  Changing the distance between the magnet and the sample allows for a range of applied magnetic fields from 0-890 G.

\subsection*{PL Images}
Following the procedure described in Ref.~\onlinecite{Exarhos2017}, composite polarized PL images as in Figure \ref{PLimages}(b) are constructed from a series of confocal PL scans recorded for 4 different linear polarization settings of the 592-nm excitation laser (0$^{\circ}$, 45$^{\circ}$, 90$^{\circ}$, 135$^{\circ}$), while collecting PL between 650-900 nm.  The individual images are colorized according to the polarization setting, registered to one another, and summed to create the composite image.

Differential PL images as in Fig.~\ref{PLimages}(c) are constructed using the value (\textit{i.e.}, brightness) coordinate from composite images acquired with and without a magnetic field. A small constant PL variation [$\approx$8\% in Fig.~\ref{PLimages}(c)], calculated by averaging over all pixels, is subtracted to account for field-induced changes to the microscope's alignment.

\subsection*{PL Spectra}
PL spectra are taken using a Princeton Instruments IsoPlane 16 spectrometer and a PIXIS 100 CCD with a spectral resolution of 0.7 nm.  Multiple exposures ($>2$) are collected, dark count subtracted and cosmic ray rejected, then averaged together.  PL spectra are not corrected for wavelength-dependent photon collection efficiencies.  A 633 nm long pass edge filter (Semrock, BLP01-633R-25) in the collection line blocks the 592 nm laser.  In Fig.~\ref{PLimages}(f), the h-BN Raman line is visible at the edge of the collection band, at $\sim 644$ nm.  The field-independent feature around 650~nm is associated with the background.

\subsection*{\textbf{PL Variation with Magnetic Field Measurements}}
To calculate the PL variation at different ($\alpha$, $\varepsilon$) orientations and at different magnetic field strengths, the background-subtracted PL is determined from a combination of Gaussian fits to PL images and measurements of the time-averaged emission rate detected by focusing directly on the emitter for 30-120~s at each setting.  At $B=0$ G, the orientation-dependent transmission of PL from the circled emitter through the collection line of the confocal microscope is measured in order to normalize for the small variations ($<6\%$) in PL that occur when the sample is rotated.  At each sample orientation, the excitation polarization is aligned with the absorptive dipole.  PL variation as a function of sample orientation (Figs. 2(a), 3(c)) are binned every 3$^{\circ}$ and PL variation as a function of magnetic field strength (Figs. 2(b-d)) are binned every 5 G, where the weighted average of the data in the bin is calculated.  Error bars represent the variance of the weighted average in a particular bin along with the average variance of points in all bins, scaled by the number of points per bin.

\subsection*{\textbf{Autocorrelation Analysis}}
Autocorrelation data is obtained using a Hanbury Brown-Twiss setup with a time correlated single-photon counting module (PicoQuant PicoHarp 300) in time-tagged, time-resolved collection mode.  The second order autocorrelation function, $g^{(2)}(t)$, is calculated from the photon arrival times using the method described in Ref. \onlinecite{Laurence2006}, and the curves are fit using empirical functions as described in the text.  The choice of model is based on the quality of weighted least-squares fits accounting for the Poissonian uncertainty of each bin.

Background-corrected amplitudes, $\tilde{C}_i$, are calculated using separate calibration measurements of the signal-to-background ratio at each field setting. 
Using an average measurement of the Poissonian PL background taken from several nearby locations on the suspended membrane, we estimate the background-correction parameter, $\rho=I/(I+I_{bkgd})$, where $I$ is the QE PL and $I_{bkgd}$ is the background PL.
The background-corrected amplitudes are then given by $\tilde{C}_i=C_i/\rho^2$. 
Best-fit parameters and their background-corrected values for all autocorrelation measurements are listed in Supplementary Table S1\cite{SI}.  Confidence intervals reflect uncertainty in the fits and the measurement of $\rho$.

The short-delay autocorrelation data in Fig.~\ref{PLimages}(e) are background corrected in a similar manner using the relation
\cite{Brouri2000}
\begin{equation}\label{eq:g2bc}
\tilde{g}^{(2)}(t)=\frac{g^{(2)}(t)-(1-\rho^2)}{\rho^2}.
\end{equation}
The underlying data are the same as in Fig.~\ref{autocorrelation}(a) (red points), rebinned over a linear scale. Since the range of delays is much smaller than the shortest bunching timescale, we fit these data using a simplified empirical function,
\begin{equation}
\tilde{g}^{(2)}(t) = 1-\tilde{C}_1e^{-|t|/\tau_1}+\tilde{C}_2,
\label{3-level}
\end{equation}
from which we determine a best-fit value $\tilde{g}^{(2)}(0)=1-\tilde{C}_1+\tilde{C}_2=-0.2\pm0.9$, satisfying the single-emitter threshold, $\tilde{g}^{(2)}(0)<0.5$, by $0.8\sigma$.  The accuracy of this measurement is limited by shot noise and detector timing jitter due to the short antibunching timescale, $\tau_1=0.8\pm0.2$~ns.

\section*{Acknowledgements}
This work was supported by the Army Research Office (W911NF-15-1-0589). MWD was supported by the Australian Research Council (DE170100169). The authors thank Jennifer Saouaf and Richard Grote for assistance in sample preparation, Tzu-Yung Huang for assistance with measurements, and Audrius Alkauskas for engaging theoretical discussions.

\bibliography{}

\begin{thebibliography}{48}%
\makeatletter
\providecommand \@ifxundefined [1]{%
 \@ifx{#1\undefined}
}%
\providecommand \@ifnum [1]{%
 \ifnum #1\expandafter \@firstoftwo
 \else \expandafter \@secondoftwo
 \fi
}%
\providecommand \@ifx [1]{%
 \ifx #1\expandafter \@firstoftwo
 \else \expandafter \@secondoftwo
 \fi
}%
\providecommand \natexlab [1]{#1}%
\providecommand \enquote  [1]{``#1''}%
\providecommand \bibnamefont  [1]{#1}%
\providecommand \bibfnamefont [1]{#1}%
\providecommand \citenamefont [1]{#1}%
\providecommand \href@noop [0]{\@secondoftwo}%
\providecommand \href [0]{\begingroup \@sanitize@url \@href}%
\providecommand \@href[1]{\@@startlink{#1}\@@href}%
\providecommand \@@href[1]{\endgroup#1\@@endlink}%
\providecommand \@sanitize@url [0]{\catcode `\\12\catcode `\$12\catcode
  `\&12\catcode `\#12\catcode `\^12\catcode `\_12\catcode `\%12\relax}%
\providecommand \@@startlink[1]{}%
\providecommand \@@endlink[0]{}%
\providecommand \url  [0]{\begingroup\@sanitize@url \@url }%
\providecommand \@url [1]{\endgroup\@href {#1}{\urlprefix }}%
\providecommand \urlprefix  [0]{URL }%
\providecommand \Eprint [0]{\href }%
\providecommand \doibase [0]{http://dx.doi.org/}%
\providecommand \selectlanguage [0]{\@gobble}%
\providecommand \bibinfo  [0]{\@secondoftwo}%
\providecommand \bibfield  [0]{\@secondoftwo}%
\providecommand \translation [1]{[#1]}%
\providecommand \BibitemOpen [0]{}%
\providecommand \bibitemStop [0]{}%
\providecommand \bibitemNoStop [0]{.\EOS\space}%
\providecommand \EOS [0]{\spacefactor3000\relax}%
\providecommand \BibitemShut  [1]{\csname bibitem#1\endcsname}%
\let\auto@bib@innerbib\@empty
\bibitem [{\citenamefont {Awschalom}\ \emph {et~al.}(2013)\citenamefont
  {Awschalom}, \citenamefont {Bassett}, \citenamefont {Dzurak}, \citenamefont
  {Hu},\ and\ \citenamefont {Petta}}]{Awschalom2013}%
  \BibitemOpen
  \bibfield  {author} {\bibinfo {author} {\bibfnamefont {D.~D.}\ \bibnamefont
  {Awschalom}}, \bibinfo {author} {\bibfnamefont {L.~C.}\ \bibnamefont
  {Bassett}}, \bibinfo {author} {\bibfnamefont {A.~S.}\ \bibnamefont {Dzurak}},
  \bibinfo {author} {\bibfnamefont {E.~L.}\ \bibnamefont {Hu}}, \ and\ \bibinfo
  {author} {\bibfnamefont {J.~R.}\ \bibnamefont {Petta}},\ }\href {\doibase
  10.1126/science.1231364} {\bibfield  {journal} {\bibinfo  {journal}
  {Science}\ }\textbf {\bibinfo {volume} {339}},\ \bibinfo {pages} {1174}
  (\bibinfo {year} {2013})}\BibitemShut {NoStop}%
\bibitem [{\citenamefont {Sipahigil}\ \emph {et~al.}(2016)\citenamefont
  {Sipahigil}, \citenamefont {Evans}, \citenamefont {Sukachev}, \citenamefont
  {Burek}, \citenamefont {Borregaard}, \citenamefont {Bhaskar}, \citenamefont
  {Nguyen}, \citenamefont {Pacheco}, \citenamefont {Atikian}, \citenamefont
  {Meuwly}, \citenamefont {Camacho}, \citenamefont {Jelezko}, \citenamefont
  {Bielejec}, \citenamefont {Park}, \citenamefont {Lon{\v c}ar},\ and\
  \citenamefont {Lukin}}]{Sipahigil2016}%
  \BibitemOpen
  \bibfield  {author} {\bibinfo {author} {\bibfnamefont {A.}~\bibnamefont
  {Sipahigil}}, \bibinfo {author} {\bibfnamefont {R.~E.}\ \bibnamefont
  {Evans}}, \bibinfo {author} {\bibfnamefont {D.~D.}\ \bibnamefont {Sukachev}},
  \bibinfo {author} {\bibfnamefont {M.~J.}\ \bibnamefont {Burek}}, \bibinfo
  {author} {\bibfnamefont {J.}~\bibnamefont {Borregaard}}, \bibinfo {author}
  {\bibfnamefont {M.~K.}\ \bibnamefont {Bhaskar}}, \bibinfo {author}
  {\bibfnamefont {C.~T.}\ \bibnamefont {Nguyen}}, \bibinfo {author}
  {\bibfnamefont {J.~L.}\ \bibnamefont {Pacheco}}, \bibinfo {author}
  {\bibfnamefont {H.~A.}\ \bibnamefont {Atikian}}, \bibinfo {author}
  {\bibfnamefont {C.}~\bibnamefont {Meuwly}}, \bibinfo {author} {\bibfnamefont
  {R.~M.}\ \bibnamefont {Camacho}}, \bibinfo {author} {\bibfnamefont
  {F.}~\bibnamefont {Jelezko}}, \bibinfo {author} {\bibfnamefont
  {E.}~\bibnamefont {Bielejec}}, \bibinfo {author} {\bibfnamefont
  {H.}~\bibnamefont {Park}}, \bibinfo {author} {\bibfnamefont {M.}~\bibnamefont
  {Lon{\v c}ar}}, \ and\ \bibinfo {author} {\bibfnamefont {M.~D.}\ \bibnamefont
  {Lukin}},\ }\href {\doibase 10.1126/science.aah6875} {\bibfield  {journal}
  {\bibinfo  {journal} {Science}\ }\textbf {\bibinfo {volume} {354}},\ \bibinfo
  {pages} {847} (\bibinfo {year} {2016})}\BibitemShut {NoStop}%
\bibitem [{\citenamefont {Kalb}\ \emph {et~al.}(2017)\citenamefont {Kalb},
  \citenamefont {Reiserer}, \citenamefont {Humphreys}, \citenamefont
  {Bakermans}, \citenamefont {Kamerling}, \citenamefont {Nickerson},
  \citenamefont {Benjamin}, \citenamefont {Twitchen}, \citenamefont {Markham},\
  and\ \citenamefont {Hanson}}]{Kalb2017}%
  \BibitemOpen
  \bibfield  {author} {\bibinfo {author} {\bibfnamefont {N.}~\bibnamefont
  {Kalb}}, \bibinfo {author} {\bibfnamefont {A.~A.}\ \bibnamefont {Reiserer}},
  \bibinfo {author} {\bibfnamefont {P.~C.}\ \bibnamefont {Humphreys}}, \bibinfo
  {author} {\bibfnamefont {J.~J.~W.}\ \bibnamefont {Bakermans}}, \bibinfo
  {author} {\bibfnamefont {S.~J.}\ \bibnamefont {Kamerling}}, \bibinfo {author}
  {\bibfnamefont {N.~H.}\ \bibnamefont {Nickerson}}, \bibinfo {author}
  {\bibfnamefont {S.~C.}\ \bibnamefont {Benjamin}}, \bibinfo {author}
  {\bibfnamefont {D.~J.}\ \bibnamefont {Twitchen}}, \bibinfo {author}
  {\bibfnamefont {M.}~\bibnamefont {Markham}}, \ and\ \bibinfo {author}
  {\bibfnamefont {R.}~\bibnamefont {Hanson}},\ }\href {\doibase
  10.1126/science.aan0070} {\bibfield  {journal} {\bibinfo  {journal}
  {Science}\ }\textbf {\bibinfo {volume} {356}},\ \bibinfo {pages} {928}
  (\bibinfo {year} {2017})}\BibitemShut {NoStop}%
\bibitem [{\citenamefont {Gao}\ \emph {et~al.}(2015)\citenamefont {Gao},
  \citenamefont {Imamoglu}, \citenamefont {Bernien},\ and\ \citenamefont
  {Hanson}}]{Gao2015}%
  \BibitemOpen
  \bibfield  {author} {\bibinfo {author} {\bibfnamefont {W.~B.}\ \bibnamefont
  {Gao}}, \bibinfo {author} {\bibfnamefont {A.}~\bibnamefont {Imamoglu}},
  \bibinfo {author} {\bibfnamefont {H.}~\bibnamefont {Bernien}}, \ and\
  \bibinfo {author} {\bibfnamefont {R.}~\bibnamefont {Hanson}},\ }\href
  {http://dx.doi.org/10.1038/nphoton.2015.58} {\bibfield  {journal} {\bibinfo
  {journal} {Nature Photonics}\ }\textbf {\bibinfo {volume} {9}},\ \bibinfo
  {pages} {363} (\bibinfo {year} {2015})}\BibitemShut {NoStop}%
\bibitem [{\citenamefont {Lovchinsky}\ \emph {et~al.}(2016)\citenamefont
  {Lovchinsky}, \citenamefont {Sushkov}, \citenamefont {Urbach}, \citenamefont
  {de~Leon}, \citenamefont {Choi}, \citenamefont {De~Greve}, \citenamefont
  {Evans}, \citenamefont {Gertner}, \citenamefont {Bersin}, \citenamefont
  {M{\"u}ller}, \citenamefont {McGuinness}, \citenamefont {Jelezko},
  \citenamefont {Walsworth}, \citenamefont {Park},\ and\ \citenamefont
  {Lukin}}]{Lovchinsky2016}%
  \BibitemOpen
  \bibfield  {author} {\bibinfo {author} {\bibfnamefont {I.}~\bibnamefont
  {Lovchinsky}}, \bibinfo {author} {\bibfnamefont {A.~O.}\ \bibnamefont
  {Sushkov}}, \bibinfo {author} {\bibfnamefont {E.}~\bibnamefont {Urbach}},
  \bibinfo {author} {\bibfnamefont {N.~P.}\ \bibnamefont {de~Leon}}, \bibinfo
  {author} {\bibfnamefont {S.}~\bibnamefont {Choi}}, \bibinfo {author}
  {\bibfnamefont {K.}~\bibnamefont {De~Greve}}, \bibinfo {author}
  {\bibfnamefont {R.}~\bibnamefont {Evans}}, \bibinfo {author} {\bibfnamefont
  {R.}~\bibnamefont {Gertner}}, \bibinfo {author} {\bibfnamefont
  {E.}~\bibnamefont {Bersin}}, \bibinfo {author} {\bibfnamefont
  {C.}~\bibnamefont {M{\"u}ller}}, \bibinfo {author} {\bibfnamefont
  {L.}~\bibnamefont {McGuinness}}, \bibinfo {author} {\bibfnamefont
  {F.}~\bibnamefont {Jelezko}}, \bibinfo {author} {\bibfnamefont {R.~L.}\
  \bibnamefont {Walsworth}}, \bibinfo {author} {\bibfnamefont {H.}~\bibnamefont
  {Park}}, \ and\ \bibinfo {author} {\bibfnamefont {M.~D.}\ \bibnamefont
  {Lukin}},\ }\href {\doibase 10.1126/science.aad8022} {\bibfield  {journal}
  {\bibinfo  {journal} {Science}\ }\textbf {\bibinfo {volume} {351}},\ \bibinfo
  {pages} {836} (\bibinfo {year} {2016})},\ \Eprint
  {http://arxiv.org/abs/http://science.sciencemag.org/content/351/6275/836.full.pdf}
  {http://science.sciencemag.org/content/351/6275/836.full.pdf} \BibitemShut
  {NoStop}%
\bibitem [{\citenamefont {Aslam}\ \emph {et~al.}(2017)\citenamefont {Aslam},
  \citenamefont {Pfender}, \citenamefont {Neumann}, \citenamefont {Reuter},
  \citenamefont {Zappe}, \citenamefont {F{\'a}varo~de Oliveira}, \citenamefont
  {Denisenko}, \citenamefont {Sumiya}, \citenamefont {Onoda}, \citenamefont
  {Isoya},\ and\ \citenamefont {Wrachtrup}}]{Aslam2017}%
  \BibitemOpen
  \bibfield  {author} {\bibinfo {author} {\bibfnamefont {N.}~\bibnamefont
  {Aslam}}, \bibinfo {author} {\bibfnamefont {M.}~\bibnamefont {Pfender}},
  \bibinfo {author} {\bibfnamefont {P.}~\bibnamefont {Neumann}}, \bibinfo
  {author} {\bibfnamefont {R.}~\bibnamefont {Reuter}}, \bibinfo {author}
  {\bibfnamefont {A.}~\bibnamefont {Zappe}}, \bibinfo {author} {\bibfnamefont
  {F.}~\bibnamefont {F{\'a}varo~de Oliveira}}, \bibinfo {author} {\bibfnamefont
  {A.}~\bibnamefont {Denisenko}}, \bibinfo {author} {\bibfnamefont
  {H.}~\bibnamefont {Sumiya}}, \bibinfo {author} {\bibfnamefont
  {S.}~\bibnamefont {Onoda}}, \bibinfo {author} {\bibfnamefont
  {J.}~\bibnamefont {Isoya}}, \ and\ \bibinfo {author} {\bibfnamefont
  {J.}~\bibnamefont {Wrachtrup}},\ }\href {\doibase 10.1126/science.aam8697}
  {\bibfield  {journal} {\bibinfo  {journal} {Science}\ }\textbf {\bibinfo
  {volume} {357}},\ \bibinfo {pages} {67} (\bibinfo {year} {2017})}\BibitemShut
  {NoStop}%
\bibitem [{\citenamefont {Le~Sage}\ \emph {et~al.}(2013)\citenamefont
  {Le~Sage}, \citenamefont {Arai}, \citenamefont {Glenn}, \citenamefont
  {DeVience}, \citenamefont {Pham}, \citenamefont {Rahn-Lee}, \citenamefont
  {Lukin}, \citenamefont {Yacoby}, \citenamefont {Komeili},\ and\ \citenamefont
  {Walsworth}}]{LeSage2013}%
  \BibitemOpen
  \bibfield  {author} {\bibinfo {author} {\bibfnamefont {D.}~\bibnamefont
  {Le~Sage}}, \bibinfo {author} {\bibfnamefont {K.}~\bibnamefont {Arai}},
  \bibinfo {author} {\bibfnamefont {D.~R.}\ \bibnamefont {Glenn}}, \bibinfo
  {author} {\bibfnamefont {S.~J.}\ \bibnamefont {DeVience}}, \bibinfo {author}
  {\bibfnamefont {L.~M.}\ \bibnamefont {Pham}}, \bibinfo {author}
  {\bibfnamefont {L.}~\bibnamefont {Rahn-Lee}}, \bibinfo {author}
  {\bibfnamefont {M.~D.}\ \bibnamefont {Lukin}}, \bibinfo {author}
  {\bibfnamefont {A.}~\bibnamefont {Yacoby}}, \bibinfo {author} {\bibfnamefont
  {A.}~\bibnamefont {Komeili}}, \ and\ \bibinfo {author} {\bibfnamefont
  {R.~L.}\ \bibnamefont {Walsworth}},\ }\href
  {http://dx.doi.org/10.1038/nature12072} {\bibfield  {journal} {\bibinfo
  {journal} {Nature}\ }\textbf {\bibinfo {volume} {496}},\ \bibinfo {pages}
  {486} (\bibinfo {year} {2013})}\BibitemShut {NoStop}%
\bibitem [{\citenamefont {Doherty}\ \emph {et~al.}(2013)\citenamefont
  {Doherty}, \citenamefont {Manson}, \citenamefont {Delaney}, \citenamefont
  {Jelezko}, \citenamefont {Wrachtrup},\ and\ \citenamefont
  {Hollenberg}}]{Doherty2013}%
  \BibitemOpen
  \bibfield  {author} {\bibinfo {author} {\bibfnamefont {M.~W.}\ \bibnamefont
  {Doherty}}, \bibinfo {author} {\bibfnamefont {N.~B.}\ \bibnamefont {Manson}},
  \bibinfo {author} {\bibfnamefont {P.}~\bibnamefont {Delaney}}, \bibinfo
  {author} {\bibfnamefont {F.}~\bibnamefont {Jelezko}}, \bibinfo {author}
  {\bibfnamefont {J.}~\bibnamefont {Wrachtrup}}, \ and\ \bibinfo {author}
  {\bibfnamefont {L.~C.}\ \bibnamefont {Hollenberg}},\ }\href {\doibase
  http://dx.doi.org/10.1016/j.physrep.2013.02.001} {\bibfield  {journal}
  {\bibinfo  {journal} {Physics Reports}\ }\textbf {\bibinfo {volume} {528}},\
  \bibinfo {pages} {1 } (\bibinfo {year} {2013})}\BibitemShut {NoStop}%
\bibitem [{\citenamefont {Sukachev}\ \emph {et~al.}(2017)\citenamefont
  {Sukachev}, \citenamefont {Sipahigil}, \citenamefont {Nguyen}, \citenamefont
  {Bhaskar}, \citenamefont {Evans}, \citenamefont {Jelezko},\ and\
  \citenamefont {Lukin}}]{Sipahigil2017}%
  \BibitemOpen
  \bibfield  {author} {\bibinfo {author} {\bibfnamefont {D.~D.}\ \bibnamefont
  {Sukachev}}, \bibinfo {author} {\bibfnamefont {A.}~\bibnamefont {Sipahigil}},
  \bibinfo {author} {\bibfnamefont {C.~T.}\ \bibnamefont {Nguyen}}, \bibinfo
  {author} {\bibfnamefont {M.~K.}\ \bibnamefont {Bhaskar}}, \bibinfo {author}
  {\bibfnamefont {R.~E.}\ \bibnamefont {Evans}}, \bibinfo {author}
  {\bibfnamefont {F.}~\bibnamefont {Jelezko}}, \ and\ \bibinfo {author}
  {\bibfnamefont {M.~D.}\ \bibnamefont {Lukin}},\ }\href {\doibase
  10.1103/PhysRevLett.119.223602} {\bibfield  {journal} {\bibinfo  {journal}
  {Phys. Rev. Lett.}\ }\textbf {\bibinfo {volume} {119}},\ \bibinfo {pages}
  {223602} (\bibinfo {year} {2017})}\BibitemShut {NoStop}%
\bibitem [{\citenamefont {Rose}\ \emph {et~al.}(2017)\citenamefont {Rose},
  \citenamefont {Huang}, \citenamefont {Zhang}, \citenamefont {Tyryshkin},
  \citenamefont {Sangtawesin}, \citenamefont {Srinivasan}, \citenamefont
  {Loudin}, \citenamefont {Markham}, \citenamefont {Edmonds}, \citenamefont
  {Twitchen}, \citenamefont {Lyon},\ and\ \citenamefont
  {de~Leon}}]{Rose2017arxiv}%
  \BibitemOpen
  \bibfield  {author} {\bibinfo {author} {\bibfnamefont {B.~C.}\ \bibnamefont
  {Rose}}, \bibinfo {author} {\bibfnamefont {D.}~\bibnamefont {Huang}},
  \bibinfo {author} {\bibfnamefont {Z.-H.}\ \bibnamefont {Zhang}}, \bibinfo
  {author} {\bibfnamefont {A.~M.}\ \bibnamefont {Tyryshkin}}, \bibinfo {author}
  {\bibfnamefont {S.}~\bibnamefont {Sangtawesin}}, \bibinfo {author}
  {\bibfnamefont {S.}~\bibnamefont {Srinivasan}}, \bibinfo {author}
  {\bibfnamefont {L.}~\bibnamefont {Loudin}}, \bibinfo {author} {\bibfnamefont
  {M.~L.}\ \bibnamefont {Markham}}, \bibinfo {author} {\bibfnamefont {A.~M.}\
  \bibnamefont {Edmonds}}, \bibinfo {author} {\bibfnamefont {D.~J.}\
  \bibnamefont {Twitchen}}, \bibinfo {author} {\bibfnamefont {S.~A.}\
  \bibnamefont {Lyon}}, \ and\ \bibinfo {author} {\bibfnamefont {N.~P.}\
  \bibnamefont {de~Leon}},\ }\href {https://arxiv.org/abs/1706.01555} {\enquote
  {\bibinfo {title} {Observation of an environmentally insensitive solid state
  spin defect in diamond},}\ } (\bibinfo {year} {2017}),\ \bibinfo {note}
  {arXiv:1706.01555 [cond-mat.mtrl-sci]}\BibitemShut {NoStop}%
\bibitem [{\citenamefont {Christle}\ \emph {et~al.}(2015)\citenamefont
  {Christle}, \citenamefont {Falk}, \citenamefont {Andrich}, \citenamefont
  {Klimov}, \citenamefont {Hassan}, \citenamefont {Son}, \citenamefont
  {Janz\'en}, \citenamefont {Ohshima},\ and\ \citenamefont
  {Awschalom}}]{Christle2015}%
  \BibitemOpen
  \bibfield  {author} {\bibinfo {author} {\bibfnamefont {D.~J.}\ \bibnamefont
  {Christle}}, \bibinfo {author} {\bibfnamefont {A.~L.}\ \bibnamefont {Falk}},
  \bibinfo {author} {\bibfnamefont {P.}~\bibnamefont {Andrich}}, \bibinfo
  {author} {\bibfnamefont {P.~V.}\ \bibnamefont {Klimov}}, \bibinfo {author}
  {\bibfnamefont {J.~U.}\ \bibnamefont {Hassan}}, \bibinfo {author}
  {\bibfnamefont {N.}~\bibnamefont {Son}}, \bibinfo {author} {\bibfnamefont
  {E.}~\bibnamefont {Janz\'en}}, \bibinfo {author} {\bibfnamefont
  {T.}~\bibnamefont {Ohshima}}, \ and\ \bibinfo {author} {\bibfnamefont
  {D.~D.}\ \bibnamefont {Awschalom}},\ }\href
  {http://dx.doi.org/10.1038/nmat4144} {\bibfield  {journal} {\bibinfo
  {journal} {Nature Materials}\ }\textbf {\bibinfo {volume} {14}},\ \bibinfo
  {pages} {160} (\bibinfo {year} {2015})}\BibitemShut {NoStop}%
\bibitem [{\citenamefont {Widmann}\ \emph {et~al.}(2015)\citenamefont
  {Widmann}, \citenamefont {Lee}, \citenamefont {Rendler}, \citenamefont {Son},
  \citenamefont {Fedder}, \citenamefont {Paik}, \citenamefont {Yang},
  \citenamefont {Zhao}, \citenamefont {Yang}, \citenamefont {Booker},
  \citenamefont {Denisenko}, \citenamefont {Jamali}, \citenamefont
  {Momenzadeh}, \citenamefont {Gerhardt}, \citenamefont {Ohshima},
  \citenamefont {Gali}, \citenamefont {Janz\'en},\ and\ \citenamefont
  {Wrachtrup}}]{Widmann2015}%
  \BibitemOpen
  \bibfield  {author} {\bibinfo {author} {\bibfnamefont {M.}~\bibnamefont
  {Widmann}}, \bibinfo {author} {\bibfnamefont {S.-Y.}\ \bibnamefont {Lee}},
  \bibinfo {author} {\bibfnamefont {T.}~\bibnamefont {Rendler}}, \bibinfo
  {author} {\bibfnamefont {N.~T.}\ \bibnamefont {Son}}, \bibinfo {author}
  {\bibfnamefont {H.}~\bibnamefont {Fedder}}, \bibinfo {author} {\bibfnamefont
  {S.}~\bibnamefont {Paik}}, \bibinfo {author} {\bibfnamefont {L.-P.}\
  \bibnamefont {Yang}}, \bibinfo {author} {\bibfnamefont {N.}~\bibnamefont
  {Zhao}}, \bibinfo {author} {\bibfnamefont {S.}~\bibnamefont {Yang}}, \bibinfo
  {author} {\bibfnamefont {I.}~\bibnamefont {Booker}}, \bibinfo {author}
  {\bibfnamefont {A.}~\bibnamefont {Denisenko}}, \bibinfo {author}
  {\bibfnamefont {M.}~\bibnamefont {Jamali}}, \bibinfo {author} {\bibfnamefont
  {S.~A.}\ \bibnamefont {Momenzadeh}}, \bibinfo {author} {\bibfnamefont
  {I.}~\bibnamefont {Gerhardt}}, \bibinfo {author} {\bibfnamefont
  {T.}~\bibnamefont {Ohshima}}, \bibinfo {author} {\bibfnamefont
  {A.}~\bibnamefont {Gali}}, \bibinfo {author} {\bibfnamefont {E.}~\bibnamefont
  {Janz\'en}}, \ and\ \bibinfo {author} {\bibfnamefont {J.}~\bibnamefont
  {Wrachtrup}},\ }\href {http://dx.doi.org/10.1038/nmat4145} {\bibfield
  {journal} {\bibinfo  {journal} {Nature Materials}\ }\textbf {\bibinfo
  {volume} {14}},\ \bibinfo {pages} {164} (\bibinfo {year} {2015})}\BibitemShut
  {NoStop}%
\bibitem [{\citenamefont {Abobeih}\ \emph {et~al.}(2018)\citenamefont
  {Abobeih}, \citenamefont {Cramer}, \citenamefont {Bakker}, \citenamefont
  {Kalb}, \citenamefont {Twitchen}, \citenamefont {Markham},\ and\
  \citenamefont {Taminiau}}]{Abobeih2018}%
  \BibitemOpen
  \bibfield  {author} {\bibinfo {author} {\bibfnamefont {M.~H.}\ \bibnamefont
  {Abobeih}}, \bibinfo {author} {\bibfnamefont {J.}~\bibnamefont {Cramer}},
  \bibinfo {author} {\bibfnamefont {M.~A.}\ \bibnamefont {Bakker}}, \bibinfo
  {author} {\bibfnamefont {N.}~\bibnamefont {Kalb}}, \bibinfo {author}
  {\bibfnamefont {D.~J.}\ \bibnamefont {Twitchen}}, \bibinfo {author}
  {\bibfnamefont {M.}~\bibnamefont {Markham}}, \ and\ \bibinfo {author}
  {\bibfnamefont {T.~H.}\ \bibnamefont {Taminiau}},\ }\href
  {https://arxiv.org/abs/1801.01196} {\enquote {\bibinfo {title} {One-second
  coherence for a single electron spin coupled to a multi-qubit nuclear-spin
  environment},}\ } (\bibinfo {year} {2018}),\ \bibinfo {note}
  {arXiv:1801.01196 [quant-ph]}\BibitemShut {NoStop}%
\bibitem [{\citenamefont {Heremans}, \citenamefont {Yale},\ and\ \citenamefont
  {Awschalom}(2016)}]{Heremans2016}%
  \BibitemOpen
  \bibfield  {author} {\bibinfo {author} {\bibfnamefont {F.~J.}\ \bibnamefont
  {Heremans}}, \bibinfo {author} {\bibfnamefont {C.~G.}\ \bibnamefont {Yale}},
  \ and\ \bibinfo {author} {\bibfnamefont {D.~D.}\ \bibnamefont {Awschalom}},\
  }\href {\doibase 10.1109/JPROC.2016.2561274} {\bibfield  {journal} {\bibinfo
  {journal} {Proceedings of the IEEE}\ }\textbf {\bibinfo {volume} {104}},\
  \bibinfo {pages} {2009} (\bibinfo {year} {2016})}\BibitemShut {NoStop}%
\bibitem [{\citenamefont {Buckley}\ \emph {et~al.}(2010)\citenamefont
  {Buckley}, \citenamefont {Fuchs}, \citenamefont {Bassett},\ and\
  \citenamefont {Awschalom}}]{Buckley2010}%
  \BibitemOpen
  \bibfield  {author} {\bibinfo {author} {\bibfnamefont {B.~B.}\ \bibnamefont
  {Buckley}}, \bibinfo {author} {\bibfnamefont {G.~D.}\ \bibnamefont {Fuchs}},
  \bibinfo {author} {\bibfnamefont {L.~C.}\ \bibnamefont {Bassett}}, \ and\
  \bibinfo {author} {\bibfnamefont {D.~D.}\ \bibnamefont {Awschalom}},\ }\href
  {\doibase 10.1126/science.1196436} {\bibfield  {journal} {\bibinfo  {journal}
  {Science}\ }\textbf {\bibinfo {volume} {330}},\ \bibinfo {pages} {1212}
  (\bibinfo {year} {2010})}\BibitemShut {NoStop}%
\bibitem [{\citenamefont {Tran}\ \emph
  {et~al.}(2016{\natexlab{a}})\citenamefont {Tran}, \citenamefont {Bray},
  \citenamefont {Ford}, \citenamefont {Toth},\ and\ \citenamefont
  {Aharonovich}}]{Tran2016}%
  \BibitemOpen
  \bibfield  {author} {\bibinfo {author} {\bibfnamefont {T.~T.}\ \bibnamefont
  {Tran}}, \bibinfo {author} {\bibfnamefont {K.}~\bibnamefont {Bray}}, \bibinfo
  {author} {\bibfnamefont {M.~J.}\ \bibnamefont {Ford}}, \bibinfo {author}
  {\bibfnamefont {M.}~\bibnamefont {Toth}}, \ and\ \bibinfo {author}
  {\bibfnamefont {I.}~\bibnamefont {Aharonovich}},\ }\href
  {http://dx.doi.org/10.1038/nnano.2015.242} {\bibfield  {journal} {\bibinfo
  {journal} {Nature Nanotechnology}\ }\textbf {\bibinfo {volume} {11}},\
  \bibinfo {pages} {37} (\bibinfo {year} {2016}{\natexlab{a}})}\BibitemShut
  {NoStop}%
\bibitem [{\citenamefont {Tran}\ \emph
  {et~al.}(2016{\natexlab{b}})\citenamefont {Tran}, \citenamefont {Elbadawi},
  \citenamefont {Totonjian}, \citenamefont {Lobo}, \citenamefont {Grosso},
  \citenamefont {Moon}, \citenamefont {Englund}, \citenamefont {Ford},
  \citenamefont {Aharonovich},\ and\ \citenamefont {Toth}}]{Tran2016a}%
  \BibitemOpen
  \bibfield  {author} {\bibinfo {author} {\bibfnamefont {T.~T.}\ \bibnamefont
  {Tran}}, \bibinfo {author} {\bibfnamefont {C.}~\bibnamefont {Elbadawi}},
  \bibinfo {author} {\bibfnamefont {D.}~\bibnamefont {Totonjian}}, \bibinfo
  {author} {\bibfnamefont {C.~J.}\ \bibnamefont {Lobo}}, \bibinfo {author}
  {\bibfnamefont {G.}~\bibnamefont {Grosso}}, \bibinfo {author} {\bibfnamefont
  {H.}~\bibnamefont {Moon}}, \bibinfo {author} {\bibfnamefont {D.~R.}\
  \bibnamefont {Englund}}, \bibinfo {author} {\bibfnamefont {M.~J.}\
  \bibnamefont {Ford}}, \bibinfo {author} {\bibfnamefont {I.}~\bibnamefont
  {Aharonovich}}, \ and\ \bibinfo {author} {\bibfnamefont {M.}~\bibnamefont
  {Toth}},\ }\href {\doibase 10.1021/acsnano.6b03602} {\bibfield  {journal}
  {\bibinfo  {journal} {ACS Nano}\ }\textbf {\bibinfo {volume} {10}},\ \bibinfo
  {pages} {7331} (\bibinfo {year} {2016}{\natexlab{b}})}\BibitemShut {NoStop}%
\bibitem [{\citenamefont {Mart\'{\i}nez}\ \emph {et~al.}(2016)\citenamefont
  {Mart\'{\i}nez}, \citenamefont {Pelini}, \citenamefont {Waselowski},
  \citenamefont {Maze}, \citenamefont {Gil}, \citenamefont {Cassabois},\ and\
  \citenamefont {Jacques}}]{Martinez2016}%
  \BibitemOpen
  \bibfield  {author} {\bibinfo {author} {\bibfnamefont {L.~J.}\ \bibnamefont
  {Mart\'{\i}nez}}, \bibinfo {author} {\bibfnamefont {T.}~\bibnamefont
  {Pelini}}, \bibinfo {author} {\bibfnamefont {V.}~\bibnamefont {Waselowski}},
  \bibinfo {author} {\bibfnamefont {J.~R.}\ \bibnamefont {Maze}}, \bibinfo
  {author} {\bibfnamefont {B.}~\bibnamefont {Gil}}, \bibinfo {author}
  {\bibfnamefont {G.}~\bibnamefont {Cassabois}}, \ and\ \bibinfo {author}
  {\bibfnamefont {V.}~\bibnamefont {Jacques}},\ }\href {\doibase
  10.1103/PhysRevB.94.121405} {\bibfield  {journal} {\bibinfo  {journal} {Phys.
  Rev. B}\ }\textbf {\bibinfo {volume} {94}},\ \bibinfo {pages} {121405}
  (\bibinfo {year} {2016})}\BibitemShut {NoStop}%
\bibitem [{\citenamefont {Jungwirth}\ \emph {et~al.}(2016)\citenamefont
  {Jungwirth}, \citenamefont {Calderon}, \citenamefont {Ji}, \citenamefont
  {Spencer}, \citenamefont {Flatté},\ and\ \citenamefont
  {Fuchs}}]{Jungwirth2016}%
  \BibitemOpen
  \bibfield  {author} {\bibinfo {author} {\bibfnamefont {N.~R.}\ \bibnamefont
  {Jungwirth}}, \bibinfo {author} {\bibfnamefont {B.}~\bibnamefont {Calderon}},
  \bibinfo {author} {\bibfnamefont {Y.}~\bibnamefont {Ji}}, \bibinfo {author}
  {\bibfnamefont {M.~G.}\ \bibnamefont {Spencer}}, \bibinfo {author}
  {\bibfnamefont {M.~E.}\ \bibnamefont {Flatté}}, \ and\ \bibinfo {author}
  {\bibfnamefont {G.~D.}\ \bibnamefont {Fuchs}},\ }\href {\doibase
  10.1021/acs.nanolett.6b01987} {\bibfield  {journal} {\bibinfo  {journal}
  {Nano Letters}\ }\textbf {\bibinfo {volume} {16}},\ \bibinfo {pages} {6052}
  (\bibinfo {year} {2016})}\BibitemShut {NoStop}%
\bibitem [{\citenamefont {Chejanovsky}\ \emph {et~al.}(2016)\citenamefont
  {Chejanovsky}, \citenamefont {Rezai}, \citenamefont {Paolucci}, \citenamefont
  {Kim}, \citenamefont {Rendler}, \citenamefont {Rouabeh}, \citenamefont
  {Fávaro~de Oliveira}, \citenamefont {Herlinger}, \citenamefont {Denisenko},
  \citenamefont {Yang}, \citenamefont {Gerhardt}, \citenamefont {Finkler},
  \citenamefont {Smet},\ and\ \citenamefont {Wrachtrup}}]{Chejanovsky2016}%
  \BibitemOpen
  \bibfield  {author} {\bibinfo {author} {\bibfnamefont {N.}~\bibnamefont
  {Chejanovsky}}, \bibinfo {author} {\bibfnamefont {M.}~\bibnamefont {Rezai}},
  \bibinfo {author} {\bibfnamefont {F.}~\bibnamefont {Paolucci}}, \bibinfo
  {author} {\bibfnamefont {Y.}~\bibnamefont {Kim}}, \bibinfo {author}
  {\bibfnamefont {T.}~\bibnamefont {Rendler}}, \bibinfo {author} {\bibfnamefont
  {W.}~\bibnamefont {Rouabeh}}, \bibinfo {author} {\bibfnamefont
  {F.}~\bibnamefont {Fávaro~de Oliveira}}, \bibinfo {author} {\bibfnamefont
  {P.}~\bibnamefont {Herlinger}}, \bibinfo {author} {\bibfnamefont
  {A.}~\bibnamefont {Denisenko}}, \bibinfo {author} {\bibfnamefont
  {S.}~\bibnamefont {Yang}}, \bibinfo {author} {\bibfnamefont {I.}~\bibnamefont
  {Gerhardt}}, \bibinfo {author} {\bibfnamefont {A.}~\bibnamefont {Finkler}},
  \bibinfo {author} {\bibfnamefont {J.~H.}\ \bibnamefont {Smet}}, \ and\
  \bibinfo {author} {\bibfnamefont {J.}~\bibnamefont {Wrachtrup}},\ }\href
  {\doibase 10.1021/acs.nanolett.6b03268} {\bibfield  {journal} {\bibinfo
  {journal} {Nano Letters}\ }\textbf {\bibinfo {volume} {16}},\ \bibinfo
  {pages} {7037} (\bibinfo {year} {2016})}\BibitemShut {NoStop}%
\bibitem [{\citenamefont {Exarhos}\ \emph {et~al.}(2017)\citenamefont
  {Exarhos}, \citenamefont {Hopper}, \citenamefont {Grote}, \citenamefont
  {Alkauskas},\ and\ \citenamefont {Bassett}}]{Exarhos2017}%
  \BibitemOpen
  \bibfield  {author} {\bibinfo {author} {\bibfnamefont {A.~L.}\ \bibnamefont
  {Exarhos}}, \bibinfo {author} {\bibfnamefont {D.~A.}\ \bibnamefont {Hopper}},
  \bibinfo {author} {\bibfnamefont {R.~R.}\ \bibnamefont {Grote}}, \bibinfo
  {author} {\bibfnamefont {A.}~\bibnamefont {Alkauskas}}, \ and\ \bibinfo
  {author} {\bibfnamefont {L.~C.}\ \bibnamefont {Bassett}},\ }\href {\doibase
  10.1021/acsnano.7b00665} {\bibfield  {journal} {\bibinfo  {journal} {ACS
  Nano}\ }\textbf {\bibinfo {volume} {11}},\ \bibinfo {pages} {3328} (\bibinfo
  {year} {2017})}\BibitemShut {NoStop}%
\bibitem [{\citenamefont {Jungwirth}\ and\ \citenamefont
  {Fuchs}(2017)}]{Jungwirth2017}%
  \BibitemOpen
  \bibfield  {author} {\bibinfo {author} {\bibfnamefont {N.~R.}\ \bibnamefont
  {Jungwirth}}\ and\ \bibinfo {author} {\bibfnamefont {G.~D.}\ \bibnamefont
  {Fuchs}},\ }\href {\doibase 10.1103/PhysRevLett.119.057401} {\bibfield
  {journal} {\bibinfo  {journal} {Phys. Rev. Lett.}\ }\textbf {\bibinfo
  {volume} {119}},\ \bibinfo {pages} {057401} (\bibinfo {year}
  {2017})}\BibitemShut {NoStop}%
\bibitem [{\citenamefont {Aharonovich}, \citenamefont {Englund},\ and\
  \citenamefont {Toth}(2016)}]{Aharonovich2016}%
  \BibitemOpen
  \bibfield  {author} {\bibinfo {author} {\bibfnamefont {I.}~\bibnamefont
  {Aharonovich}}, \bibinfo {author} {\bibfnamefont {D.}~\bibnamefont
  {Englund}}, \ and\ \bibinfo {author} {\bibfnamefont {M.}~\bibnamefont
  {Toth}},\ }\href {http://dx.doi.org/10.1038/nphoton.2016.186} {\bibfield
  {journal} {\bibinfo  {journal} {Nature Photonics}\ }\textbf {\bibinfo
  {volume} {10}},\ \bibinfo {pages} {631} (\bibinfo {year} {2016})}\BibitemShut
  {NoStop}%
\bibitem [{\citenamefont {Zaitsev}(2001)}]{Zaitsev2001}%
  \BibitemOpen
  \bibfield  {author} {\bibinfo {author} {\bibfnamefont {A.~M.}\ \bibnamefont
  {Zaitsev}},\ }\href@noop {} {\emph {\bibinfo {title} {Optical properties of
  diamond: a data handbook}}}\ (\bibinfo  {publisher} {Springer-Verlag},\
  \bibinfo {year} {2001})\BibitemShut {NoStop}%
\bibitem [{\citenamefont {Geist}\ and\ \citenamefont
  {Römelt}(1964)}]{Geist1964}%
  \BibitemOpen
  \bibfield  {author} {\bibinfo {author} {\bibfnamefont {D.}~\bibnamefont
  {Geist}}\ and\ \bibinfo {author} {\bibfnamefont {G.}~\bibnamefont
  {Römelt}},\ }\href {\doibase https://doi.org/10.1016/0038-1098(64)90404-1}
  {\bibfield  {journal} {\bibinfo  {journal} {Solid State Communications}\
  }\textbf {\bibinfo {volume} {2}},\ \bibinfo {pages} {149} (\bibinfo {year}
  {1964})}\BibitemShut {NoStop}%
\bibitem [{\citenamefont {Fanciulli}(1997)}]{Fanciulli1997}%
  \BibitemOpen
  \bibfield  {author} {\bibinfo {author} {\bibfnamefont {M.}~\bibnamefont
  {Fanciulli}},\ }\href {\doibase 10.1080/01418639708241100} {\bibfield
  {journal} {\bibinfo  {journal} {Philosophical Magazine Part B}\ }\textbf
  {\bibinfo {volume} {76}},\ \bibinfo {pages} {363} (\bibinfo {year}
  {1997})}\BibitemShut {NoStop}%
\bibitem [{\citenamefont {Abdi}\ \emph
  {et~al.}(2017{\natexlab{a}})\citenamefont {Abdi}, \citenamefont {Hwang},
  \citenamefont {Aghtar},\ and\ \citenamefont {Plenio}}]{Abdi2017}%
  \BibitemOpen
  \bibfield  {author} {\bibinfo {author} {\bibfnamefont {M.}~\bibnamefont
  {Abdi}}, \bibinfo {author} {\bibfnamefont {M.-J.}\ \bibnamefont {Hwang}},
  \bibinfo {author} {\bibfnamefont {M.}~\bibnamefont {Aghtar}}, \ and\ \bibinfo
  {author} {\bibfnamefont {M.~B.}\ \bibnamefont {Plenio}},\ }\href {\doibase
  10.1103/PhysRevLett.119.233602} {\bibfield  {journal} {\bibinfo  {journal}
  {Phys. Rev. Lett.}\ }\textbf {\bibinfo {volume} {119}},\ \bibinfo {pages}
  {233602} (\bibinfo {year} {2017}{\natexlab{a}})}\BibitemShut {NoStop}%
\bibitem [{\citenamefont {Li}\ \emph {et~al.}(2017)\citenamefont {Li},
  \citenamefont {Shepard}, \citenamefont {Cupo}, \citenamefont {Camporeale},
  \citenamefont {Shayan}, \citenamefont {Luo}, \citenamefont {Meunier},\ and\
  \citenamefont {Strauf}}]{Li2017}%
  \BibitemOpen
  \bibfield  {author} {\bibinfo {author} {\bibfnamefont {X.}~\bibnamefont
  {Li}}, \bibinfo {author} {\bibfnamefont {G.~D.}\ \bibnamefont {Shepard}},
  \bibinfo {author} {\bibfnamefont {A.}~\bibnamefont {Cupo}}, \bibinfo {author}
  {\bibfnamefont {N.}~\bibnamefont {Camporeale}}, \bibinfo {author}
  {\bibfnamefont {K.}~\bibnamefont {Shayan}}, \bibinfo {author} {\bibfnamefont
  {Y.}~\bibnamefont {Luo}}, \bibinfo {author} {\bibfnamefont {V.}~\bibnamefont
  {Meunier}}, \ and\ \bibinfo {author} {\bibfnamefont {S.}~\bibnamefont
  {Strauf}},\ }\href {\doibase 10.1021/acsnano.7b00638} {\bibfield  {journal}
  {\bibinfo  {journal} {ACS Nano}\ }\textbf {\bibinfo {volume} {11}},\ \bibinfo
  {pages} {6652} (\bibinfo {year} {2017})}\BibitemShut {NoStop}%
\bibitem [{\citenamefont {Koperski}, \citenamefont {Nogajewski},\ and\
  \citenamefont {Potemski}(2018)}]{Koperski2017}%
  \BibitemOpen
  \bibfield  {author} {\bibinfo {author} {\bibfnamefont {M.}~\bibnamefont
  {Koperski}}, \bibinfo {author} {\bibfnamefont {K.}~\bibnamefont
  {Nogajewski}}, \ and\ \bibinfo {author} {\bibfnamefont {M.}~\bibnamefont
  {Potemski}},\ }\href {\doibase https://doi.org/10.1016/j.optcom.2017.10.083}
  {\bibfield  {journal} {\bibinfo  {journal} {Optics Communications}\ }\textbf
  {\bibinfo {volume} {411}},\ \bibinfo {pages} {158 } (\bibinfo {year}
  {2018})}\BibitemShut {NoStop}%
\bibitem [{\citenamefont {Tawfik}\ \emph {et~al.}(2017)\citenamefont {Tawfik},
  \citenamefont {Ali}, \citenamefont {Fronzi}, \citenamefont {Mianinia},
  \citenamefont {Tran}, \citenamefont {Stampfl}, \citenamefont {Aharonovich},
  \citenamefont {Toth},\ and\ \citenamefont {Ford}}]{Tawfik2017}%
  \BibitemOpen
  \bibfield  {author} {\bibinfo {author} {\bibfnamefont {S.~A.}\ \bibnamefont
  {Tawfik}}, \bibinfo {author} {\bibfnamefont {S.}~\bibnamefont {Ali}},
  \bibinfo {author} {\bibfnamefont {M.}~\bibnamefont {Fronzi}}, \bibinfo
  {author} {\bibfnamefont {M.}~\bibnamefont {Mianinia}}, \bibinfo {author}
  {\bibfnamefont {T.~T.}\ \bibnamefont {Tran}}, \bibinfo {author}
  {\bibfnamefont {C.}~\bibnamefont {Stampfl}}, \bibinfo {author} {\bibfnamefont
  {I.}~\bibnamefont {Aharonovich}}, \bibinfo {author} {\bibfnamefont
  {M.}~\bibnamefont {Toth}}, \ and\ \bibinfo {author} {\bibfnamefont {M.~J.}\
  \bibnamefont {Ford}},\ }\href {\doibase 10.1039/C7NR04270A} {\bibfield
  {journal} {\bibinfo  {journal} {Nanoscale}\ }\textbf {\bibinfo {volume}
  {9}},\ \bibinfo {pages} {13575} (\bibinfo {year} {2017})}\BibitemShut
  {NoStop}%
\bibitem [{\citenamefont {Wu}\ \emph {et~al.}(2017)\citenamefont {Wu},
  \citenamefont {Galatas}, \citenamefont {Sundararaman}, \citenamefont
  {Rocca},\ and\ \citenamefont {Ping}}]{Wu2017}%
  \BibitemOpen
  \bibfield  {author} {\bibinfo {author} {\bibfnamefont {F.}~\bibnamefont
  {Wu}}, \bibinfo {author} {\bibfnamefont {A.}~\bibnamefont {Galatas}},
  \bibinfo {author} {\bibfnamefont {R.}~\bibnamefont {Sundararaman}}, \bibinfo
  {author} {\bibfnamefont {D.}~\bibnamefont {Rocca}}, \ and\ \bibinfo {author}
  {\bibfnamefont {Y.}~\bibnamefont {Ping}},\ }\href {\doibase
  10.1103/PhysRevMaterials.1.071001} {\bibfield  {journal} {\bibinfo  {journal}
  {Phys. Rev. Materials}\ }\textbf {\bibinfo {volume} {1}},\ \bibinfo {pages}
  {071001} (\bibinfo {year} {2017})}\BibitemShut {NoStop}%
\bibitem [{\citenamefont {Abdi}\ \emph
  {et~al.}(2017{\natexlab{b}})\citenamefont {Abdi}, \citenamefont {Chou},
  \citenamefont {Gali},\ and\ \citenamefont {Plenio}}]{Abdi2017_arXiv}%
  \BibitemOpen
  \bibfield  {author} {\bibinfo {author} {\bibfnamefont {M.}~\bibnamefont
  {Abdi}}, \bibinfo {author} {\bibfnamefont {J.-P.}\ \bibnamefont {Chou}},
  \bibinfo {author} {\bibfnamefont {A.}~\bibnamefont {Gali}}, \ and\ \bibinfo
  {author} {\bibfnamefont {M.~B.}\ \bibnamefont {Plenio}},\ }\href
  {https://arxiv.org/abs/1709.05414} {\bibfield  {journal} {\bibinfo  {journal}
  {arXiv:1709.05414 [cond-mat.mes-hall]}\ } (\bibinfo {year}
  {2017}{\natexlab{b}})}\BibitemShut {NoStop}%
\bibitem [{\citenamefont {van Oort}, \citenamefont {Manson},\ and\
  \citenamefont {Glasbeek}(1988)}]{Oort1988}%
  \BibitemOpen
  \bibfield  {author} {\bibinfo {author} {\bibfnamefont {E.}~\bibnamefont {van
  Oort}}, \bibinfo {author} {\bibfnamefont {N.~B.}\ \bibnamefont {Manson}}, \
  and\ \bibinfo {author} {\bibfnamefont {M.}~\bibnamefont {Glasbeek}},\ }\href
  {http://stacks.iop.org/0022-3719/21/i=23/a=020} {\bibfield  {journal}
  {\bibinfo  {journal} {Journal of Physics C: Solid State Physics}\ }\textbf
  {\bibinfo {volume} {21}},\ \bibinfo {pages} {4385} (\bibinfo {year}
  {1988})}\BibitemShut {NoStop}%
\bibitem [{SI()}]{SI}%
  \BibitemOpen
  \href@noop {} {}\bibinfo {note} {See the supplemental information online for
  further details.}\BibitemShut {Stop}%
\bibitem [{\citenamefont {Epstein}\ \emph {et~al.}(2005)\citenamefont
  {Epstein}, \citenamefont {Mendoza}, \citenamefont {Kato},\ and\ \citenamefont
  {Awschalom}}]{Epstein2005}%
  \BibitemOpen
  \bibfield  {author} {\bibinfo {author} {\bibfnamefont {R.~J.}\ \bibnamefont
  {Epstein}}, \bibinfo {author} {\bibfnamefont {F.~M.}\ \bibnamefont
  {Mendoza}}, \bibinfo {author} {\bibfnamefont {Y.~K.}\ \bibnamefont {Kato}}, \
  and\ \bibinfo {author} {\bibfnamefont {D.~D.}\ \bibnamefont {Awschalom}},\
  }\href {http://dx.doi.org/10.1038/nphys141} {\bibfield  {journal} {\bibinfo
  {journal} {Nature Physics}\ }\textbf {\bibinfo {volume} {1}},\ \bibinfo
  {pages} {94} (\bibinfo {year} {2005})}\BibitemShut {NoStop}%
\bibitem [{\citenamefont {Hong}\ \emph {et~al.}(2017)\citenamefont {Hong},
  \citenamefont {Jin}, \citenamefont {Yuan},\ and\ \citenamefont
  {Zhang}}]{Hong2017_defectsin2dreview}%
  \BibitemOpen
  \bibfield  {author} {\bibinfo {author} {\bibfnamefont {J.}~\bibnamefont
  {Hong}}, \bibinfo {author} {\bibfnamefont {C.}~\bibnamefont {Jin}}, \bibinfo
  {author} {\bibfnamefont {J.}~\bibnamefont {Yuan}}, \ and\ \bibinfo {author}
  {\bibfnamefont {Z.}~\bibnamefont {Zhang}},\ }\href {\doibase
  10.1002/adma.201606434} {\bibfield  {journal} {\bibinfo  {journal} {Advanced
  Materials}\ }\textbf {\bibinfo {volume} {29}},\ \bibinfo {pages} {1606434}
  (\bibinfo {year} {2017})}\BibitemShut {NoStop}%
\bibitem [{\citenamefont {Bassett}\ \emph {et~al.}(2014)\citenamefont
  {Bassett}, \citenamefont {Heremans}, \citenamefont {Christle}, \citenamefont
  {Yale}, \citenamefont {Burkard}, \citenamefont {Buckley},\ and\ \citenamefont
  {Awschalom}}]{Bassett2014}%
  \BibitemOpen
  \bibfield  {author} {\bibinfo {author} {\bibfnamefont {L.~C.}\ \bibnamefont
  {Bassett}}, \bibinfo {author} {\bibfnamefont {F.~J.}\ \bibnamefont
  {Heremans}}, \bibinfo {author} {\bibfnamefont {D.~J.}\ \bibnamefont
  {Christle}}, \bibinfo {author} {\bibfnamefont {C.~G.}\ \bibnamefont {Yale}},
  \bibinfo {author} {\bibfnamefont {G.}~\bibnamefont {Burkard}}, \bibinfo
  {author} {\bibfnamefont {B.~B.}\ \bibnamefont {Buckley}}, \ and\ \bibinfo
  {author} {\bibfnamefont {D.~D.}\ \bibnamefont {Awschalom}},\ }\href {\doibase
  10.1126/science.1255541} {\bibfield  {journal} {\bibinfo  {journal}
  {Science}\ }\textbf {\bibinfo {volume} {345}},\ \bibinfo {pages} {1333}
  (\bibinfo {year} {2014})}\BibitemShut {NoStop}%
\bibitem [{\citenamefont {Chen}, \citenamefont {MacQuarrie},\ and\
  \citenamefont {Fuchs}(2018)}]{Chen2018}%
  \BibitemOpen
  \bibfield  {author} {\bibinfo {author} {\bibfnamefont {H.~Y.}\ \bibnamefont
  {Chen}}, \bibinfo {author} {\bibfnamefont {E.~R.}\ \bibnamefont
  {MacQuarrie}}, \ and\ \bibinfo {author} {\bibfnamefont {G.~D.}\ \bibnamefont
  {Fuchs}},\ }\href {\doibase 10.1103/PhysRevLett.120.167401} {\bibfield
  {journal} {\bibinfo  {journal} {Phys. Rev. Lett.}\ }\textbf {\bibinfo
  {volume} {120}},\ \bibinfo {pages} {167401} (\bibinfo {year}
  {2018})}\BibitemShut {NoStop}%
\bibitem [{\citenamefont {Lovchinsky}\ \emph {et~al.}(2017)\citenamefont
  {Lovchinsky}, \citenamefont {Sanchez-Yamagishi}, \citenamefont {Urbach},
  \citenamefont {Choi}, \citenamefont {Fang}, \citenamefont {Andersen},
  \citenamefont {Watanabe}, \citenamefont {Taniguchi}, \citenamefont
  {Bylinskii}, \citenamefont {Kaxiras}, \citenamefont {Kim}, \citenamefont
  {Park},\ and\ \citenamefont {Lukin}}]{Lovchinsky2017}%
  \BibitemOpen
  \bibfield  {author} {\bibinfo {author} {\bibfnamefont {I.}~\bibnamefont
  {Lovchinsky}}, \bibinfo {author} {\bibfnamefont {J.~D.}\ \bibnamefont
  {Sanchez-Yamagishi}}, \bibinfo {author} {\bibfnamefont {E.~K.}\ \bibnamefont
  {Urbach}}, \bibinfo {author} {\bibfnamefont {S.}~\bibnamefont {Choi}},
  \bibinfo {author} {\bibfnamefont {S.}~\bibnamefont {Fang}}, \bibinfo {author}
  {\bibfnamefont {T.~I.}\ \bibnamefont {Andersen}}, \bibinfo {author}
  {\bibfnamefont {K.}~\bibnamefont {Watanabe}}, \bibinfo {author}
  {\bibfnamefont {T.}~\bibnamefont {Taniguchi}}, \bibinfo {author}
  {\bibfnamefont {A.}~\bibnamefont {Bylinskii}}, \bibinfo {author}
  {\bibfnamefont {E.}~\bibnamefont {Kaxiras}}, \bibinfo {author} {\bibfnamefont
  {P.}~\bibnamefont {Kim}}, \bibinfo {author} {\bibfnamefont {H.}~\bibnamefont
  {Park}}, \ and\ \bibinfo {author} {\bibfnamefont {M.~D.}\ \bibnamefont
  {Lukin}},\ }\href
  {http://science.sciencemag.org/content/early/2017/01/18/science.aal2538}
  {\bibfield  {journal} {\bibinfo  {journal} {Science}\ } (\bibinfo {year}
  {2017})}\BibitemShut {NoStop}%
\bibitem [{\citenamefont {Britton}\ \emph {et~al.}(2012)\citenamefont
  {Britton}, \citenamefont {Sawyer}, \citenamefont {Keith}, \citenamefont
  {Wang}, \citenamefont {Freericks}, \citenamefont {Uys}, \citenamefont
  {Biercuk},\ and\ \citenamefont {Bollinger}}]{Britton2012}%
  \BibitemOpen
  \bibfield  {author} {\bibinfo {author} {\bibfnamefont {J.~W.}\ \bibnamefont
  {Britton}}, \bibinfo {author} {\bibfnamefont {B.~C.}\ \bibnamefont {Sawyer}},
  \bibinfo {author} {\bibfnamefont {A.~C.}\ \bibnamefont {Keith}}, \bibinfo
  {author} {\bibfnamefont {C.-C.~J.}\ \bibnamefont {Wang}}, \bibinfo {author}
  {\bibfnamefont {J.~K.}\ \bibnamefont {Freericks}}, \bibinfo {author}
  {\bibfnamefont {H.}~\bibnamefont {Uys}}, \bibinfo {author} {\bibfnamefont
  {M.~J.}\ \bibnamefont {Biercuk}}, \ and\ \bibinfo {author} {\bibfnamefont
  {J.~J.}\ \bibnamefont {Bollinger}},\ }\href
  {http://dx.doi.org/10.1038/nature10981} {\bibfield  {journal} {\bibinfo
  {journal} {Nature}\ }\textbf {\bibinfo {volume} {484}},\ \bibinfo {pages}
  {489} (\bibinfo {year} {2012})}\BibitemShut {NoStop}%
\bibitem [{\citenamefont {Cai}\ \emph {et~al.}(2013)\citenamefont {Cai},
  \citenamefont {Retzker}, \citenamefont {Jelezko},\ and\ \citenamefont
  {Plenio}}]{Cai2013}%
  \BibitemOpen
  \bibfield  {author} {\bibinfo {author} {\bibfnamefont {J.}~\bibnamefont
  {Cai}}, \bibinfo {author} {\bibfnamefont {A.}~\bibnamefont {Retzker}},
  \bibinfo {author} {\bibfnamefont {F.}~\bibnamefont {Jelezko}}, \ and\
  \bibinfo {author} {\bibfnamefont {M.~B.}\ \bibnamefont {Plenio}},\ }\href
  {http://dx.doi.org/10.1038/nphys2519} {\bibfield  {journal} {\bibinfo
  {journal} {Nature Physics}\ }\textbf {\bibinfo {volume} {9}},\ \bibinfo
  {pages} {168} (\bibinfo {year} {2013})}\BibitemShut {NoStop}%
\bibitem [{\citenamefont {Pelliccione}\ \emph {et~al.}(2016)\citenamefont
  {Pelliccione}, \citenamefont {Jenkins}, \citenamefont {Ovartchaiyapong},
  \citenamefont {Reetz}, \citenamefont {Emmanouilidou}, \citenamefont {Ni},\
  and\ \citenamefont {Bleszynski~Jayich}}]{Pelliccione2016}%
  \BibitemOpen
  \bibfield  {author} {\bibinfo {author} {\bibfnamefont {M.}~\bibnamefont
  {Pelliccione}}, \bibinfo {author} {\bibfnamefont {A.}~\bibnamefont
  {Jenkins}}, \bibinfo {author} {\bibfnamefont {P.}~\bibnamefont
  {Ovartchaiyapong}}, \bibinfo {author} {\bibfnamefont {C.}~\bibnamefont
  {Reetz}}, \bibinfo {author} {\bibfnamefont {E.}~\bibnamefont
  {Emmanouilidou}}, \bibinfo {author} {\bibfnamefont {N.}~\bibnamefont {Ni}}, \
  and\ \bibinfo {author} {\bibfnamefont {A.~C.}\ \bibnamefont
  {Bleszynski~Jayich}},\ }\href {http://dx.doi.org/10.1038/nnano.2016.68}
  {\bibfield  {journal} {\bibinfo  {journal} {Nature Nanotechnology}\ }\textbf
  {\bibinfo {volume} {11}},\ \bibinfo {pages} {700} (\bibinfo {year}
  {2016})}\BibitemShut {NoStop}%
\bibitem [{\citenamefont {Gross}\ \emph {et~al.}(2017)\citenamefont {Gross},
  \citenamefont {Akhtar}, \citenamefont {Garcia}, \citenamefont {Martínez},
  \citenamefont {Chouaieb}, \citenamefont {Garcia}, \citenamefont
  {Carrétéro}, \citenamefont {Barthélémy}, \citenamefont {Appel},
  \citenamefont {Maletinsky}, \citenamefont {Kim}, \citenamefont {Chauleau},
  \citenamefont {Jaouen}, \citenamefont {Viret}, \citenamefont {Bibes},
  \citenamefont {Fusil},\ and\ \citenamefont {Jacques}}]{Gross2017}%
  \BibitemOpen
  \bibfield  {author} {\bibinfo {author} {\bibfnamefont {I.}~\bibnamefont
  {Gross}}, \bibinfo {author} {\bibfnamefont {W.}~\bibnamefont {Akhtar}},
  \bibinfo {author} {\bibfnamefont {V.}~\bibnamefont {Garcia}}, \bibinfo
  {author} {\bibfnamefont {L.~J.}\ \bibnamefont {Martínez}}, \bibinfo {author}
  {\bibfnamefont {S.}~\bibnamefont {Chouaieb}}, \bibinfo {author}
  {\bibfnamefont {K.}~\bibnamefont {Garcia}}, \bibinfo {author} {\bibfnamefont
  {C.}~\bibnamefont {Carrétéro}}, \bibinfo {author} {\bibfnamefont
  {A.}~\bibnamefont {Barthélémy}}, \bibinfo {author} {\bibfnamefont
  {P.}~\bibnamefont {Appel}}, \bibinfo {author} {\bibfnamefont
  {P.}~\bibnamefont {Maletinsky}}, \bibinfo {author} {\bibfnamefont {J.-V.}\
  \bibnamefont {Kim}}, \bibinfo {author} {\bibfnamefont {J.~Y.}\ \bibnamefont
  {Chauleau}}, \bibinfo {author} {\bibfnamefont {N.}~\bibnamefont {Jaouen}},
  \bibinfo {author} {\bibfnamefont {M.}~\bibnamefont {Viret}}, \bibinfo
  {author} {\bibfnamefont {M.}~\bibnamefont {Bibes}}, \bibinfo {author}
  {\bibfnamefont {S.}~\bibnamefont {Fusil}}, \ and\ \bibinfo {author}
  {\bibfnamefont {V.}~\bibnamefont {Jacques}},\ }\href
  {http://dx.doi.org/10.1038/nature23656} {\bibfield  {journal} {\bibinfo
  {journal} {Nature}\ }\textbf {\bibinfo {volume} {549}},\ \bibinfo {pages}
  {252} (\bibinfo {year} {2017})}\BibitemShut {NoStop}%
\bibitem [{\citenamefont {Lee}\ \emph {et~al.}(2013)\citenamefont {Lee},
  \citenamefont {Widmann}, \citenamefont {Rendler}, \citenamefont {Doherty},
  \citenamefont {Babinec}, \citenamefont {Yang}, \citenamefont {Eyer},
  \citenamefont {Siyushev}, \citenamefont {Hausmann}, \citenamefont {Loncar},
  \citenamefont {Bodrog}, \citenamefont {Gali}, \citenamefont {Manson},
  \citenamefont {Fedder},\ and\ \citenamefont {Wrachtrup}}]{Lee2013}%
  \BibitemOpen
  \bibfield  {author} {\bibinfo {author} {\bibfnamefont {S.-Y.}\ \bibnamefont
  {Lee}}, \bibinfo {author} {\bibfnamefont {M.}~\bibnamefont {Widmann}},
  \bibinfo {author} {\bibfnamefont {T.}~\bibnamefont {Rendler}}, \bibinfo
  {author} {\bibfnamefont {M.~W.}\ \bibnamefont {Doherty}}, \bibinfo {author}
  {\bibfnamefont {T.~M.}\ \bibnamefont {Babinec}}, \bibinfo {author}
  {\bibfnamefont {S.}~\bibnamefont {Yang}}, \bibinfo {author} {\bibfnamefont
  {M.}~\bibnamefont {Eyer}}, \bibinfo {author} {\bibfnamefont {P.}~\bibnamefont
  {Siyushev}}, \bibinfo {author} {\bibfnamefont {B.~J.~M.}\ \bibnamefont
  {Hausmann}}, \bibinfo {author} {\bibfnamefont {M.}~\bibnamefont {Loncar}},
  \bibinfo {author} {\bibfnamefont {Z.}~\bibnamefont {Bodrog}}, \bibinfo
  {author} {\bibfnamefont {A.}~\bibnamefont {Gali}}, \bibinfo {author}
  {\bibfnamefont {N.~B.}\ \bibnamefont {Manson}}, \bibinfo {author}
  {\bibfnamefont {H.}~\bibnamefont {Fedder}}, \ and\ \bibinfo {author}
  {\bibfnamefont {J.}~\bibnamefont {Wrachtrup}},\ }\href
  {http://dx.doi.org/10.1038/nnano.2013.104} {\bibfield  {journal} {\bibinfo
  {journal} {Nat Nano}\ }\textbf {\bibinfo {volume} {8}},\ \bibinfo {pages}
  {487} (\bibinfo {year} {2013})}\BibitemShut {NoStop}%
\bibitem [{\citenamefont {Doherty}\ \emph {et~al.}(2016)\citenamefont
  {Doherty}, \citenamefont {Meriles}, \citenamefont {Alkauskas}, \citenamefont
  {Fedder}, \citenamefont {Sellars},\ and\ \citenamefont
  {Manson}}]{Doherty2016}%
  \BibitemOpen
  \bibfield  {author} {\bibinfo {author} {\bibfnamefont {M.~W.}\ \bibnamefont
  {Doherty}}, \bibinfo {author} {\bibfnamefont {C.~A.}\ \bibnamefont
  {Meriles}}, \bibinfo {author} {\bibfnamefont {A.}~\bibnamefont {Alkauskas}},
  \bibinfo {author} {\bibfnamefont {H.}~\bibnamefont {Fedder}}, \bibinfo
  {author} {\bibfnamefont {M.~J.}\ \bibnamefont {Sellars}}, \ and\ \bibinfo
  {author} {\bibfnamefont {N.~B.}\ \bibnamefont {Manson}},\ }\href {\doibase
  10.1103/PhysRevX.6.041035} {\bibfield  {journal} {\bibinfo  {journal} {Phys.
  Rev. X}\ }\textbf {\bibinfo {volume} {6}},\ \bibinfo {pages} {041035}
  (\bibinfo {year} {2016})}\BibitemShut {NoStop}%
\bibitem [{\citenamefont {Luo}\ \emph {et~al.}(2017)\citenamefont {Luo},
  \citenamefont {Xu}, \citenamefont {Zhu}, \citenamefont {Wu}, \citenamefont
  {McCormick}, \citenamefont {Zhan}, \citenamefont {Neupane},\ and\
  \citenamefont {Kawakami}}]{Luo2017_mos2graphenespintransfer}%
  \BibitemOpen
  \bibfield  {author} {\bibinfo {author} {\bibfnamefont {Y.~K.}\ \bibnamefont
  {Luo}}, \bibinfo {author} {\bibfnamefont {J.}~\bibnamefont {Xu}}, \bibinfo
  {author} {\bibfnamefont {T.}~\bibnamefont {Zhu}}, \bibinfo {author}
  {\bibfnamefont {G.}~\bibnamefont {Wu}}, \bibinfo {author} {\bibfnamefont
  {E.~J.}\ \bibnamefont {McCormick}}, \bibinfo {author} {\bibfnamefont
  {W.}~\bibnamefont {Zhan}}, \bibinfo {author} {\bibfnamefont {M.~R.}\
  \bibnamefont {Neupane}}, \ and\ \bibinfo {author} {\bibfnamefont {R.~K.}\
  \bibnamefont {Kawakami}},\ }\href
  {http://dx.doi.org/10.1021/acs.nanolett.7b01393} {\bibfield  {journal}
  {\bibinfo  {journal} {Nano Letters}\ }\textbf {\bibinfo {volume} {17}},\
  \bibinfo {pages} {3877} (\bibinfo {year} {2017})}\BibitemShut {NoStop}%
\bibitem [{\citenamefont {Laurence}, \citenamefont {Fore},\ and\ \citenamefont
  {Huser}(2006)}]{Laurence2006}%
  \BibitemOpen
  \bibfield  {author} {\bibinfo {author} {\bibfnamefont {T.~A.}\ \bibnamefont
  {Laurence}}, \bibinfo {author} {\bibfnamefont {S.}~\bibnamefont {Fore}}, \
  and\ \bibinfo {author} {\bibfnamefont {T.}~\bibnamefont {Huser}},\ }\href
  {\doibase 10.1364/OL.31.000829} {\bibfield  {journal} {\bibinfo  {journal}
  {Opt. Lett.}\ }\textbf {\bibinfo {volume} {31}},\ \bibinfo {pages} {829}
  (\bibinfo {year} {2006})}\BibitemShut {NoStop}%
\bibitem [{\citenamefont {Brouri}\ \emph {et~al.}(2000)\citenamefont {Brouri},
  \citenamefont {Beveratos}, \citenamefont {Poizat},\ and\ \citenamefont
  {Grangier}}]{Brouri2000}%
  \BibitemOpen
  \bibfield  {author} {\bibinfo {author} {\bibfnamefont {R.}~\bibnamefont
  {Brouri}}, \bibinfo {author} {\bibfnamefont {A.}~\bibnamefont {Beveratos}},
  \bibinfo {author} {\bibfnamefont {J.-P.}\ \bibnamefont {Poizat}}, \ and\
  \bibinfo {author} {\bibfnamefont {P.}~\bibnamefont {Grangier}},\ }\href
  {\doibase 10.1364/OL.25.001294} {\bibfield  {journal} {\bibinfo  {journal}
  {Opt. Lett.}\ }\textbf {\bibinfo {volume} {25}},\ \bibinfo {pages} {1294}
  (\bibinfo {year} {2000})}\BibitemShut {NoStop}%
\end{thebibliography}%


\begin{thebibliography}{12}%
\makeatletter
\providecommand \@ifxundefined [1]{%
 \@ifx{#1\undefined}
}%
\providecommand \@ifnum [1]{%
 \ifnum #1\expandafter \@firstoftwo
 \else \expandafter \@secondoftwo
 \fi
}%
\providecommand \@ifx [1]{%
 \ifx #1\expandafter \@firstoftwo
 \else \expandafter \@secondoftwo
 \fi
}%
\providecommand \natexlab [1]{#1}%
\providecommand \enquote  [1]{``#1''}%
\providecommand \bibnamefont  [1]{#1}%
\providecommand \bibfnamefont [1]{#1}%
\providecommand \citenamefont [1]{#1}%
\providecommand \href@noop [0]{\@secondoftwo}%
\providecommand \href [0]{\begingroup \@sanitize@url \@href}%
\providecommand \@href[1]{\@@startlink{#1}\@@href}%
\providecommand \@@href[1]{\endgroup#1\@@endlink}%
\providecommand \@sanitize@url [0]{\catcode `\\12\catcode `\$12\catcode
  `\&12\catcode `\#12\catcode `\^12\catcode `\_12\catcode `\%12\relax}%
\providecommand \@@startlink[1]{}%
\providecommand \@@endlink[0]{}%
\providecommand \url  [0]{\begingroup\@sanitize@url \@url }%
\providecommand \@url [1]{\endgroup\@href {#1}{\urlprefix }}%
\providecommand \urlprefix  [0]{URL }%
\providecommand \Eprint [0]{\href }%
\providecommand \doibase [0]{http://dx.doi.org/}%
\providecommand \selectlanguage [0]{\@gobble}%
\providecommand \bibinfo  [0]{\@secondoftwo}%
\providecommand \bibfield  [0]{\@secondoftwo}%
\providecommand \translation [1]{[#1]}%
\providecommand \BibitemOpen [0]{}%
\providecommand \bibitemStop [0]{}%
\providecommand \bibitemNoStop [0]{.\EOS\space}%
\providecommand \EOS [0]{\spacefactor3000\relax}%
\providecommand \BibitemShut  [1]{\csname bibitem#1\endcsname}%
\let\auto@bib@innerbib\@empty
\bibitem [{\citenamefont {Exarhos}\ \emph {et~al.}(2017)\citenamefont
  {Exarhos}, \citenamefont {Hopper}, \citenamefont {Grote}, \citenamefont
  {Alkauskas},\ and\ \citenamefont {Bassett}}]{Exarhos2017}%
  \BibitemOpen
  \bibfield  {author} {\bibinfo {author} {\bibfnamefont {A.~L.}\ \bibnamefont
  {Exarhos}}, \bibinfo {author} {\bibfnamefont {D.~A.}\ \bibnamefont {Hopper}},
  \bibinfo {author} {\bibfnamefont {R.~R.}\ \bibnamefont {Grote}}, \bibinfo
  {author} {\bibfnamefont {A.}~\bibnamefont {Alkauskas}}, \ and\ \bibinfo
  {author} {\bibfnamefont {L.~C.}\ \bibnamefont {Bassett}},\ }\href {\doibase
  10.1021/acsnano.7b00665} {\bibfield  {journal} {\bibinfo  {journal} {ACS
  Nano}\ }\textbf {\bibinfo {volume} {11}},\ \bibinfo {pages} {3328} (\bibinfo
  {year} {2017})}\BibitemShut {NoStop}%
\bibitem [{\citenamefont {Stoneham}(1975)}]{Stoneham1975}%
  \BibitemOpen
  \bibfield  {author} {\bibinfo {author} {\bibfnamefont {A.~M.}\ \bibnamefont
  {Stoneham}},\ }\href {\doibase 10.1093/acprof:oso/9780198507802.001.0001}
  {\emph {\bibinfo {title} {Theory of Defects in Solids: Electronic Structure
  of Defects in Insulators and Semiconductors}}}\ (\bibinfo  {publisher}
  {Oxford University Press},\ \bibinfo {year} {1975})\BibitemShut {NoStop}%
\bibitem [{\citenamefont {Maze}\ \emph {et~al.}(2011)\citenamefont {Maze},
  \citenamefont {Gali}, \citenamefont {Togan}, \citenamefont {Chu},
  \citenamefont {Trifonov}, \citenamefont {Kaxiras},\ and\ \citenamefont
  {Lukin}}]{Maze2011}%
  \BibitemOpen
  \bibfield  {author} {\bibinfo {author} {\bibfnamefont {J.~R.}\ \bibnamefont
  {Maze}}, \bibinfo {author} {\bibfnamefont {A.}~\bibnamefont {Gali}}, \bibinfo
  {author} {\bibfnamefont {E.}~\bibnamefont {Togan}}, \bibinfo {author}
  {\bibfnamefont {Y.}~\bibnamefont {Chu}}, \bibinfo {author} {\bibfnamefont
  {A.}~\bibnamefont {Trifonov}}, \bibinfo {author} {\bibfnamefont
  {E.}~\bibnamefont {Kaxiras}}, \ and\ \bibinfo {author} {\bibfnamefont
  {M.~D.}\ \bibnamefont {Lukin}},\ }\href
  {http://stacks.iop.org/1367-2630/13/i=2/a=025025} {\bibfield  {journal}
  {\bibinfo  {journal} {New J. Phys.}\ }\textbf {\bibinfo {volume} {13}},\
  \bibinfo {pages} {025025} (\bibinfo {year} {2011})}\BibitemShut {NoStop}%
\bibitem [{\citenamefont {Doherty}\ \emph {et~al.}(2011)\citenamefont
  {Doherty}, \citenamefont {Manson}, \citenamefont {Delaney},\ and\
  \citenamefont {Hollenberg}}]{Doherty2011}%
  \BibitemOpen
  \bibfield  {author} {\bibinfo {author} {\bibfnamefont {M.~W.}\ \bibnamefont
  {Doherty}}, \bibinfo {author} {\bibfnamefont {N.~B.}\ \bibnamefont {Manson}},
  \bibinfo {author} {\bibfnamefont {P.}~\bibnamefont {Delaney}}, \ and\
  \bibinfo {author} {\bibfnamefont {L.~C.~L.}\ \bibnamefont {Hollenberg}},\
  }\href {http://stacks.iop.org/1367-2630/13/i=2/a=025019} {\bibfield
  {journal} {\bibinfo  {journal} {New J. Phys.}\ }\textbf {\bibinfo {volume}
  {13}},\ \bibinfo {pages} {025019} (\bibinfo {year} {2011})}\BibitemShut
  {NoStop}%
\bibitem [{\citenamefont {Abdi}\ \emph {et~al.}(2017)\citenamefont {Abdi},
  \citenamefont {Chou}, \citenamefont {Gali},\ and\ \citenamefont
  {Plenio}}]{Abdi2017_arXiv}%
  \BibitemOpen
  \bibfield  {author} {\bibinfo {author} {\bibfnamefont {M.}~\bibnamefont
  {Abdi}}, \bibinfo {author} {\bibfnamefont {J.-P.}\ \bibnamefont {Chou}},
  \bibinfo {author} {\bibfnamefont {A.}~\bibnamefont {Gali}}, \ and\ \bibinfo
  {author} {\bibfnamefont {M.~B.}\ \bibnamefont {Plenio}},\ }\href
  {https://arxiv.org/abs/1709.05414} {\bibfield  {journal} {\bibinfo  {journal}
  {arXiv:1709.05414 [cond-mat.mes-hall]}\ } (\bibinfo {year}
  {2017})}\BibitemShut {NoStop}%
\bibitem [{\citenamefont {Wu}\ \emph {et~al.}(2017)\citenamefont {Wu},
  \citenamefont {Galatas}, \citenamefont {Sundararaman}, \citenamefont
  {Rocca},\ and\ \citenamefont {Ping}}]{Wu2017}%
  \BibitemOpen
  \bibfield  {author} {\bibinfo {author} {\bibfnamefont {F.}~\bibnamefont
  {Wu}}, \bibinfo {author} {\bibfnamefont {A.}~\bibnamefont {Galatas}},
  \bibinfo {author} {\bibfnamefont {R.}~\bibnamefont {Sundararaman}}, \bibinfo
  {author} {\bibfnamefont {D.}~\bibnamefont {Rocca}}, \ and\ \bibinfo {author}
  {\bibfnamefont {Y.}~\bibnamefont {Ping}},\ }\href {\doibase
  10.1103/PhysRevMaterials.1.071001} {\bibfield  {journal} {\bibinfo  {journal}
  {Phys. Rev. Materials}\ }\textbf {\bibinfo {volume} {1}},\ \bibinfo {pages}
  {071001} (\bibinfo {year} {2017})}\BibitemShut {NoStop}%
\bibitem [{\citenamefont {Tawfik}\ \emph {et~al.}(2017)\citenamefont {Tawfik},
  \citenamefont {Ali}, \citenamefont {Fronzi}, \citenamefont {Mianinia},
  \citenamefont {Tran}, \citenamefont {Stampfl}, \citenamefont {Aharonovich},
  \citenamefont {Toth},\ and\ \citenamefont {Ford}}]{Tawfik2017}%
  \BibitemOpen
  \bibfield  {author} {\bibinfo {author} {\bibfnamefont {S.~A.}\ \bibnamefont
  {Tawfik}}, \bibinfo {author} {\bibfnamefont {S.}~\bibnamefont {Ali}},
  \bibinfo {author} {\bibfnamefont {M.}~\bibnamefont {Fronzi}}, \bibinfo
  {author} {\bibfnamefont {M.}~\bibnamefont {Mianinia}}, \bibinfo {author}
  {\bibfnamefont {T.~T.}\ \bibnamefont {Tran}}, \bibinfo {author}
  {\bibfnamefont {C.}~\bibnamefont {Stampfl}}, \bibinfo {author} {\bibfnamefont
  {I.}~\bibnamefont {Aharonovich}}, \bibinfo {author} {\bibfnamefont
  {M.}~\bibnamefont {Toth}}, \ and\ \bibinfo {author} {\bibfnamefont {M.~J.}\
  \bibnamefont {Ford}},\ }\href {\doibase 10.1039/C7NR04270A} {\bibfield
  {journal} {\bibinfo  {journal} {Nanoscale}\ }\textbf {\bibinfo {volume}
  {9}},\ \bibinfo {pages} {13575} (\bibinfo {year} {2017})}\BibitemShut
  {NoStop}%
\bibitem [{\citenamefont {Koster}\ \emph {et~al.}(1963)\citenamefont {Koster},
  \citenamefont {Dimmock}, \citenamefont {Wheeler},\ and\ \citenamefont
  {Statz}}]{Koster1963}%
  \BibitemOpen
  \bibfield  {author} {\bibinfo {author} {\bibfnamefont {G.~F.}\ \bibnamefont
  {Koster}}, \bibinfo {author} {\bibfnamefont {J.~O.}\ \bibnamefont {Dimmock}},
  \bibinfo {author} {\bibfnamefont {R.~G.}\ \bibnamefont {Wheeler}}, \ and\
  \bibinfo {author} {\bibfnamefont {H.}~\bibnamefont {Statz}},\ }\href@noop {}
  {\emph {\bibinfo {title} {Properties of the thirty-two point groups}}}\
  (\bibinfo  {publisher} {MIT Press},\ \bibinfo {year} {1963})\BibitemShut
  {NoStop}%
\bibitem [{\citenamefont {Basch\'{e}}\ \emph {et~al.}(1996)\citenamefont
  {Basch\'{e}}, \citenamefont {Kummer},\ and\ \citenamefont
  {Br\"{a}uchle}}]{Basche1996}%
  \BibitemOpen
  \bibfield  {author} {\bibinfo {author} {\bibfnamefont {T.}~\bibnamefont
  {Basch\'{e}}}, \bibinfo {author} {\bibfnamefont {S.}~\bibnamefont {Kummer}},
  \ and\ \bibinfo {author} {\bibfnamefont {C.}~\bibnamefont {Br\"{a}uchle}},\
  }\enquote {\bibinfo {title} {Single-molecule optical detection, imaging and
  spectroscopy},}\ \ (\bibinfo  {publisher} {VCH Verlagsgesellschaft mbH,
  Weinheim, Germany},\ \bibinfo {year} {1996})\ Chap.\ \bibinfo {chapter} {2:
  Excitation and Emission Spectroscopy and Quantum Optical
  Measurements}\BibitemShut {NoStop}%
\bibitem [{\citenamefont {Geist}\ and\ \citenamefont
  {Römelt}(1964)}]{Geist1964}%
  \BibitemOpen
  \bibfield  {author} {\bibinfo {author} {\bibfnamefont {D.}~\bibnamefont
  {Geist}}\ and\ \bibinfo {author} {\bibfnamefont {G.}~\bibnamefont
  {Römelt}},\ }\href {\doibase https://doi.org/10.1016/0038-1098(64)90404-1}
  {\bibfield  {journal} {\bibinfo  {journal} {Solid State Communications}\
  }\textbf {\bibinfo {volume} {2}},\ \bibinfo {pages} {149} (\bibinfo {year}
  {1964})}\BibitemShut {NoStop}%
\bibitem [{\citenamefont {Fanciulli}(1997)}]{Fanciulli1997}%
  \BibitemOpen
  \bibfield  {author} {\bibinfo {author} {\bibfnamefont {M.}~\bibnamefont
  {Fanciulli}},\ }\href {\doibase 10.1080/01418639708241100} {\bibfield
  {journal} {\bibinfo  {journal} {Philosophical Magazine Part B}\ }\textbf
  {\bibinfo {volume} {76}},\ \bibinfo {pages} {363} (\bibinfo {year}
  {1997})}\BibitemShut {NoStop}%
\bibitem [{\citenamefont {Morton}\ and\ \citenamefont
  {Preston}(1978)}]{Morton1978}%
  \BibitemOpen
  \bibfield  {author} {\bibinfo {author} {\bibfnamefont {J.}~\bibnamefont
  {Morton}}\ and\ \bibinfo {author} {\bibfnamefont {K.}~\bibnamefont
  {Preston}},\ }\href {\doibase https://doi.org/10.1016/0022-2364(78)90284-6}
  {\bibfield  {journal} {\bibinfo  {journal} {Journal of Magnetic Resonance
  (1969)}\ }\textbf {\bibinfo {volume} {30}},\ \bibinfo {pages} {577 }
  (\bibinfo {year} {1978})}\BibitemShut {NoStop}%
\end{thebibliography}%

\end{document}


\title[SI: Spin-dependent fluorescence in h-BN]
  {Supporting Information:\\  Spin-Dependent Quantum Emission in Hexagonal Boron Nitride \\ at Room Temperature}
  
\author{Annemarie L. Exarhos} 
\altaffiliation[Present address: ]{Department of Physics, Lawrence University, Appleton, WI 54911, USA}
\affiliation{Quantum Engineering Laboratory, Department of Electrical and Systems Engineering, University of Pennsylvania, Philadelphia, Pennsylvania 19104, United States}

\author{David A. Hopper}
\affiliation{Quantum Engineering Laboratory, Department of Electrical and Systems Engineering, University of Pennsylvania, Philadelphia, Pennsylvania 19104, United States}
\affiliation{Department of Physics and Astronomy, University of Pennsylvania, Philadelphia, Pennsylvania 19104, United States}

\author{Raj N. Patel}
\affiliation{Quantum Engineering Laboratory, Department of Electrical and Systems Engineering, University of Pennsylvania, Philadelphia, Pennsylvania 19104, United States}

\author{Marcus W. Doherty}
\affiliation{Laser Physics Centre, Research School of Physics and Engineering, Australian National University, Canberra ACT 2601, Australia}

\author{Lee C. Bassett}
\email{lbassett@seas.upenn.edu}
\affiliation{Quantum Engineering Laboratory, Department of Electrical and Systems Engineering, University of Pennsylvania, Philadelphia, Pennsylvania 19104, United States}

\date{\today}

\maketitle 


\renewcommand{\thefigure}{S\arabic{figure}}
\renewcommand{\figurename}{Figure}
\renewcommand{\thesection}{S\arabic{section}}
\renewcommand{\thetable}{S\arabic{table}}

\def \experiment {
\begin{figure*}[!b]
\includegraphics[width=6.7in]{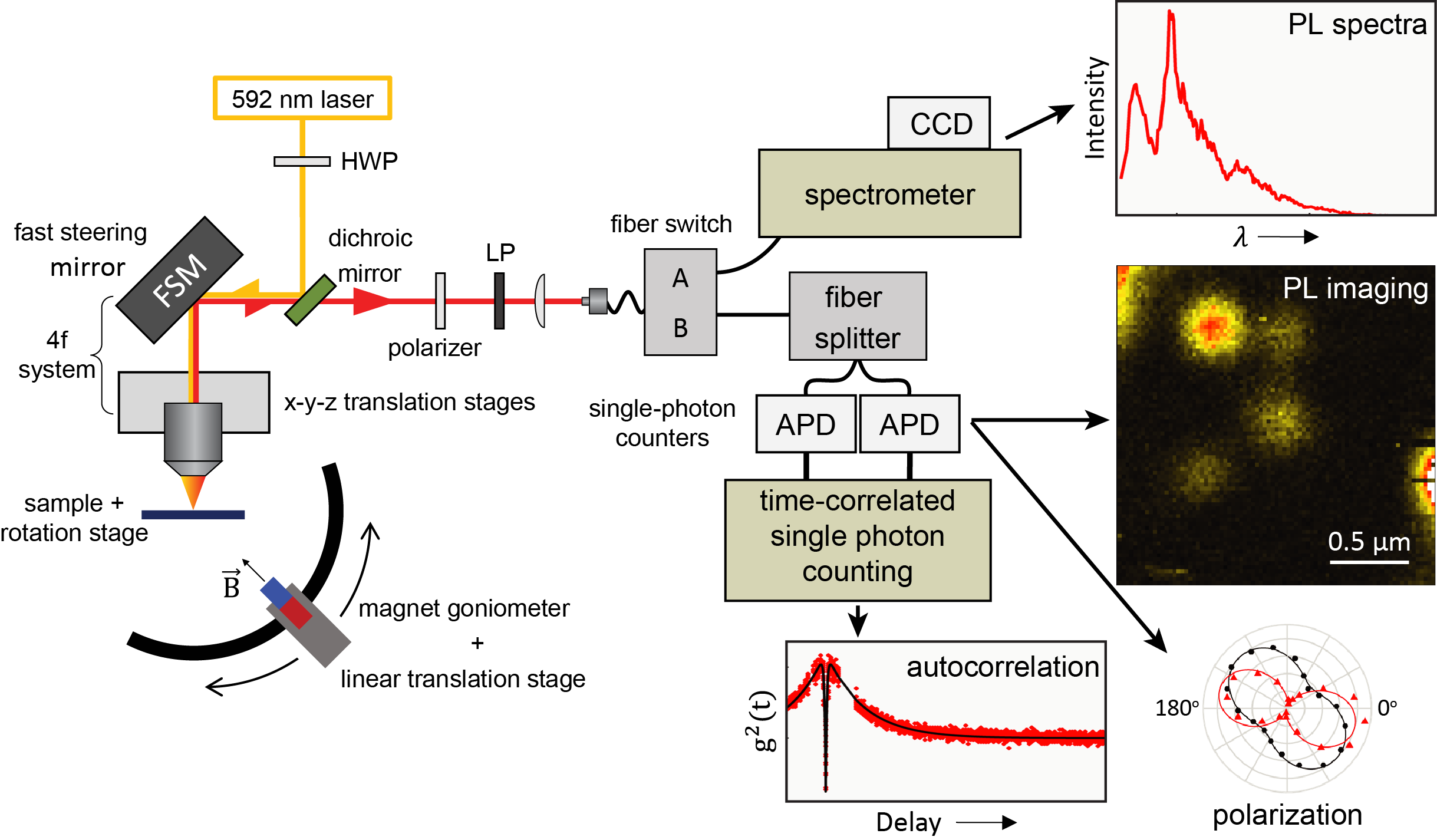}
\caption{\textbf{Scanning confocal fluorescence microscope.} Experimental setup for studying quantum emitters in h-BN.  Abbreviations:  HWP - half wave plate, LP - long pass filter, APD - avalanche photodiode.
\label{experiment}}
\end{figure*}
}

\def\AFM{
\begin{figure*}[t]
  \includegraphics[width=6.58in]{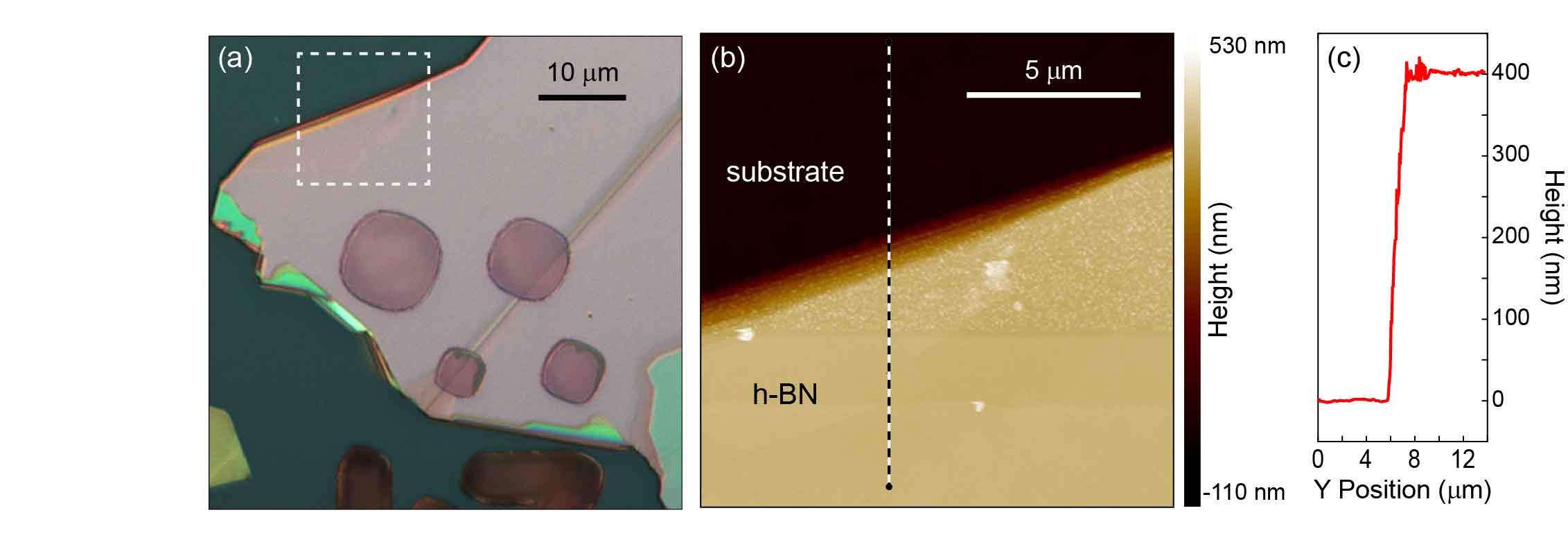}
  \caption{\textbf{Optical and AFM characterization.} (a) Optical and (b) AFM images for the h-BN flake.  The region denoted by the dotted line in (a) corresponds to the area of the AFM scan shown in (b).  (c) Line cut through the AFM scan at the dotted line in (b) showing the height of the h-BN flake. }
  \label{AFM}
\end{figure*}}

\def\power{
\begin{figure*}[htb]
  \includegraphics[width=3.33in]{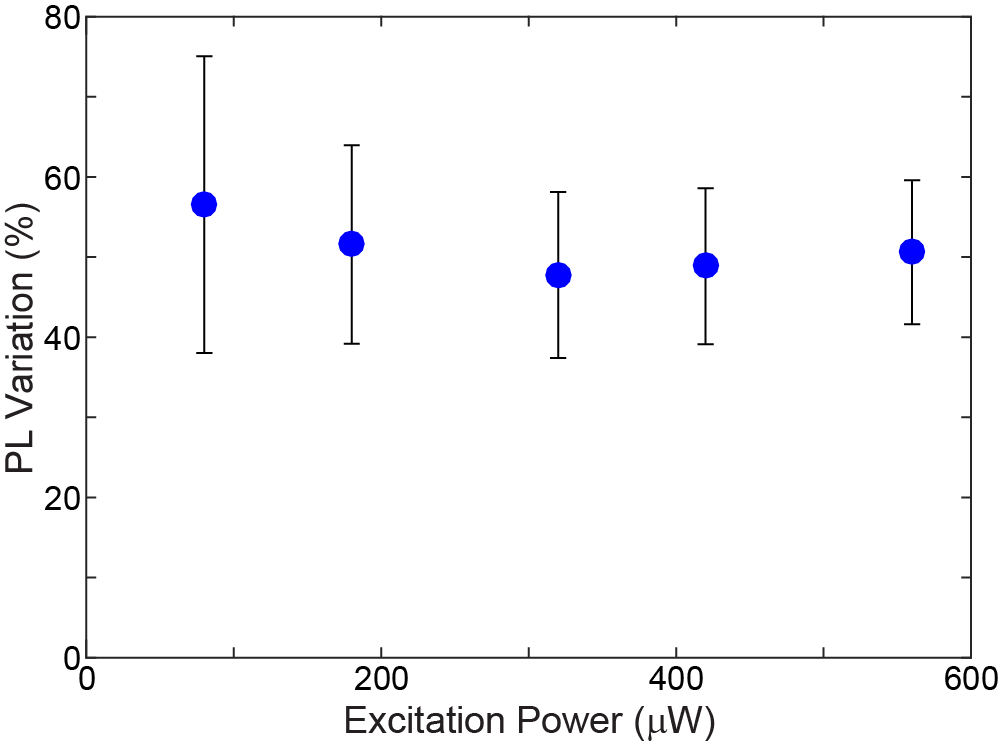}
  \caption{\textbf{Power dependence of PL variation.} PL variation \% as a function of excitation power for the main defect from the manuscript.  The absorptive dipole is oriented parallel to the applied 240 G magnetic field.}
  \label{power}
\end{figure*}}

\def\sqSpot{
\begin{figure*}[b]
\includegraphics[width=5.5in]{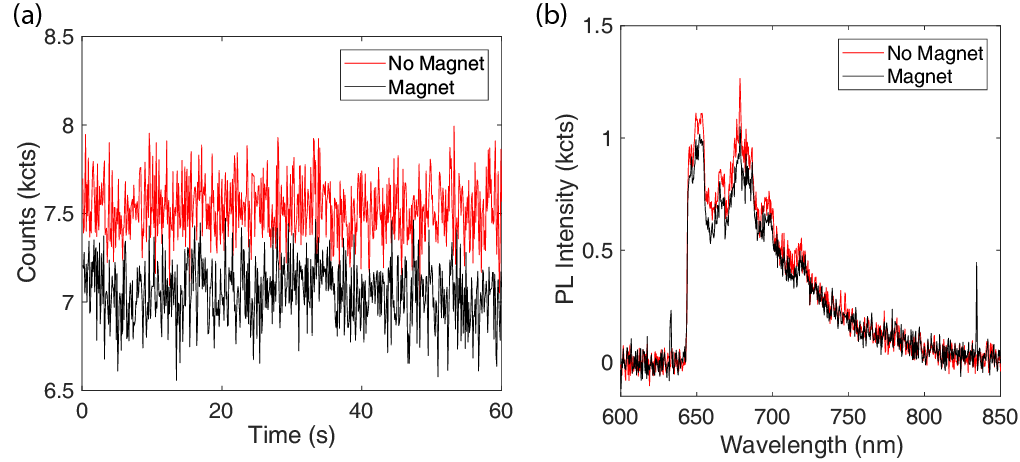}
\caption{\textbf{Magnetic field dependence of emitter marked square in main text.} 
(a) Emission in absence and presence of magnetic field averaged over three scans showing reduced emission when magnetic field is applied. 
(b) PL spectra showing reduced emission when magnetic field is applied.}
\label{sqSpot}
\end{figure*}}

\def\redSpot{
\begin{figure*}[htb]
  \includegraphics[width=7in]{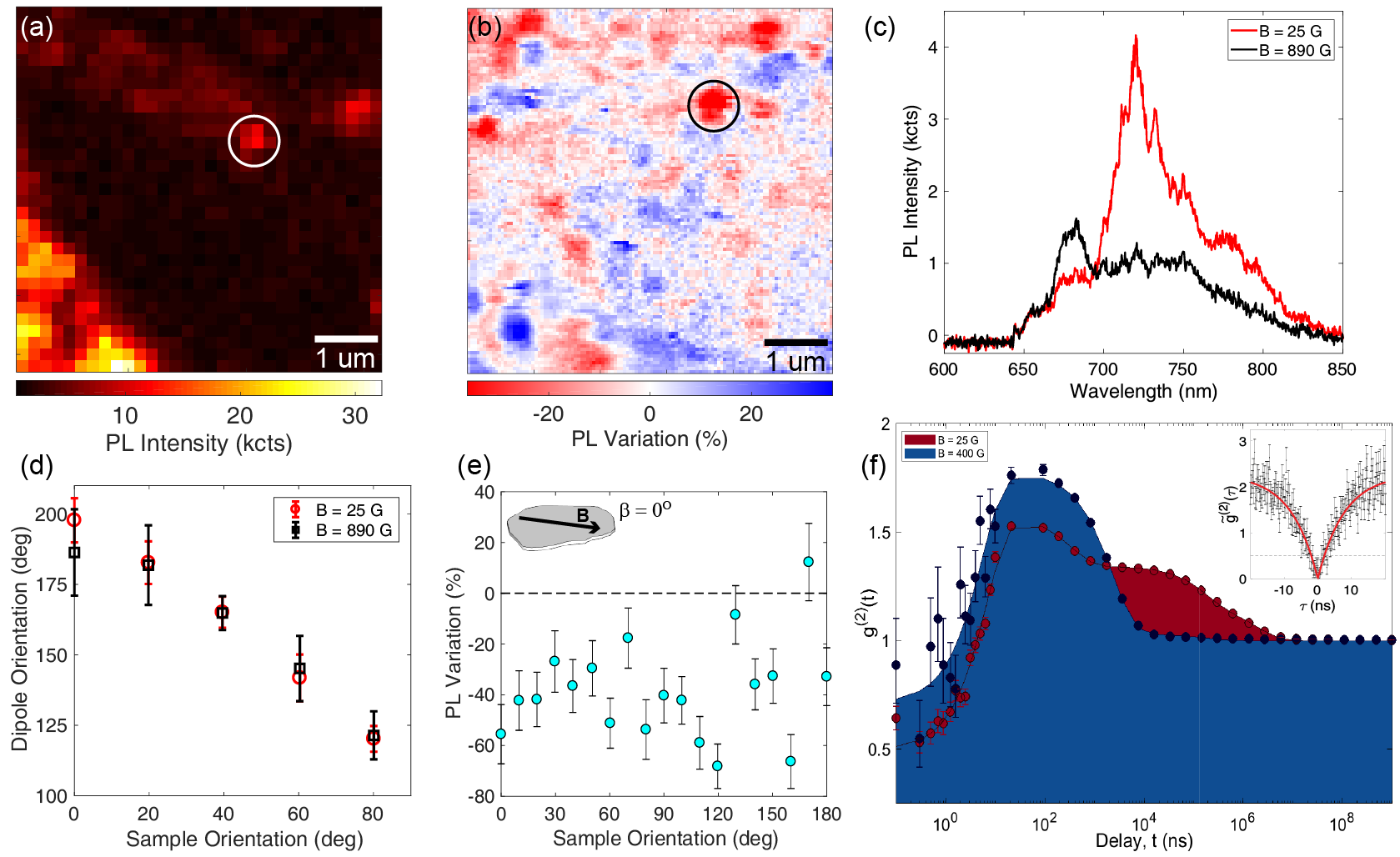}
  \caption{\textbf{Additional magnetic field-dependent quantum emitter}
  (a) PL image of another region on the suspended h-BN flake shown in Figure~1(a) of main text. The circled spot is the additional emitter that showed magnetic field-dependent PL. (b) Background-subtracted differential PL variation image from the same region shown in (a) identifying changes due to an applied in-plane magnetic field. Circled red spot is the same region as the circle in (a). (c) PL spectra for the defect circled in (a,b) for in-plane magnetic fields of 25~G and 890~G, oriented such that $\mathbf{B}$ is parallel to the absorptive dipole. (d) The measured absorptive dipole for 25~G and 890~G as a function of sample orientation. (e) PL variation comparing 890~G and 25~G for various sample orientations. The dashed line shows the average PL variation. (f) Measured photon autocorrelation function (points) for in-plane applied magnetic fields ($\beta = 0^{\circ}$) of 25~G and 400~G. Fits (black curves) are described in the text. The inset shows the background-corrected short-delay photon autocorrelation function. The dashed line shows the single photon emission criterion.}
  \label{redSpot}
\end{figure*}}

\def\dipole{
\begin{figure*}[htp]
  \includegraphics[width=7in]{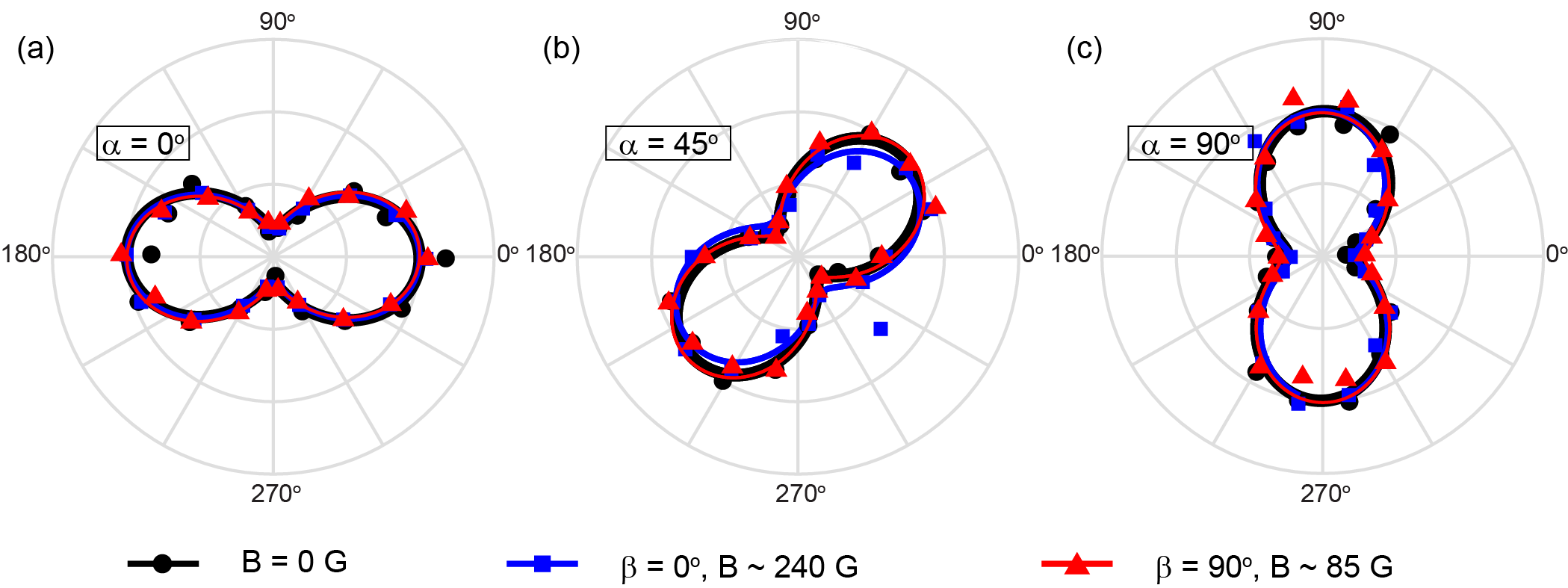}
  \caption{\textbf{Absorptive optical dipole orientation field dependence.} Background-subtracted and normalized PL as a function of excitation polarization axes with fits to the absorptive dipole orientation for zero, in-plane, and out-of-plane applied magnetic fields at (a) $\alpha = 0^{\circ}$, (b) $\alpha =45^{\circ}$, and (c) $\alpha = 90^{\circ}$.}
  \label{dipole}
\end{figure*}}


 \def\g2table{\begin{table*} 
 \centering
 \caption{\textbf{Magnetic-Field-Dependent Photon Emission Statistics}.  Best-fit parameters to photon autocorrelation curves from the emitter reported in the main text at different orientations of an applied magnetic field.}
 \label{tab:g2parameters}
 \begin{tabularx}{\textwidth}{YYYYYYYYYY}
 $(\alpha,\beta)$ & $C_1$ & $C_2$ & $C_3$ & $\tau_1$ (ns)$^\dagger$ & $\tau_2$ ($\mu$s) & $\tau_3$ ($\mu$s) & $\tilde{C}_1^{\ddagger}$ & $\tilde{C}_2^{\ddagger}$ & $\tilde{C}_3^{\ddagger}$ \vspace{.025in} \\  \hline \hline 
  \noalign{\vskip 3pt} 
 (0$^{\circ}$,-)$^\ast$ & 1.58$\pm$0.02 & 1.7$\pm$0.5 & 0.09$\pm$0.01 & 1.2$\pm$0.7 & 1.48$\pm$0.03 & 16$\pm$2 & 5.6$^{+0.8}_{-0.6}$ & 6.0$^{+0.9}_{-1.2}$ & 0.33$\pm 0.04$ \\ 
 \noalign{\vskip 3pt} 
 (45$^{\circ}$,-)$^\ast$ & 1.48$\pm$0.03 & 1.5$\pm$0.6 & 0.08$\pm$0.01 & 1.1$\pm$0.8 & 1.33$\pm$0.04 & 17$\pm$2 & 5.4$^{+0.8}_{-0.6}$ & 5.3$^{+1.3}_{-1.8}$ & 0.28$\pm 0.01$ \\  \hline
 \noalign{\vskip 3pt} 
 (0$^{\circ}$,0$^{\circ}$) & 1.31$\pm$0.02 & 1.3$\pm$0.6 & - & 1.5$\pm$1.1 & 1.43$\pm$0.03 & - & 3.00$^{+0.14}_{-0.13}$ & 2.9$^{+1.0}_{-1.1}$ & -\\ 
 \noalign{\vskip 3pt} 
 (45$^{\circ}$,0$^{\circ}$) & 1.65$\pm$0.02 & 2.1$\pm$0.6 & - & 0.8$\pm$0.4 & 1.41$\pm$0.02 & - & 6.0$^{+0.9}_{-0.7}$ & 7.7$^{+1.0}_{-1.3}$ & -\\ 
 \noalign{\vskip 3pt} 
 (90$^{\circ}$,0$^{\circ}$) & 1.36$\pm$0.04 & 1.5$\pm$1.1 & - & 0.9$\pm$1.1 & 1.21$\pm$0.04 & - & 3.26$^{+0.16}_{-0.15}$ & 4$\pm$2 & -\\ \hline 
 \multicolumn{8}{l}{$^\ast$\footnotesize{$(\alpha,\beta) = (\alpha, -)$ indicates no applied magnetic field.  When applied, the field strength is 240~G.}}\\
 \multicolumn{8}{l}{$^\dagger$\footnotesize{$\tau_1$ lifetimes are likely limited by detector timing jitter.}}\\
 \multicolumn{8}{l}{$^{\ddagger}$\footnotesize{Background corrected bunching amplitudes}}\\
 \end{tabularx}
 \end{table*}}

\def\charactertable{\begin{table*}[tb] 
 \centering
 \caption{\textbf{Character table for $C_{2v}$}}
 \label{tab:c2v_character}
 \setlength{\tabcolsep}{7.5pt}
 \setlength{\extrarowheight}{7.5pt}
 \begin{tabularx}{0.73\textwidth}{ l | c  c  c  c | l  l}
 
 \hline
       & \textbf{E} & \textbf{C$_{2}$} & $\sigma_{\textbf{v}}$(\textbf{xy}) & $\sigma_{\textbf{v}}$(\textbf{xz}) & \textbf{Linear functions} & \textbf{Quadratic functions} \\  
\hline
$A_{1}$  & 1 & 1  & 1  & 1  & x & x$^{2}$, y$^{2}$, z$^{2}$, S$_{x}^2$, S$_{y}^2$, S$_{z}^2$ \\ 

$A_{2}$  & 1 & 1  & -1 & -1 & S$_{x}$ & yz, S$_{y}$S$_{z}$ \\ 

$B_{1}$  & 1 & -1 & 1  & -1 & y, S$_{z}$ & xy, S$_{x}$S$_{y}$ \\ 

$B_{2}$  & 1 & -1 & -1 & 1  & z, S$_{y}$ & xz, S$_{x}$S$_{z}$ \\ 
\hline
$E_{1/2}^\ast$   & 2 & 0  & 0  & 2  &   &   \\ 
\hline
 \multicolumn{7}{l}{$^\ast$\footnotesize{Double group representation.}}\\
\end{tabularx}
\end{table*}}

\def\multiplicationtable{\begin{table*}[tb] 
\centering
\caption{\textbf{Group multiplication table}}
\label{tab:c2v_multiplication}
\setlength{\tabcolsep}{7.5pt}
\setlength{\extrarowheight}{7.5pt}
\begin{tabularx}{0.4\textwidth}{ l  l  l  l | l }
 
 $A_{1}$ & $A_{2}$ & $B_{1}$ & $B_{2}$ & \\  \hline

$A_{1}$ ($\hat{X}$)$^\ast$ & $A_2$ (-) & $B_1$ ($\hat{Y}$) & $B_2$ ($\hat{Z}$) & $A_{1}$ \\ 

 & $A_{1}$ ($\hat{X}$) & $B_2$ ($\hat{Z}$) & $B_1$ ($\hat{Y}$) & $A_{2}$ \\ 

 & & $A_{1}$ ($\hat{X}$)& $A_2$ (-) & $B_{1}$ \\ 

 & & & $A_{1}$ ($\hat{X}$) & $B_{2}$ \\
  \multicolumn{5}{l}{$^\ast$\footnotesize{Optical selection rules shown in parentheses.}}\\
\end{tabularx}
\end{table*}}


\section{Experimental Setup}
\experiment
Figure \ref{experiment} shows the general layout of the scanning confocal fluorescence microscope.  We use a 0.9 NA objective (Olympus) and 592 nm continuous (CW) excitation with $175 - 550$ $\mu$W at the sample.  The optics shown are used for all photoluminescence (PL) imaging, autocorrelation measurements, and PL spectra, with the exception of the polarizer in the collection line, which is only in place for emission polarization dependence measurements.  Two single-photon counters are used for PL imaging and autocorrelation measurements:  Excelitas (SPCM-AQRH-14-FC) and MPD (PDM-R) detectors.  Photon autocorrelation measurements are performed using a Hanbury Brown-Twiss setup with a PicoQuant PicoHarp 300 time correlated single-photon counting module.  PL spectra are obtained using a Princeton Instruments IsoPlane 160 spectrometer and PIXIS 100 CCD with a spectral resolution of 0.7 nm.  Spectra are uncorrected for the wavelength-dependent transmission efficiency of the collection line.

Excitation polarization dependence is measured by rotating the linear polarization of the excitation laser using a half waveplate.  The PL is not polarization-selected; all emitted PL ($\lambda_{PL} > 633$ nm), regardless of polarization, is collected.  For the emission polarization measurements, the excitation polarization is fixed (typically at the angle which maximizes the collected PL for the defect in question) and a linear polarizer is placed in front of the detector and rotated to the desired emission polarization angles.  Polarized emission with wavelengths between $\sim$710 nm and $\sim$745 nm is measured.  Both the excitation and collection lines are corrected to account for the birefringence of the dichroic mirror and other optics in the microscope.

\section{Sample Preparation}
Single-crystal h-BN purchased from HQ Graphene is exfoliated onto patterned silicon wafers topped with a 90-nm-thick thermal SiO$_2$ layer as described in Ref. \onlinecite{Exarhos2017}.  Following exfoliation, samples undergo an O$_2$ plasma clean in an oxygen barrel asher (Anatech SCE 108) and are  annealed in Ar at 850$^{\circ}$ for 30 minutes.  Subsequently, the samples were imaged in a scanning electron microscope (SEM) operating at 3 kV (FEI Strata DB235 FIB SEM).  The flake studied in this work was imaged for $< 5$ minutes, and the sample was in the SEM chamber for $<$30 minutes. Following SEM, the samples are again annealed in Ar at 850$^{\circ}$ for 30 minutes.  

\AFM
Figure \ref{AFM} shows a white light optical microscope image of the h-BN flake studied along with an atomic force microscope (AFM) image taken over a portion of the supported sample near the suspended region under study (dotted box in the optical image).  A line cut through the AFM data shows that the sample is $\sim 400$ nm thick. 


\section{\textbf{Power Dependence of Magnetic Field-Dependent Emission}}
\power
Figure~\ref{power} shows the PL variation of the emitter studied in the main text due to an in-plane magnetic field as a function of the optical excitation power.  The variation appears to be independent of power across the range of settings used in our experiments.

\section{\textbf{Field Independence of the Absorptive Dipole Orientation}}
The apparent orientation of the absorptive dipole for the defect featured in Figures 2-3 of the main text is unaffected by an applied magnetic field, as shown in Figure \ref{dipole}.  Excitation polarization scans taken at three different sample orientations ($\alpha = 0^{\circ}$, $45^{\circ}$, and $90^{\circ}$) with applied magnetic fields both in-plane and out-of-plane display a fixed absorptive dipole orientation.
\dipole

\section{Analysis and Fits of Photon Counting Statistics}

Photon autocorrelation data acquired for different settings of the applied magnetic field are fitted using an empirical model as described in the main text and corrected for the measured background.  Best-fit parameters are listed in Table~\ref{tab:g2parameters}.

\g2table

\section{\textbf{Additional Magnetic-Field-Dependent Emitters}}

\sqSpot

We observed reproducible PL variations in response to in-plane magnetic fields for a handful of other emitters in the same suspended region of the h-BN flake.  For example, the emitter marked with a square in Fig.~1(b,c) exhibited a $\approx-20$\% PL variation (Figure~\ref{sqSpot}) in the orientation shown.  Other spots brightened in response to a field.  Unfortunately, many of these emitters bleached away upon further study.  Blinking and other instabilities on long timescales also complicate the field-dependent measurements.  We estimate that $<$5\% of emitters observed in this sample exhibit a reproducible field dependence, although we note that our searches for field-dependent emitters via differential images were taken at one particular magnetic field orientation and could have excluded emitters for which the field orientation resulted in no PL variation.  Nonetheless, the number of field-dependent emitters compared to all emitters is small.

\redSpot

Figure~\ref{redSpot} summarizes a series of measurements concerning a field-dependent emitter in a different region of the same suspended flake considered in the main text. Figure~\ref{redSpot}(a) shows a PL image of a suspended part of the flake which contains the emitter (circled). The PL image is obtained with linearly polarized 592~nm illumination (linear polarization setting of 20$^\circ$ relative to the horizontal) that maximizes emission, with an applied in-plane magnetic field in the horizontal direction of 25~G. All measurements are performed with an illumination power of 130 $\mu$W (measured before the objective).

Figure~\ref{redSpot}(b) shows the PL variation as defined in the main text, $(I_{B}-I_{0})/I_{0}$, where $I_{B}$, $I_{0}$ refer to the PL intensity for an in-plane field of 890~G and 25~G, respectively. The circled emitter's PL intensity decreases by $\sim$40$\%$. Figure~\ref{redSpot} (c) shows the emitter's PL spectrum with an applied in-plane magnetic field of 25~G and 890~G. The main component of the low-field PL spectrum between $\approx$700-800~nm clearly reflects the field-dependent decrease in emission rate, although interestingly there is a blue-shifted increase in PL around 670~nm when the field is applied. The field-independent feature around 650~nm is associated with the background. The peak at $\sim$730~nm at low field (25~G) as well as most emission in the range 700~nm - 750~nm is similar to that of the emitter discussed in main text. However, we note that other field-dependent emitters observed in this sample (\textit{e.g.,} the emitter whose spectrum is plotted in Fig.~\ref{sqSpot}) exhibit qualitatively different spectra.

Figure~\ref{redSpot}(d) shows the absorptive dipole orientation at different sample orientations for 25~G and 890~G in-plane magnetic field. The apparent dipole orientation is independent of applied magnetic field, similar to the emitter presented in the main text (see Fig.~\ref{dipole}). Figure~\ref{redSpot}(e) depicts the PL variation as a function of sample orientation, comparing in-plane magnetic field strengths of 25~G and 890~G. In contrast to the 90$^\circ$-periodic pattern of brightening and dimming observed for the emitter studied in the main text, this defect exhibits reduced PL for all in-plane field directions.  The dashed line signifies the average PL variation of $\sim$38$\%$. Blinking is observed especially at low fields, and is responsible for the large spread in observations.  The PL variation in response to an out-of-plane field (not shown) is also negative.

Figure~\ref{redSpot}(f) depicts the photon autocorrelation function for in-plane magnetic field strengths of 25~G and 400~G. At both fields, the defect exhibits antibunching with a timescale, $\tau_1\approx 7$~ns, and bunching over longer times. The data is fitted with the empirical model discussed in the text.  The inset shows the photon autocorrelation data for 25~G, after correction for a Poissonian background (measured at a nearby location on the flake) and binned by delay on a linear scale. A fit to the data using a single exponential antibunching decay drops below the single-emitter criterion with a best-fit minimum value $\tilde{g}^{(2)}=0.00\pm 0.13 <0.5$. 

At low fields, at least three bunching decay terms are required to achieve a suitable fit to the data (i.e., $n=4$ in the model), with characteristic bunching timescales $\tau_2\approx350$~ns, $\tau_3\approx180$~$\mu$s, and $\tau_3\approx2$~ms.  When the field is applied, the behavior is well-approximated by a single decay at $\tau_2\approx2.3$~$\mu$s. Again, this is qualitatively similar to the behavior of the defect in the main text, in that the long-timescale bunching vanishes with the field is applied, but it also is quantitatively different in several important respects.  We believe that these differences can be captured within our model\,---\,even using the same basic level structures\,---\,through variations in the local defect parameters due, \textit{e.g.}, to strain that shifts the triplet zero-field splitting terms, variations in inter-system crossing (ISC) rates due to energy offsets between the triplet and singlet states, and variations in the triplet spin-$T_1$ relaxation time. Clearly, further investigation is required to understand the relevant perturbations and the range of possible field-dependent behaviors for different emitters.

\section{Molecular orbital theory for h-BN defects and optical dynamics}

The goal of our theoretical study is to use molecular orbital (MO) theory to enumerate a set of simplified models for defect electronic structure based on symmetry considerations \cite{Stoneham1975}, and then to perform semiclassical calculations to simulate their optical and spin dynamics under steady-state illumination, for comparisons with experimental results.
To that end, we do not start with a particular defect model and study it in detail; rather we explore the qualitative similarities and differences between various electronic configurations in an effort to narrow the space of possibilities. We hope this will motivate future efforts to compare these qualitative predictions with quantitative, \textit{ab initio}, calculations of prospective defect configurations in h-BN.

\subsection{Choice of Point Group}

The starting point for any MO calculation is the identification of the relevant point group describing the symmetry of the molecule or defect system. Here, we focus on the point group $C_{2v}$ based on the following considerations: 
\begin{enumerate}
\item We universally observe optical selection rules in absorption and emission for linearly-polarized photons in the $(x,y)$ plane, \textit{i.e.}, parallel to the hBN membrane. These selection rules naturally result from the symmetry-allowed orbitals in $C_{2v}$ as shown below.
\item Our observations of field-dependent PL are consistent with underlying reflection symmetry about the $(x,y)$ plane, whereas we observe fourfold (90$^\circ$) rotational symmetry for in-plane fields.
\item $C_{2v}$ symmetry is the expected point group for many defect complexes in h-BN, including distorted vacancies such as N$_{B}$V$_{N}$, and vacancy-impurity complexes such as C$_{B}$V$_{N}$, both of which have been proposed as models for hBN's visible quantum emission.
\end{enumerate}
We view (1) and (2) as the most important considerations, and briefly discuss later why other possible point groups such as D$_{3h}$ or C$_{s}$ are unlikely to yield behavior consistent with our observations.

Details for the point group $C_{2v}$ and its irreducible representations (IRs) are provided in Tables~\ref{tab:c2v_character} and \ref{tab:c2v_multiplication}. Note that we choose a coordinate system with $x$ as the principal symmetry axis, lying in the h-BN plane, with the $z$-axis oriented normal to the h-BN plane.

\charactertable

\multiplicationtable

\subsection*{Spin Hamiltonian}
To begin, we consider only the electronic degrees of freedom, \textit{i.e.}, neglecting hyperfine coupling with nuclear spins. In this case, we start with the generalized Hamiltonian for an electronic configuration with total spin $S$:
\begin{equation}\label{eq:SpinH}
\begin{split}
H &= \mu_\mathrm{B} \mathbf{S} \cdot g \cdot \mathbf{B} - \mathbf{S} \cdot \Lambda \cdot \mathbf{S} \\
&= \mu_\mathrm{B} (g_{xx}B_{x}S_{x} + g_{yy}B_{y}S_{y} + g_{zz}B_{z}S_{z}) + D \Bigg(S_{x}^2 - \frac{1}{3}S\big(S+1\big) \Bigg) + E\big(S_{y}^2 - S_{z}^2\big) 
\end{split}
\end{equation}
Since spin-orbit coupling in h-BN is relatively weak, we assume that the components of the $g$-tensor are nearly equal to the bare value, $g\sim 2$, since spin-orbit corrections to the $g$-factor are of order $\frac{\lambda}{\Delta}\ll 1$, where ${\lambda}$ is the spin-orbit strength and ${\Delta}$ is the orbital crystal field splitting. In $C_{2v}$ we need to include two zero-field splitting (ZFS) parameters, $D$ and $E$, in the fine structure term of $H$.  This is in contrast to higher-symmetry cases such as $C_{3v}$ or $D_{3h}$ where $E$ vanishes due to symmetry. The fact that $E$ must be included is easy to see from the fact that all of the terms ($S_{x}^2, S_{y}^2, S_{z}^2$) transform like the trivial representation, $A_1$, and therefore are allowed to appear in the Hamiltonian, which also transforms as $A_1$ by definition. In our general treatment, $D$ and $E$ are empirical parameters; their origin can be either first-order spin-spin or second-order spin-orbit interactions, although spin-spin interactions are likely to dominate due to the weak spin-orbit coupling in h-BN.  In this case, their values can be calculated explicitly in terms of two electron integrals given a specific orbital configuration \cite{Maze2011,Doherty2011,Abdi2017_arXiv}. Hence the Hamiltonian takes the simplified form:
\begin{equation}
H = g \mu_\mathrm{B} \mathbf{B} \cdot \mathbf{S} + D \Bigg(S_{x}^2 - \frac{1}{3}S\big(S+1\big) \Bigg) + E\big(S_{y}^2 - S_{z}^2\big) 
\end{equation}

We consider cases with total spin S = 0, $\frac{1}{2}$, 1, and $\frac{3}{2}$. Since there is no orbital degeneracy in $C_{2v}$, configurations with S$>$$\frac{1}{2}$ are likely to occur as metastable excited states, but they can also become the ground states in the case of closely-spaced orbitals and sufficient spin-exchange interactions, as predicted for some defects in hBN by Refs.~\cite{Wu2017,Tawfik2017,Abdi2017_arXiv}. However, states with higher spin require additional closely-spaced orbitals (\textit{e.g.,} three orbitals in the case of a quartet) and therefore are progressively more unlikely. For this reason, we do not consider spin configurations with S$>$$\frac{3}{2}$. 

We consider each case below:

\vspace{6pt}
\noindent\textit{S=0: Singlet configuration}

Here the spin Hamiltonian vanishes.  The symmetry of the spin-singlet configuration is A$_{1}$, as can be confirmed from the coupling coefficients and the double-group representation in $C_{2v}$ (\textit{e.g.} from Ref.~\cite{Koster1963}).

\vspace{6pt}
\noindent\textit{S=$\frac{1}{2}$: Doublet configuration}

The ZFS vanishes in the spin doublet configuration, $H = g\mu_\mathrm{B}  B \cdot \mathbf{S}$, where the components of $\mathbf{S}$ are 2$\times$2 Pauli spin operators.  The doublet components transform according to the double-group representation, $E_{1/2}$.

\vspace{6pt}
\noindent\textit{S=1: Triplet configuration}

The zero-field spin eigenstates in the triplet configuration are non-degenerate due to the ZFS parameters.  We can identify those eigenstates as the $\{\ket{s_x},\ket{s_y},\ket{s_z}\}$ spin basis
\begin{subequations}\label{eq:TripletBasis}
\begin{align}
\ket{s_x} & = \frac{1}{\sqrt{2}}(\ket{\uparrow\downarrow} + \ket{\downarrow\uparrow}) \sim A_{2}\\
\ket{s_y} & = \frac{1}{\sqrt{2}}(\ket{\uparrow\uparrow} - \ket{\downarrow\downarrow}) \sim B_{2}\\
\ket{s_z} & = \frac{1}{\sqrt{2}}(\ket{\uparrow\uparrow} + \ket{\downarrow\downarrow}) \sim B_{1},
\end{align}
\end{subequations}
where $\ket{\uparrow}$ and $\ket{\downarrow}$ are spin-$1/2$ eigenstates of $S_x$, and we determine the corresponding IR for each state using the symmetry coupling coefficients for the double group \cite{Koster1963}.  In this basis, the general Hamiltonian of eq.~\ref{eq:SpinH} takes the form:
\begin{equation}
H = \begin{pmatrix}
-\frac{2D}{3} & iB_{z}g_{zz}\mu_\mathrm{B} & B_{y}g_{yy}\mu_\mathrm{B} \\ \\
-iB_{z}g_{zz}\mu_\mathrm{B} & \frac{D}{3} - E & B_{x}g_{xx}\mu_\mathrm{B} \\ \\
B_{y}g_{yy}\mu_\mathrm{B} & B_{x}g_{xx}\mu_\mathrm{B} &   \frac{D}{3} + E \\ \\
\end{pmatrix}
\end{equation}
Here we notice that each Cartesian component of the magnetic field mixes one pair of spin eigenstates.  In particular, in-plane components of the field mix $\ket{s_{x}}$  and $\ket{s_y}$  with $\ket{s_z}$, \textit{i.e.}, in polar coordinates:
\begin{equation}
H|_{B_{z}=0} = \begin{pmatrix}
-\frac{2D}{3} & 0 & g_{yy}\mu_\mathrm{B}B\sin(\phi) \\ \\
0 & \frac{D}{3} - E & g_{xx}\mu_\mathrm{B}B\cos(\phi) \\ \\
g_{yy}\mu_\mathrm{B}B\sin(\phi) & g_{xx}\mu_\mathrm{B}B\cos(\phi) &   \frac{D}{3} + E \\ \\
\end{pmatrix}
\end{equation}
Such mixing is essential to our models of spin-dependent fluorescence, and the symmetry is important.  In this case, the eigenstates exhibit underlying 180$^\circ$ rotational symmetry (they depend on the relative sign between the two mixing terms but not the overall sign).  However, we will show below that in situations where the optical dynamics depend on the overall projection of the triplet eigenstates on $\ket{s_z}$, or on mixing between $\ket{s_x}$  and $\ket{s_y}$  (which for the case of in-plane fields can only occur through $\ket{s_z}$), the resulting fluorescence can exhibit 90$^\circ$ symmetry, consistent with our observations. 

\begin{figure}[tbp]
  \includegraphics[width=3in]{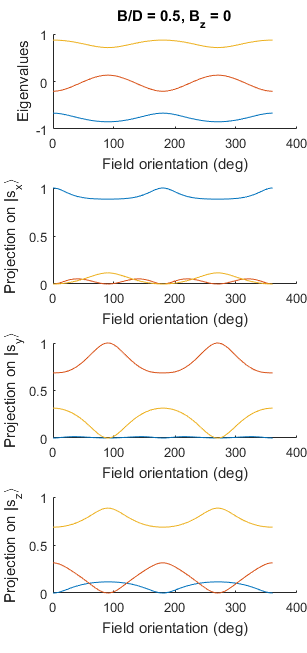}
  \caption{\textbf{Spin-Triplet Eigenstates.} Energy eigenvalues (units of $D$) and corresponding projections on the $\{\ket{s_x},\ket{s_y},\ket{s_z}\}$ basis, as a function of in-plane orientation of an applied magnetic field.}
  \label{fig:TripletEigenstates}
\end{figure}

For example, Fig.~\ref{fig:TripletEigenstates} shows the spin eigenvalues and projections on each basis vector as a function of in-plane magnetic field direction, when $E=0.2D$ and $g\mu_\mathrm{B}B/D = 0.5$.  Note that the overall pattern has 180$^\circ$ symmetry, but that the mixing between states dominated by $|s_{x}\rangle$  and $|s_{y}\rangle$  (blue and orange) exhibits 90 deg symmetry.

\vspace{6pt}
\noindent\textit{S$=\frac{3}{2}$: Quartet configuration}

The quartet is split by the ZFS into two Kramers doublets at energies  $\pm{\sqrt{D^2 + 3E^2}}$.  The zero-field Hamiltonian in the x-projection basis, \textit{m$_{s}$} = \{${\frac{3}{2},\frac{1}{2},-\frac{1}{2},-\frac{3}{2}}$\} takes the form: 
\begin{equation}
H = \begin{pmatrix}
D & 0 & \sqrt{3}E & 0 \\ \\
0 & -D & 0 & \sqrt{3}E \\ \\
\sqrt{3}E & 0 & -D & 0 \\ \\
0 & \sqrt{3}E & 0 & D \\ \\
\end{pmatrix}
\end{equation}
Assuming D$\gg$E, the doublets consist mostly of the \textit{m$_{s}$}=$\pm{\frac{1}{2}}$ and $\pm{\frac{3}{2}}$ eigenstates, respectively, but the $E$ term slightly mixes the pairs of states \{${\frac{3}{2},{-}\frac{1}{2}}$\} and \{${\frac{1}{2},{-}\frac{3}{2}}$\}.  This makes sense from a group theory perspective, since there is only one double group representation for $C_{2v}$, and so both doublets will transform under the same IR: $E_{1/2}$.  It is also easy to show using the coupling coefficients in Ref.~\onlinecite{Koster1963} that the combinations $\ket{\uparrow\uparrow\uparrow}\sim\ket{\frac{3}{2}}$ and $\ket{\uparrow\downarrow\downarrow}\sim\ket{-\frac{1}{2}}$ both transform like $E_{1/2}^{-1/2}$ whereas $\ket{\downarrow\downarrow\downarrow}\sim\ket{-\frac{3}{2}}$ and $\ket{\uparrow\uparrow\downarrow}\sim\ket{\frac{1}{2}}$ both transform like  $E_{1/2}^{+1/2}$, which is consistent with the predicted mixing even at zero field.

\begin{figure}[tbp]
  \includegraphics[width=3in]{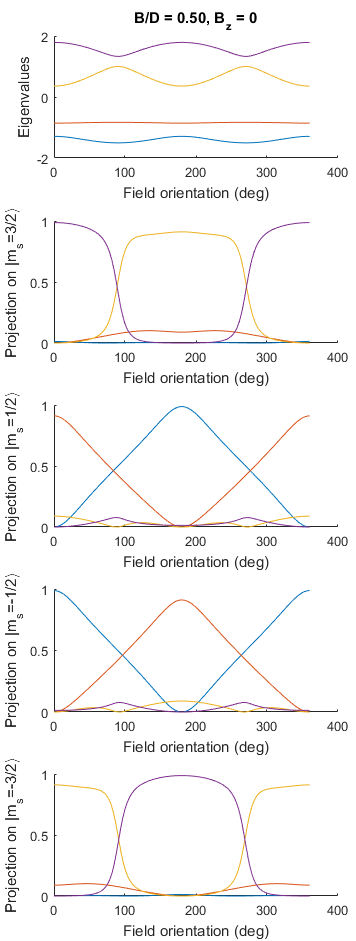}
  \caption{\textbf{Spin-Quartet Eigenstates.} Energy eigenvalues (units of $D$) and corresponding projections on the $S_x$ basis, $m_s=\{3/2,1/2,-1/2,-3/2\}$, as a function of in-plane orientation of an applied magnetic field.  Zero-field splitting parameters are the same as in Fig.~\ref{fig:TripletEigenstates}.}
  \label{fig:QuartetEigenstates}
\end{figure}

Notably, however, the rotational symmetry in response to in-plane magnetic fields is different than for the triplet case.  The Hamiltonian as a function of $B$ in polar coordinates becomes:
\begin{equation}
H|_{B_{z}=0} = \begin{pmatrix}
D + \frac{3}{2}B\cos[\phi]g_{xx}\mu_\mathrm{B} & \frac{1}{2}\sqrt{3}B\sin[\phi]g_{yy}\mu_\mathrm{B} & \sqrt{3}E & 0 \\ \\
\frac{1}{2}\sqrt{3}B\sin[\phi]g_{yy}\mu_\mathrm{B} & -D + \frac{1}{2}B\cos[\phi]g_{xx}\mu_\mathrm{B} & B\sin[\phi]g_{yy}\mu_\mathrm{B} & \sqrt{3}E \\ \\
\sqrt{3}E & B\sin[\phi]g_{yy}\mu_\mathrm{B} & -D - \frac{1}{2}B\cos[\phi]g_{xx}\mu_\mathrm{B} & \frac{1}{2}\sqrt{3}B\sin[\phi]g_{yy}\mu_\mathrm{B} \\ \\
0 & \sqrt{3}E & \frac{1}{2}\sqrt{3}Bsin[\phi]g_{yy}\mu_\mathrm{B} & D - \frac{3}{2}Bcos[\phi]g_{xx}\mu_\mathrm{B} \\ \\
\end{pmatrix}
\end{equation}
Here, whereas the eigenvalues still exhibit 180$^\circ$ rotational symmetry, the eigenvectors exhibit 360$^\circ$ deg symmetry. Figure~\ref{fig:QuartetEigenstates} gives an example using the same parameters as for the triplet case in Fig.~\ref{fig:TripletEigenstates}.
There is no indication of couplings between the states that could give rise to dynamics with 90$^\circ$ symmetry in this case.

Even more importantly, we will explore below how spin-orbit interactions give rise to spin-dependent selection rules for the ISC between singlet and triplet states, which subsequently produce changes in the PL as a function of magnetic field orientation.  For the case of the doublet-to-quartet ISC, however, such selection rules do not arise naturally on the basis of symmetry, since there is only one double-group representation in $C_{2v}$ that must describe all states in both manifolds.  For these reasons, we believe it is unlikely that a defect with S=$\frac{1}{2}$ or S=$\frac{3}{2}$ configurations can be consistent with our observations, and we do not consider these systems further.

\subsection*{Electronic Level Structure: Jablonski Diagrams}

Following the MO theory treatment \cite{Stoneham1975}, we enumerate the configurations of single-particle and multi-particle energy levels that can arise from defects with $C_{2v}$ symmetry, together with their radiative (optical dipole) and non-radiative (ISC) transition selection rules.  Based on the considerations above, we consider cases including spin singlet and triplet manifolds, either of which could contain the ground state (GS) and fluorescent excited-state (ES).

\vspace{6pt}
\noindent\textit{Spin-singlet ground state:}

\begin{figure}[tbp]
  \includegraphics[width=5in]{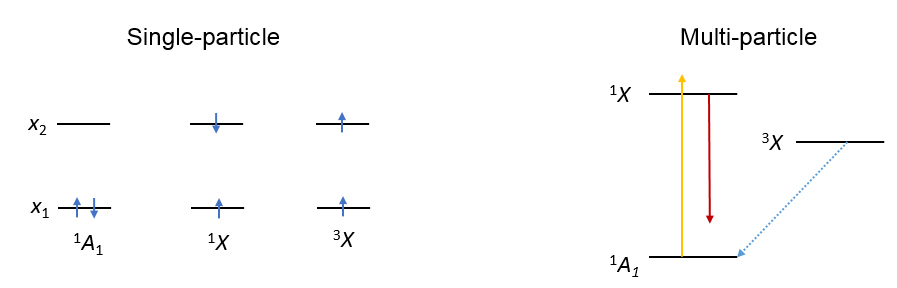}
  \caption{\textbf{Electronic configurations (Singlet-GS).} Two single-particle levels of arbitrary symmetry $x_1$ and $x_2$ occupied by two electrons give rise to three multi-particle levels as shown at right.}
  \label{fig:SingletConfigurations}
\end{figure}

The simplest case we could consider includes only two single-particle (SP) levels active in the optical dynamics, which can be arranged in a ground or excited-state singlet, or a triplet state with intermediate energy, as shown in Fig.~\ref{fig:SingletConfigurations}.
Based on the multiplication rules (Table~\ref{tab:c2v_multiplication}), we know that the ground state must have symmetry $^{1}A_{1}$ independent of the symmetry of the SP levels. The symmetry of the singlet ES and optical selection rules are similarly determined from the group multiplication table. In modeling the ISC, we assume the transition is mediated by spin-orbit coupling, with symmetric phonons accounting for the energy relaxation. The allowed transitions are those for which the total spin-orbit symmetry of the system is conserved. (This is the case, for example, for the $(^{3}E\rightarrow ^1\!A_{1})$ ISC for the diamond NV center, whose spin-selectivity arises since there exists a combination of $m_{s}=\pm1$ levels in $^{3}E$ that have spin-orbit symmetry $A_{1}$, forming an allowed transition to the singlet level which also transforms like $A_{1}$.)  Therefore, it is helpful to enumerate the spin-orbit symmetry of each spin sublevel using the symmetrized spin-triplet states of eq.~(\ref{eq:TripletBasis}) and the group multiplication rules (Table~\ref{tab:c2v_multiplication}).  The resulting spin-orbit representations are listed in Table~\ref{tab:SpinOrbitIRs}:
\begin{table*}[h!] 
\centering
\caption{\textbf{Spin-Orbit Representations}}
\label{tab:SpinOrbitIRs}
\setlength{\tabcolsep}{7.5pt}
\setlength{\extrarowheight}{7.5pt}
\begin{tabularx}{0.3\textwidth}{ c | l  l l}
\hline 
 \textbf{Orbital} & $\ket{S_x}$ & $\ket{S_y}$ & $\ket{S_z}$ \\  \hline

$^{3}A_{1}$ & $|A_{2}\rangle$ & $|B_{2}\rangle$ & $|B_{1}\rangle$  \\ 

$^{3}A_{2}$ & $|A_{1}\rangle$ & $|B_{1}\rangle$ & $|B_{2}\rangle$ \\ 

$^{3}B_{1}$ & $|B_{2}\rangle$ & $|A_{2}\rangle$ & $|A_{1}\rangle$  \\ 

$^{3}B_{2}$ & $|B_{1}\rangle$ & $|A_{1}\rangle$ & $|A_{2}\rangle$ \\ \hline
 
\end{tabularx}
\end{table*}

Using Table~\ref{tab:SpinOrbitIRs}, we can determine which spin eigenstate is allowed to pass through the ISC.  Noting that the spin-orbit symmetry of all three triplet states is distinct from that of the orbital state alone, we observe that there are no spin-orbit-allowed transitions between the states $^{1}X$ and $^{3}X$ in Fig.~\ref{fig:SingletConfigurations}.  These transitions can be allowed via other means (\textit{e.g.}, through spin-spin interactions or asymmetric phonons) but are not likely to be strongly spin selective.  Depending on the symmetry of $X$, there could be a spin-selective ISC back to the $^{1}A_{1}$ ground state.

\begin{figure}[tb]
  \includegraphics[width=4in]{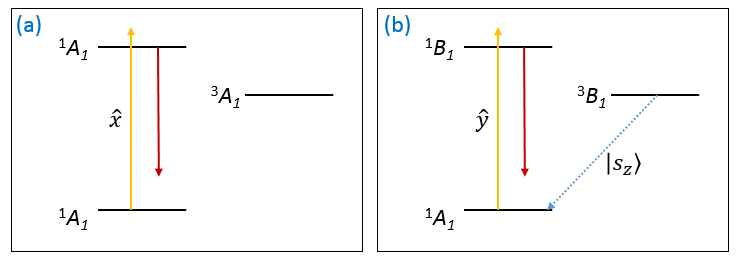}
  \caption{\textbf{Jablonski Diagrams (Singlet-GS).} Level diagrams as in Fig.~\ref{fig:SingletConfigurations} consistent with in-plane optical dipole transitions.}
  \label{fig:SingletDiagrams}
\end{figure}

In our experiments, we observe optical dipole transitions with in-plane linear polarization selection rules.  There are only two possibilities for the multi-particle level structure that are consistent with this observation, as shown in Fig.~\ref{fig:SingletDiagrams}.
In the case of Fig.~\ref{fig:SingletDiagrams}(a), the two SP orbitals are the same ($x_{1} = x_{2}$), and there are no spin-orbit-allowed ISC transitions.  We should not expect to see any spin-dependent effects in this situation. In Fig.~\ref{fig:SingletDiagrams}(b), the two SP orbitals are different (chosen from either \{$A_{1},B_{1}$\}  or \{$A_{2},B_{2}$\}), and the metastable triplet decay should be spin dependent.

\begin{figure}[tb]
  \includegraphics[width=5in]{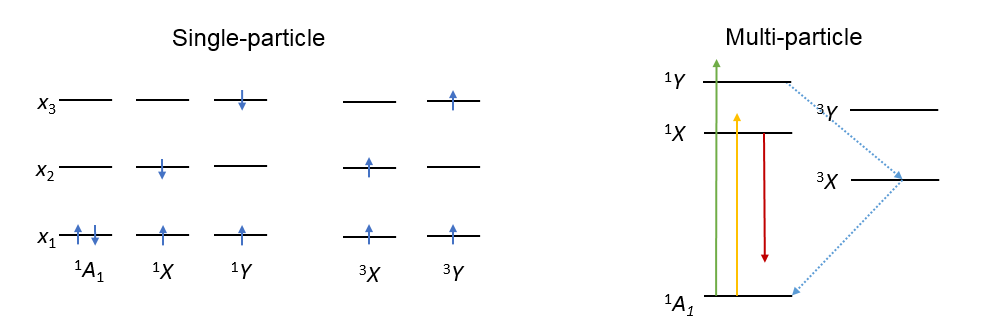}
  \caption{\textbf{Additional excited states (Singlet-GS).} Generalized electronic configurations formed from three SP states and a singlet GS.}
  \label{fig:SingletConfigurations2}
\end{figure}

It is also possible that more than two SP levels are involved.  In fact, we suspect additional excited states are likely to play a role based on the difference between the polarization dependence of the absorptive and emissive optical transitions.  For example, the next-simplest level structure is shown in Fig.~\ref{fig:SingletConfigurations2}.
Still, since the PL is strongly polarized in plane, the state $^1X$ is probably $^1A_{1}$ or $^1B_{1}$ as above, so the only qualitative change here might be that the brief occupation of the state $^{1}Y$ might contribute a spin-dependent pathway to the $^{3}X$ triplet state, which will be symmetry-allowed if $X\neq Y$. Depending on the level spacings, a further $^1 A_1$ singlet excited state resulting from double occupation of the $x_2$ SP level might also be close in energy to $^1 Y$.
However, any such states would be transiently occupied, most likely for a timescale shorter than the optical relaxation time ($\sim$1 ns) and therefore we ignore them hereafter from the point of view of simulating ISC dynamics.

\vspace{6pt}
\noindent\textit{Spin-triplet ground state:}

\begin{figure}[tb]
  \includegraphics[width=5in]{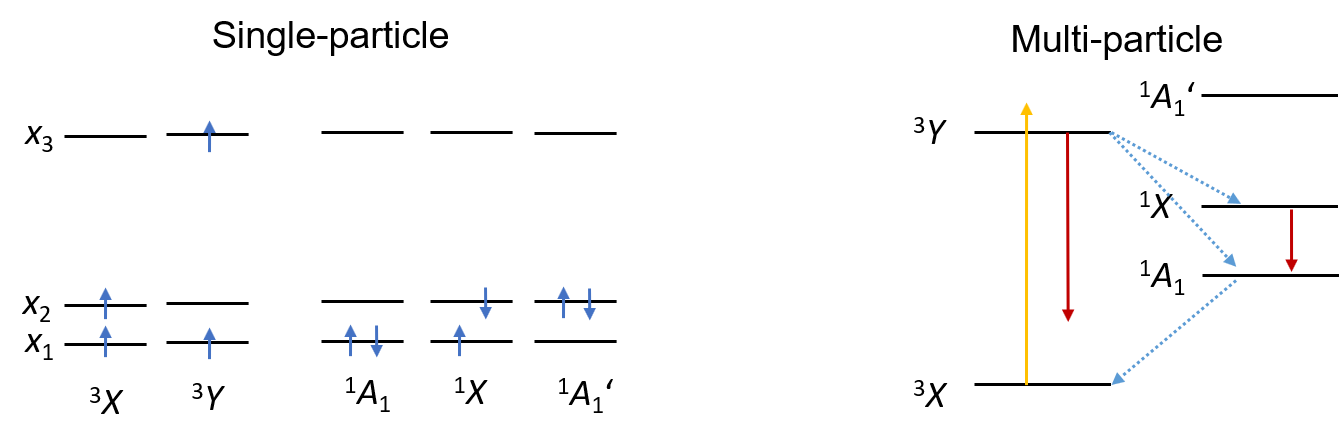}
  \caption{\textbf{Electronic configurations (Triplet-GS).} Three single-particle levels of arbitrary symmetry $x_1,x_2,x_3$ occupied by two electrons give rise to multi-particle levels as shown at right.}
  \label{fig:TripletConfigurations}
\end{figure}

For the case of a ground-state spin triplet, we need at least three SP levels to encompass the optical ground and excited states, as shown in Fig.~\ref{fig:TripletConfigurations}. As before, the lowest-lying singlet state must have symmetry $^{1}A_{1}$.  Since the orbital configurations $X$ and $Y$ can each be chosen from one of the four IRs in $C_{2v}$, this arrangement gives rise $4^{2} = 16$ potential configurations of multi-particle states. But again, we can reduce this set by considering only those configurations which exhibit an optical selection rule for in-plane polarization between $^{3}$X and $^{3}$Y.  Based on the $C_{2v}$ multiplication table, we see that:
\begin{itemize}
\item If the emission is polarized along $x$, then $X=Y$.
\item If the emission is polarized along $y$, then $\{X,Y\} \in \{A_{1}, B_{1}\}$ or $\{A_{2}, B_{2}\}$.
\end{itemize}
This reduces the number of possible combinations from 16 to 8.

We also note that there will be singlet excited states that have orbital parts of the form $|x_{1} x_{2}\rangle + |x_{2} x_{1}\rangle$, and similarly for the $x_{1}x_{3}$ combination, whose total orbital symmetry will be the same as the triplet states $X$ and $Y$, respectively. Especially if the states $x_{1}$ and $x_{2}$ are close in energy, as would be expected if exchange splitting is strong enough to make $^{3}X$ the ground state, we should expect another singlet state $^{1}X$ of the same symmetry to be close in energy to $^{1}A_{1}$, so it can also potentially contribute to the ISC. As shown in Fig.~\ref{fig:TripletConfigurations}, an additional $^1A_1^\prime$ state composed of the $(x_2)^2$ SP-level configuration could also play a role in the dynamics, although it is likely that Coulomb interactions will increase the energy spacing between two singlet levels of the same symmetry such as $^1A_1^\prime$ and $^1A_1$.  Finally, similarly to the case of the singlet-GS configurations in Fig.~\ref{fig:SingletConfigurations2}, additional triplet excited states could play a role in off-resonant absorption (\textit{e.g.}, the triplet state composed of the $(x_2x_3)$ SP-level configuration), but we assume they do not contribute to the radiative or ISC relaxation pathways.

\begin{figure}[tb]
  \includegraphics[width=4in]{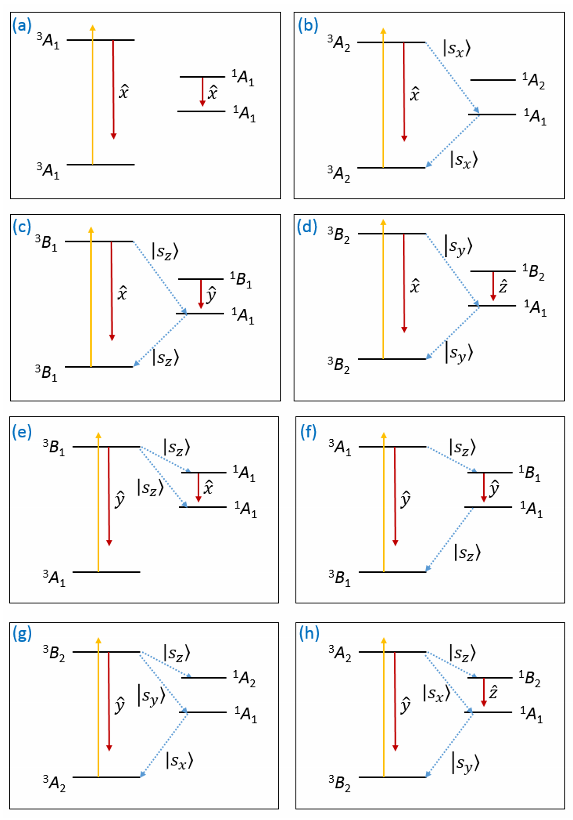}
  \caption{\textbf{Jablonski Diagrams (Triplet-GS).} Level diagrams as in Fig.~\ref{fig:TripletConfigurations} consistent with in-plane optical dipole transitions.}
  \label{fig:TripletDiagrams}
\end{figure}

We account for all of these possibilities by varying the choice of ISC selection rules. 
As before, the spin-dependent ISC selection rules are determined by the spin-orbit-symmetrized triplet configuration (Table~\ref{tab:SpinOrbitIRs}) that transforms identically to the corresponding singlet state.  We assume that optical-dipole allowed relaxation within the singlet levels means that the final ISC transition will always occur from the lowest-lying $^1A_1$ state.  
The 8 resulting Jablonski diagrams are listed in Fig.~\ref{fig:TripletDiagrams}.

\clearpage
\subsection*{Modeling Optical Dynamics:}

\begin{figure}[bp]
  \includegraphics[width=3in]{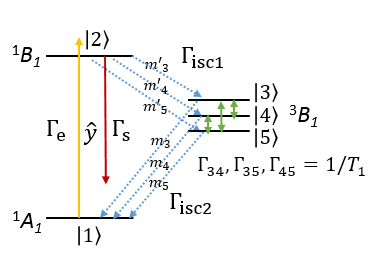}
  \caption{\textbf{Transition Diagram (Singlet GS Model).} Rates and selection rules determine the structure of a master equation for the optical dynamics of singlet-GS models.}
  \label{fig:Transitions_SingletGS}
\end{figure}

We use a semiclassical Master Equation (ME) model to simulate the orbital and spin dynamics of these systems under optical illumination.  For example, level structure (b) of the singlet-GS cases consists of five individual states, coupled by rates as indicated in Fig.~\ref{fig:Transitions_SingletGS}.
Here $\Gamma_{e}$  is the optical excitation rate, $\Gamma_{s}$  is the spontaneous radiative decay rate, and $\Gamma_\mathrm{ISC1}$, $\Gamma_\mathrm{ISC2}$  are the non-radiative ISC rates to and from the metastable triplet state, respectively.  The ISC spin selection rules are encapsulated in the coefficients $m_{i}$, $m_{i}\prime$ given by
\begin{equation}
m_{i} = \sum_{\mu}p_{\mu}|\langle{s_{\mu}|s_{i}}\rangle|^{2},
\end{equation}
where $\mu \in \{x,y,z\}$  and $p_{\mu}$  are the normalized selection rules between the singlet and the corresponding zero-field eigenstate $|s_{\mu}\rangle$.  This incoherent sum over projections onto the field-dependent eigenstates $|s_{i}\rangle$  corresponds to the usual assumption that the ISC transition are incoherent, \textit{i.e.}, that the triplet state resulting from an ISC is described by a density matrix
\begin{equation}
\rho_{triplet} = m_{1}|s_{1}\rangle \langle s_{1}| + m_{2}|s_{2}\rangle \langle s_{2}| + m_{3}|s_{3}\rangle \langle s_{3}|
\end{equation}
We can also include additional spin relaxation processes within the triplet state.  With knowledge about a specific process (e.g., hyperfine or spin-phonon coupling) one could include decoherence using the Lindblad ME approach.  Here, in order to capture the qualitative effects, we simply include spin relaxation through a set of uniform transition elements connecting all three pairs of triplet states, at the rate $1/T_{1}$.  
Hence the ME takes the form 
$\dot{\textbf{\textit{x}}}=\textit{R\textbf{x}}$, where the off-diagonal elements of the rate matrix are
\begin{equation}
\textit{R} - \mathrm{diag}(\textit{R}) = \begin{pmatrix}
\dottedbox & \Gamma_{s} & m_{3}\Gamma_\mathrm{ISC2} & m_{4}\Gamma_\mathrm{ISC2} & m_{5}\Gamma_\mathrm{ISC2} \\ \\
\Gamma_{e} & \dottedbox & 0 & 0 & 0\\ \\
0 & m_{3}{\prime}\Gamma_\mathrm{ISC1} & \dottedbox & \frac{1}{T_{1}} & \frac{1}{T_{1}} \\ \\
0 & m_{4}{\prime}\Gamma_\mathrm{ISC1} & \frac{1}{T_{1}} & \dottedbox & \frac{1}{T_{1}} \\ \\
0 & m_{5}{\prime}\Gamma_\mathrm{ISC1} & \frac{1}{T_{1}} & \frac{1}{T_{1}} & \dottedbox \\ \\
\end{pmatrix}
\end{equation}
and the diagonal components are simply $R_{ii} = -\Sigma_{j}R_{ji}$ in order to conserve total probability. 

\begin{figure}[!h]
  \includegraphics[width=2.5in]{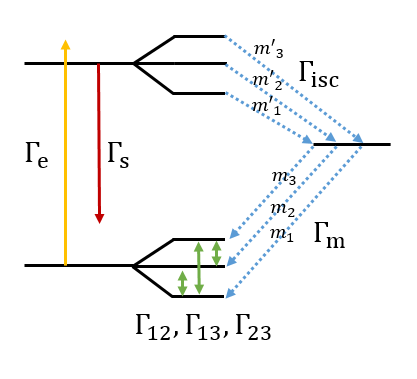}
  \caption{\textbf{Transition Diagram (Triplet GS Model).} Rates and selection rules determine the structure of a master equation for the optical dynamics of triplet-GS models.}
  \label{fig:Transitions_TripletGS}
\end{figure}

The ME for the triplet-GS cases are constructed in a similar way for a system with seven states as shown in Fig.~\ref{fig:Transitions_TripletGS}.
Here we additionally make the assumption that optical excitation and emission conserve the triplet spin state, ignoring spin-mixing transitions that might occur due to the fact that the ground and excited-state spin Hamiltonians are different. We also ignore spin relaxation in the triplet excited state due to its short ($\sim$1 ns) lifetime.

\vspace{6pt}
\noindent\textit{Steady-state PL and the g$^{(2)}$ function:}

Once the master equation is constructed for a given set of parameters and field settings, the steady-state PL is calculated from the solution of $\langle{\dot{\textbf{\textit{x}}}\rangle} = \textit{R}\langle{\textit{\textbf{x}}}\rangle = \textbf{0}$, \textit{i.e.}, from the null space of $R$.   In the general case, $\langle \mathrm{PL} \rangle = \sum_{i\in\mathrm{GS},j\in\mathrm{ES}}R_{ij}\langle{x_{j}}\rangle$, which reduces to $\langle \mathrm{PL} \rangle = \Gamma{_{s}}\sum_{i\in\mathrm{ES}}\langle{x_{i}}\rangle$ in the case of spin-conserving emission.

The autocorrelation function, meanwhile, is calculated by numerically integrating the ME starting with an initial condition corresponding to the configuration that follows emission of a photon. For the singlet-GS case, this initial state is simply $x_\mathrm{GS}(t=0) = 1$ , whereas for the triplet-GS case the initial population is distributed in the ground state according to
\begin{equation}
\mathbf{x}_\mathrm{GS}(0) = P \cdot \frac{\langle{\mathbf{x}_\mathrm{ES}}\rangle}{\sum{\langle{\mathbf{x}_\mathrm{ES}}}\rangle},
\end{equation}
where $P_{ij} = R_{ij}/\sum_{i}R_{ij}$ gives the branching ratio probabilities for decay from ES state \textit{j} to GS state \textit{i}. In the case where the transitions are spin conserving, $P = I$. [Note that, in general $\mathbf{x}_\mathrm{GS}(0) \neq \langle{\mathbf{x}_\mathrm{GS}}\rangle$].
Given this initial condition, the autocorrelation function is related to the subsequent evolution of the excited-state population \textit{via}
\begin{equation}
g^{(2)}(t) = \frac{\mathrm{PL}(t)}{\langle\mathrm{PL}\rangle} = \frac{\sum_{i\in\mathrm{GS},j\in\mathrm{ES}}R_{ij}x_{j}(t)}{\langle\mathrm{PL}\rangle}
\end{equation}

\vspace{6pt}
\noindent\textit{Free parameters in simulations}

The simulations require values for a number of parameters, some of which can be related directly to experimental observations whereas others are less constrained.  The spin-triplet Hamiltonian depends on the zero-field splitting parameters $D$ and $E$.  For typical molecules and point defects, these parameters have values ranging from a few hundred megahertz to a few gigahertz ($\sim$1-10 $\mu$eV in energy units).  Therefore, in simulations we scale the magnetic field in units of $D/g \mu_\mathrm{B}$, and explore the effect of changing both the sign and value of the ratio $E/D$.

The ISC spin-selection rules that determine the coupling coefficients $(m_{i},m_{i}\prime)$ follow from the spin-orbit allowed transitions indicated in the corresponding Jablonski diagrams (Figs.~\ref{fig:SingletDiagrams} and \ref{fig:TripletDiagrams}).  In cases where no spin-orbit transition is allowed by symmetry, we assume the transition proceeds spin-nonselectively \textit{via} other means, \textit{i.e.}, $p_{x}=p_{y}=p_{z}=\frac{1}{3}$.  We also explore the effect of relaxing the predicted spin-selectivity in some cases, \textit{e.g.} setting an $|s_{x}\rangle$-selective transition to have $p_{x} = 1 - 2\epsilon$,  while $p_{y} = p_{z} = \epsilon$, as a function of a small parameter $\epsilon {\ll} 1$.

The rates $\Gamma_{s}$, $\Gamma_{e}$, $\Gamma_\mathrm{ISC1}$, and $\Gamma_\mathrm{ISC2}$  can be estimated from an analytical three-level model for the autocorrelation function \cite{Basche1996}. Full fluorescence saturation curves were not recorded for fear of photobleaching the emitters of interest, so we introduce a free parameter to quantify the relative saturation of the optical dipole transition, $x = \Gamma_{e}/\Gamma_{s}$.  We expect that $x \approx$ 0.1--0.5 based on partial saturation curves.
In this case, and assuming $\Gamma_\mathrm{ISC2} \ll \Gamma_\mathrm{ISC1}$  (as justified by later analysis), a three-level model yields approximate analytic expressions for the underlying rates as a function of the observed antibunching ($\tau_{1}$) and bunching ($\tau_{2}$) timescales and the bunching amplitude ($C_{2}$) in the fluorescence autocorrelation function, which takes the general form:
\begin{equation}
g^{2}(t) = 1 - C_{1}e^{-t/\tau_{1}} + C_{2}e^{-t/\tau_{2}},
\end{equation}
where
\begin{subequations}
\begin{align}
\Gamma_{s} & \approx \frac{1}{\tau_{1}}\left(\frac{1}{1+x}\right),\\
\Gamma_\mathrm{ISC2} & \approx \frac{1}{\tau_{2}}\left(\frac{1}{1+C_{2}}\right),\\
\Gamma_\mathrm{ISC1} & \approx \frac{1+x}{x}\left(\frac{1}{\tau_{2}}-\Gamma_\mathrm{ISC2}\right). 
\end{align}
\end{subequations}
Of these values, only $\Gamma_\mathrm{ISC1}$  varies strongly with \textit{x} in the expected range $x\approx0.1$--0.5. This implies that the true value of $\Gamma_\mathrm{ISC1}$  remains uncertain based on our measurements, but nevertheless the simulations are robust to variations in \textit{x} when parameterized in this way.  For example, the average values of the autocorrelation parameters based on the zero-field measurements shown in Fig.~3 of the main text are (see Table~\ref{tab:g2parameters}):
\begin{quote}
$\tau_{1}$ = 1.1 ns, $\tau_{2}$ = 1.4 µs, \textit{C} = 5.4
\end{quote}
Assuming \textit{x} = 0.5, this implies that
\begin{quote}
$\Gamma_{s} \approx$ 600 MHz,  $\Gamma_\mathrm{ISC1} \approx$ 1.8 MHz,  $\Gamma_\mathrm{ISC2} \approx$ 0.11 MHz
\end{quote}

\subsection*{Results and Discussion}

We performed calculations using this model for all of the level structures described in the previous section.  Simulations of the PL as a function of in-plane magnetic field across a range of parameters are shown in the appendix.  Since we lack a concrete estimate for the triplet spin-anisotropy parameter ($E/D$) and the spin lifetime ($T_{1}$), we performed calculations spanning a range of their values.

For the case of a singlet ground-state, there is only one level-structure of interest [Fig.~\ref{fig:SingletDiagrams}(b)], since Fig.~\ref{fig:SingletDiagrams}(a) does not include spin-dependent selection rules.  The simulations of this system exhibit clear variations of steady-state PL as a function of in-plane field, with four bright lobes offset from the $x$ and $y$ axes.  The pattern exhibits near fourfold symmetry especially near $E/D\sim -0.3$, and the contrast between bright and dark regions increases with $T_{1}$.  The PL exhibits a minimum at $B=0$, and maintains this minimum value for $\mathbf{B}$ applied along the $z$ axis.  As discussed below, this model reproduces many features of our experiments, and is a leading candidate to explain the observations.

The simulations of triplet-ground-state systems (Fig.~\ref{fig:TripletDiagrams}) generally fall into distinct categories based on the arrangement of ISC selection rules, namely:
\begin{itemize}
\item \textit{Class I: No Spin Selection Rules}

In diagram (a), similar to case (a) of the singlet-GS configuration, no ISC transitions are allowed by spin-orbit coupling.  It is possible that ISC transitions can proceed via different means, but they are likely to be spin-nonselective and therefore do not depend on an external magnetic field.

\item \textit{Class II: Symmetric ISC involving $|s_{x}\rangle$ or $|s_{y}\rangle$}

In diagrams (b) and (d), both ISC transitions are selective to the same spin projection.  This configuration does produce anisotropic changes in PL as a function of field direction, however the modulation amplitude is small when the experimentally-relevant ISC rate parameters are used in simulations (the contrast is only $\sim0.1\%$ in the appendix simulations).  Moreover, the patterns general exhibit 180$^\circ$ rather than 90$^\circ$ symmetry, with a clear difference in brightness for a field oriented along $x$ or $y$.  Finally, the PL is predicted to decrease slightly as a function of increasing $B$ applied along $z$, in contrast to our experiments.

\item \textit{Class III: ISC involving $|s_{z}\rangle$ only}

Diagrams (c), (e), and (f) involve $|s_{z}\rangle$-selective ISC transitions only.  Simulations of these level structures exhibit patterns of PL variations broadly similar to the singlet-GS diagram \ref{fig:SingletDiagrams}(b) discussed above, with approximate fourfold symmetry as a function of in-plane field.  The triplet-GS patterns are inverted with respect to the singlet-GS case, with a global PL maximum at $B=0$ and four dark lobes when the field is not aligned with either the $x$ or $y$ axes.  The PL is independent of $B$ applied along $z$ (\textit{i.e.}, it remains bright).  Diagram (e), consisting of one $|s_{z}\rangle$-selective ISC transition and one spin-nonselective transition, exhibits a large PL contrast, whereas the contrast for diagrams (c) and (f) is much smaller than we observe in experiments.

\item \textit{Class IV: Asymmetric ISC involving $|s_{x}\rangle$ and $|s_{y}\rangle$}

In the case of diagrams (g) and (h), if the upper and lower ISC transitions proceed entirely through the lowest ($^{1}A_{1}$) singlet state, the result is a pattern of PL variations as a function of in-plane field orientation with approximate four-fold symmetry and\,---\,especially for increasing T$_{1}$\,---\,relatively strong contrast.  These features generally agree with our observations; however, the simulations also predict that the PL should reach a maximum at $B=0$ and decrease strongly for $B$ applied along \textit{z}, in clear contradiction to experiments.

\item \textit{Class V: Asymmetric ISC involving $|s_{x}\rangle$ or $|s_{y}\rangle$ and $|s_{z}\rangle$}

Alternative schemes for diagrams (g) and (h) invoke the ISC from the triplet excited state to the higher-lying singlet state, which is selective for the $|s_{z}\rangle$ spin projection.  Simulations for these schemes exhibit strongly anisotropic PL as a function of in-plane field orientation, but clear 180$^{\circ}$ symmetry as compared to the 90$^{\circ}$ symmetry we observe in experiments.  A further pattern emerges when comparing simulations for asymmetric ISC level structures in classes IV and V: the PL is generally bright for fields oriented along the two orthogonal directions corresponding to allowed spin transitions but strongly suppressed when the field is applied along the third axis.

\end{itemize}
On the basis of these simulations, we conclude that configurations in Class III, especially diagram (e), together with the singlet-GS case (b), exhibit the greatest similarity with our experiments. We subsequently investigated the behavior of these two models in greater detail, considering especially the predicted behavior of the photon emission autocorrelation function as a function of magnetic field.

By comparing the simulations of steady-state PL and the photon autocorrelation function to our experimental observations, we arrived at the set of model parameters listed in Table~\ref{tab:ModelParameters} that are used for the simulations presented in Figure 4 of the main text.
In both cases, we set $E/D = −0.33$, corresponding to the configuration with clearest 90$^\circ$ symmetry for in-plane fields.  We also set the saturation parameter $x = 0.1$, which subsequently determines the relative magnitude of $\Gamma_{e}$  and $\Gamma_\mathrm{ISC1}$  as discussed previously.  We confirmed that the results are generally independent of $x \in[0.1,1]$ as long as other parameters are scaled appropriately.

\begin{table*}[b] 
\centering
\caption{\textbf{Simulation Parameters.}}
\label{tab:ModelParameters}
\setlength{\tabcolsep}{7.5pt}
\setlength{\extrarowheight}{7.5pt}
\begin{tabularx}{0.8\textwidth}{c c c}
\hline 
 \textbf{Parameter} & \textbf{Singlet-GS [model (b)]} & \textbf{Triplet-GS [model (e)]}  \\  \hline

$E/D$ & -0.33 & -0.33   \\ 
$T_{1}$ ($\mu$s) & 50 & 50  \\ 
$\Gamma_{s}$ (MHz) & 820 & 820   \\ 
$\Gamma_{e}$ (MHz) & 82 & 82  \\ 
$\Gamma_\mathrm{ISC1}$ (MHz) & 7.7 & 33  \\ 
$\Gamma_\mathrm{ISC2}$ (MHz) & 0.85 & 0.13   \\ 
$\epsilon^{\ast}$ & 0.02 & 0.05  \\ 
\hline
\multicolumn{3}{l}{$^\ast$\footnotesize{The parameter $\epsilon$ relaxes the spin selectivity of the spin-orbit-allowed transition as described in the text.}}
\end{tabularx}
\end{table*}

Note that no quantitative fit or numerical optimization procedure was performed. Rather, we manually adjusted parameters in order to approximate the key observations.  We therefore do not make a claim regarding confidence intervals or the uniqueness of this solution. It is clear, in fact, that some parameters exhibit strong covariance with respect to observations (the covariance of the optical excitation rate and the ISC rate $\Gamma_\mathrm{ISC1}$  is one such example).  Furthermore, various subtle features of the data are not reproduced by our simple model.  Our primary goal is to establish the feasibility of simple electronic level structures to explain the main features in experiments and ideally to uncover the leading candidates such that they can guide future measurements and calculations.

\begin{figure}[tbp]
  \includegraphics[width=5in]{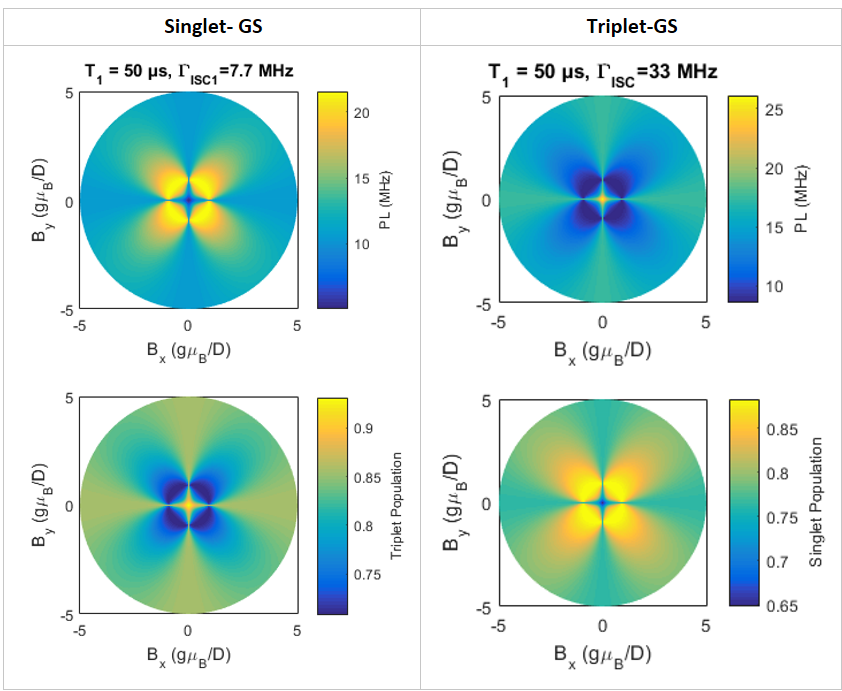}
  \caption{\textbf{PL Simulations.} Steady-state PL as a function of in-plane magnetic field (top panels) as well as the steady-state population in the metastable state (bottom panels).}
  \label{fig:PLsimulations}
\end{figure}

\begin{figure}[tbp]
  \includegraphics[width=4in]{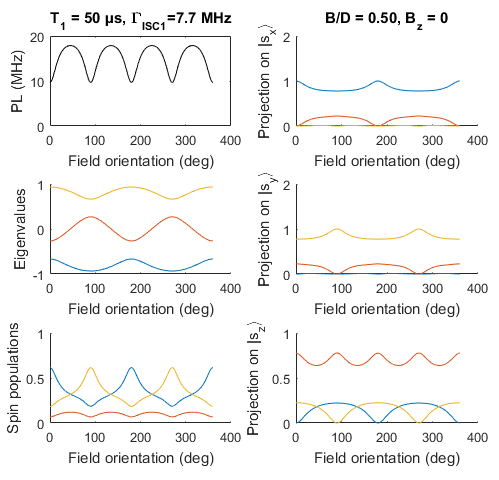}
  \caption{\textbf{Spin Simulations.} Steady-state PL as a function of in-plane magnetic field orientation (top left) for the singlet-GS model together with the eigenvalues, steady-state populations, and projections of the metastable-triplet spin eigenstates.}
  \label{fig:SpinSimulations}
\end{figure}

Figure~\ref{fig:PLsimulations} shows the simulated PL as a function of in-plane field angle together with the total steady-state population in the metastable triplet (singlet) for the case of the singlet (triplet) ground-state model.  As expected, the metastable population is inversely correlated with the PL.  
Figure~\ref{fig:SpinSimulations} shows the variation of the spin-triplet eigenstates for a fixed field amplitude ($g\mu_\mathrm{B}B/D = 0.5$) together with the PL for the singlet-GS case. The triplet Hamiltonian is identical in the triplet-GS case. 

\begin{figure}[tbp]
  \includegraphics[width=7in]{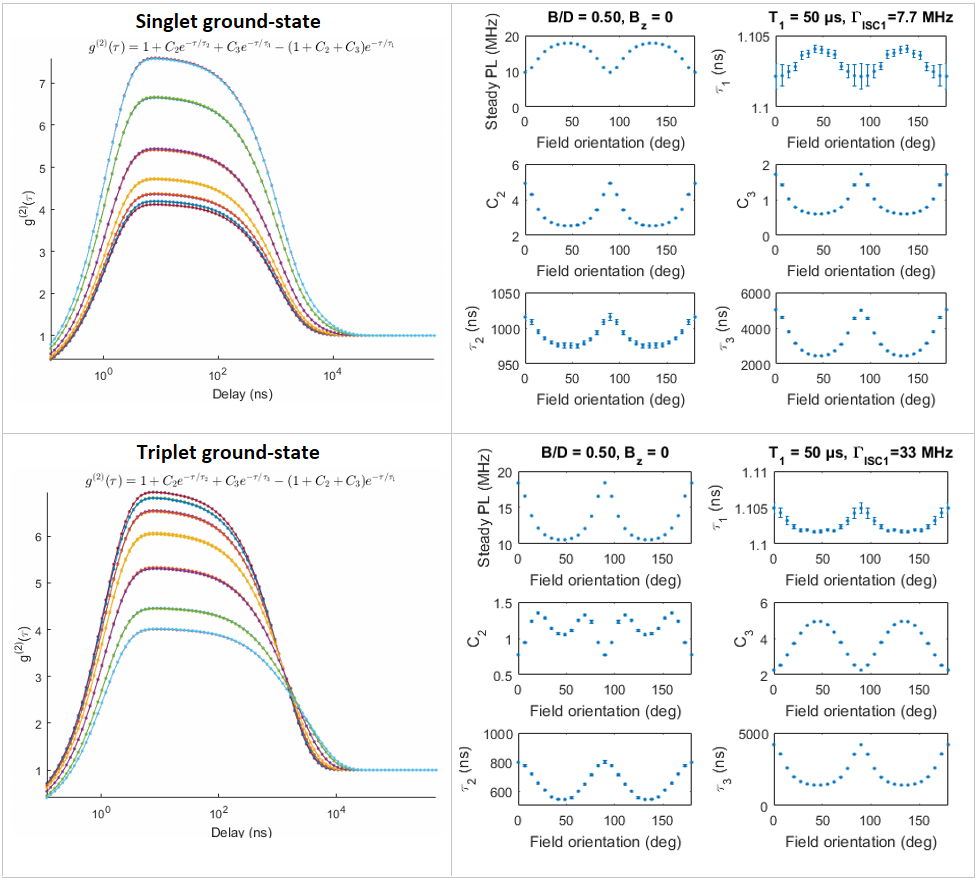}
  \caption{\textbf{Autocorrelation Simulations.} Left panels: Simulated photon autocorrelation function for different orientations of an in-plane magnetic field. Right panels: Best-fit parameters from a three-rate empirical model.}
  \label{fig:g2simulations}
\end{figure}

Finally, we simulate the photon autocorrelation function for various settings of in-plane field angle, and fit the result to the empirical model discussed in the text, with either two or three rates ($n=2$ or 3).  The simulated curves are shown in Fig.~\ref{fig:g2simulations} along with three-rate fits.  The corresponding results of two-rate fits are plotted in Fig. 4 of the main text.

In comparing these models, we observe the following qualitative differences that lead to our conclusion that the singlet-GS model is the best match to our experimental observation
\begin{itemize}
\item The singlet-GS model exhibits dark PL at $B=0$ and for $B$ oriented along $x$, $y$, or $z$, whereas the PL response in the triplet-GS case is inverted.  In our observations the PL is at a minimum when the field is aligned or orthogonal to the observed emission dipole axis, in agreement with the singlet-GS case.  Note that the observed excitation dipole is rotated from the emission dipole by $\sim 53^{\circ}$ , in seeming agreement with the triplet-GS case.  However, we expect that the emission dipole is a truer representation of the defect's symmetry axis on the basis of its higher visibility and the likely possibility of off-resonant excitation addressing additional short-lived excited states.

\item Although both models exhibit similar steady-state PL contrast as a function of in-plane field angle, the corresponding optical dynamics exhibited by the photon autocorrelation function is qualitatively different.  In the singlet-GS case, the bunching lifetime stays nearly constant whereas the amplitude changes strongly, in agreement with our observations.  On the other hand, the triplet-GS model exhibits strong variations in both the amplitude and lifetime of bunching.  Qualitatively, this makes sense since the bunching lifetime in the strong-excitation regime considered here is related to the effective ISC rate (averaged across all spin levels) from the excited state to the metastable state, whereas the bunching amplitude reflects both ISC rates.  In the singlet-GS case, the excited-state ISC is independent of field whereas the ground-state ISC varies strongly, and vice versa for the triplet-GS case.  The clear difference in photon emission statistics is strong evidence in favor of the singlet-GS case as a model for our observations.
\end{itemize}

Nonetheless, some features of the experiments are not reproduced by either simulation.  For example, the autocorrelation function in the singlet-GS case in the presence of an in-plane field is better described by a three-rate model, compared to the two-rate shape observed in experiments.  Furthermore, neither model predicts the increase in PL we observe as a function of B applied along $z$ or the non-monotonic changes at small values of B applied along $x$ or $y$.  We hypothesize that these discrepancies might be related to other physical effects such as hyperfine or spin-orbit coupling between levels that are not included in our model, but further experiments will be required to fully answer these and other questions raised by this work. In the next section we explore possible hyperfine interactions and their effects.

\clearpage
\subsection*{Hyperfine interactions}

Hyperfine interactions in hBN are expected to be complex because multiple nuclear spins will couple to any defect electron spin due to every atomic site possessing a nuclear spin. Such complicated interactions have been observed in electron paramagnetic resonance (EPR) spectroscopy of bulk hBN \cite{Geist1964,Fanciulli1997}. The size of the observed hyperfine structure was on the order of $30$ MHz. However, much larger interactions (up to $\approx 1$ GHz) are theoretically possible, as we describe in the subsection below. Also below, we estimate the magnitude of the electron ZFS parameters to be 1--6 GHz. 

If the hyperfine and ZFS parameters are of a similar size, then the hyperfine interactions may explain the features currently not captured by our purely electronic optical dynamics model: (1) the increase in the PL with the application of a magnetic field along $z$ that saturates after $\approx 200$ G, and (2) the non-monotonic changes at small values of magnetic fields applied along $x$ or $y$ with turning points at $\approx 70$ G. A crude explanation is that the hyperfine interactions define an effective in-plane internal magnetic field that is experienced by the electron spin. Only when this internal field is overcome by the applied field does our purely electronic modeling apply. This would explain (1) and (2) and indicate that the internal fields correspond to hyperfine interactions of the order of $\approx 560$ MHz and $\approx 196$ MHz, respectively, which are values reasonably consistent with our estimates below. However, we emphasize that more precise calculations of the effects of hyperfine interactions are required to confirm this explanation and these can only be performed once further information about the defect (\textit{i.e.}, its structure, ZFS parameters, etc.) is established.

\subsubsection*{Estimation of electronic ZFS parameters}
Assuming that the ZFS parameters arise from spin-spin interaction between two unpaired electrons of a triplet level, approximate expressions for the parameters are \cite{Stoneham1975}
\begin{subequations}
\begin{align}
D & = \frac{3}{2}\frac{\mu_0}{4\pi}g^2\mu_B^2\langle\frac{1-3x_{12}^2/r_{12}^2}{r_{12}^3}\rangle\\
E & = \frac{3}{2}\frac{\mu_0}{4\pi}g^2\mu_B^2\langle\frac{z_{12}^2-y_{12}^2}{r_{12}^5}\rangle 
\end{align}
\end{subequations}
where $\mu_0$ is the vacuum permeability, $x_{12}$, $y_{12}$, $z_{12}$ are the coordinates of the displacement vector $\vec{r}_{12}$ with magnitude $r_{12}$ that connects the positions of the two electrons, and the angle brackets denote the direct integral over the unpaired spin density. Note that the exchange integral has been neglected in the above.

Rough estimates of the parameters can be made by assuming that the defect axis is aligned with one of the in-plane bonds and that the spin density is contained in-plane and distributed over next-to-nearest neighbor atoms (\textit{i.e.} as for a vacancy-centered defect). In this case, we can approximate the average separation of the unpaired electrons by the displacement vector between next-to-nearest neighbor lattice sites, such that $|x_{12}|\approx 3l_B/2 = 2.18 \angstrom$, $|y_{12}|\approx \sqrt{3}l_B/2 = 1.26 \angstrom$ and $|z_{12}|\approx 0$, where $l_B=1.45 \angstrom$ is the bond length of hBN. Using these values, we find $D\approx -6$ GHz and $E\approx -1$ GHz. The relative sign of $E$ and $D$ would change if the spin density was rather primarily contained out-of-plane (\textit{i.e.} in $\pi-$orbitals). 

\subsubsection*{Estimation of hyperfine interaction in hBN}
The hyperfine interaction between the defect's electronic spin and a given nuclear spin can be described by two parameters
\begin{subequations}
\begin{align}
A_\parallel & =  f+d \\
A_\perp & =  f-2d
\end{align}
\end{subequations}
which are defined by the Fermi contact $f$ and dipolar $d$ interactions between the unpaired electron spin density associated with the atom and its nuclear spin \cite{Stoneham1975}. If the unpaired electrons occupy an atomic orbital of the form
\begin{equation}
\psi = c_s \phi_s + c_p \phi_p
\end{equation}
where $\phi_s$ and $\phi_p$ are the $2s$ and $2p$ orbitals, respectively, and $c_s$ and $c_p$ are linear coefficients satisfying the normalization condition $|c_s|^2+|c_p|^2=1$, then the interactions are
\begin{subequations}
\begin{align}
f & = \frac{8\pi}{3}\frac{\mu_0}{4\pi}g\mu_Bg_n\mu_n|c_s|^2\eta|\phi_s(0)|^2 \\
d & = \frac{2}{5}\frac{\mu_0}{4\pi}g\mu_Bg_n\mu_n|c_p|^2\eta\langle\phi_p|\frac{1}{r^3}|\phi_p\rangle
\end{align}
\end{subequations}
where $\eta$ is the portion of the total spin density at the atom, $g_n$ and $\mu_n$ are the nuclear g-factor and magneton, respectively, and $r$ is the distance of the electron spin from the nucleus. The values of $|\phi_s(0)|^2$ and $\langle\phi_p|\frac{1}{r^3}|\phi_p\rangle$ are distinct for B and N and can be obtained from ab initio calculations of the atoms. 

For the in-plane $sp^2$ $\sigma-$orbitals of hBN, the expected values for the linear coefficients are $|c_s|^2=1/3$ and $|c_p|^2=2/3$, and so these orbitals result in both contact and dipolar interactions. For the out-of-plane $p$ $\pi-$orbitals, the expected values are instead $|c_s|^2=0$ and $|c_p|^2=1$, and so these orbitals only yield a dipolar interaction. Using the values for $|\phi_s(0)|^2$ and $\langle\phi_p|\frac{1}{r^3}|\phi_p\rangle$ reported by Ref \onlinecite{Morton1978} for $^{11}$B and $^{14}$N (the most prevalent isotopes with spin), we have estimated the hyperfine parameters in table \ref{tab:hyperfine_parameters} for both $\sigma-$ and $\pi-$orbitals.

\begin{table*}[tb] 
\centering
\caption{\textbf{Estimated hyperfine parameters of different atomic orbitals in hBN}}
\label{tab:hyperfine_parameters}
\setlength{\tabcolsep}{7.5pt}
\setlength{\extrarowheight}{7.5pt}
\begin{tabularx}{0.75\textwidth}{ l | l  l || l | l  l }
\hline
$\pi$-orbital & $A_\parallel$ (MHz) & $A_\perp$ (MHz) & $\sigma-$orbital & $A_\parallel$ (MHz) & $A_\perp$ (MHz) \\
\hline
$^{11}$B($I=1/2$) & 64 & -127 & $^{11}$B($I=1/2$) & 891 & 764 \\
$^{14}$N($I=1$) & 56 & -111 & $^{14}$N($I=1$) & 641 & 530 \\
\hline
\end{tabularx}
\end{table*}

The values in table \ref{tab:hyperfine_parameters} correspond to the hyperfine interactions that would result if all of the electron spin density occupied a single atomic orbital (\textit{i.e.}, $\eta=1$). In reality, the density will be spread over multiple atomic orbitals.  Since at least three atomic orbitals are likely to contribute, we can estimate an upper bound of the hyperfine interactions by dividing the above by 3. This will yield interactions on the order of $\approx$30 MHz for $\pi-$orbitals and $\approx$300 MHz for $\sigma-$orbitals.

\subsection*{ODMR Simulations}

A crucial next step in the study and application of h-BN's spin defects is to attempt ODMR spectroscopy and time-domain spin control.  In order to inform these future measurements, we use our master-equation model to estimate the predicted ODMR response as a function of applied dc magnetic field. These estimates highlight the optimal conditions for experiments, and further inform the technological potential of these defects for quantum information processing and quantum sensing.

To simulate the ODMR response within our model, we add additional spin-mixing terms within the triplet manifold.  For example, in the singlet-GS model of Fig.~\ref{fig:Transitions_SingletGS}, also used in Fig.~4 of the main text, states $\{\ket{3}, \ket{4},\ket{5}\}$ are the triplet eigenstates for the given settings of applied dc magnetic field, and we need only include an additional rate $\Gamma_{ij}^\mathrm{ODMR}\gg\Gamma_\mathrm{ISC2}$ in order to ensure complete mixing of the states $\ket{i}$ and $\ket{j}$.

\begin{figure}[tbp]
  \includegraphics[width=7in]{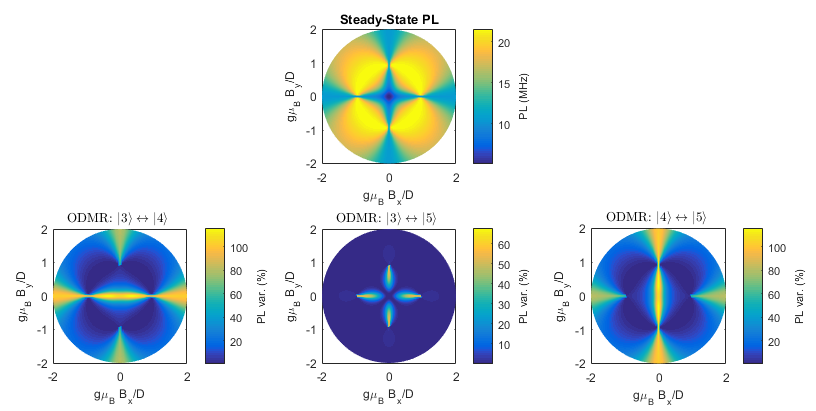}
  \caption{\textbf{ODMR Simulations.} Predicted ODMR PL variation (bottom panels) as a function of in-plane magnetic field when any two of the triplet eigenstates are resonantly driven by an ac magnetic field.  The corresponding steady-state PL is shown in the top panel. }
  \label{fig:ODMRsims}
\end{figure}

Simulations of the ODMR PL variation, \textit{i.e.}, $(I_\mathrm{MR}-I_0)/I_0$ where $I_\mathrm{MR}$ ($I_0$) is the PL in the presence (absence) of a resonant driving field, are shown in Fig.~\ref{fig:ODMRsims} for the singlet-GS model (b).  The other parameters in the simulation remain those listed in Table~\ref{tab:ModelParameters}.  As expected, the effect of a resonant field that mixes the triplet states is to reduce the steady-state population trapped in the triplet, causing an increase in PL.  The effect is most pronounced when the steady-state PL is near a minimum, especially near zero field.  The PL variation is also strongest when transitions are driven between states $\ket{3}\leftrightarrow\ket{4}$  or $\ket{4}\leftrightarrow\ket{5}$. This makes sense since state $\ket{4}$ has the largest projection on $\ket{s_z}$  for this range of field settings, and the lower ISC decay from the triplet selects for the $\ket{s_z}$ projection in this level diagram.

Here we are assuming that the ac field direction and magnitude can be chosen to strongly drive the selected pair of triplet eigenstates (we used a fixed value of $\Gamma^\mathrm{ODMR}=100$ MHz to calculate the PL variation).  Based on the spin-triplet Hamiltonian, we see that in this case the optimum transitions between state $\ket{4}\sim\ket{s_z}$  and states $\{\ket{3},\ket{5}\} \sim \{\ket{s_x},\ket{s_y}\}$
will be driven by in-plane fields oriented along $\hat{y}$ and $\hat{x}$, respectively. In the case of the strong driving fields adopted in the simulations, the ODMR linewidth will be power broadened.  Assuming the spin lifetime is limited by the decay of the metastable triplet state, estimated in the simulations to occur with a rate $\Gamma_\mathrm{ISC2}=0.85$~MHz, the minimum ODMR linewidth would be $\approx\Gamma_\mathrm{ISC2}/2\pi = 140$~kHz.

\bibliography{}

\clearpage
\appendix
\subsection{Appendix: PL Simulations}

\begin{figure}[h!]
  \includegraphics[width=5in]{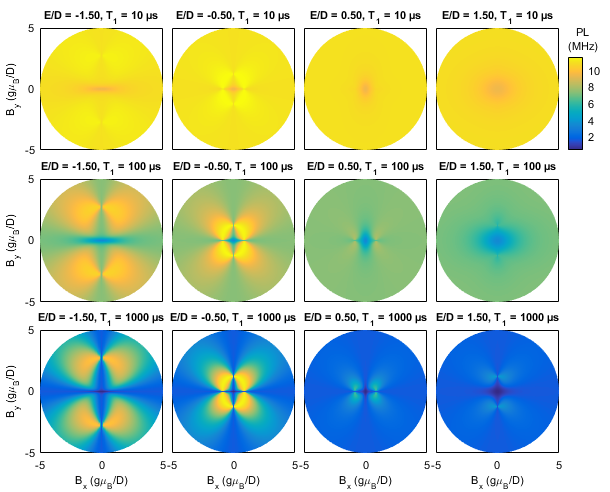}
  \caption{PL as a function of in-plane magnetic field for singlet-GS level diagram (b), assuming coupling coefficients $\bm{m}^{\prime}= (\frac{1}{3},\frac{1}{3},\frac{1}{3})$  and $\bm{m} = (0,0,1)$. }
  \label{fig:singlet_b}
\end{figure}

\begin{figure}[tbp]
  \includegraphics[width=5in]{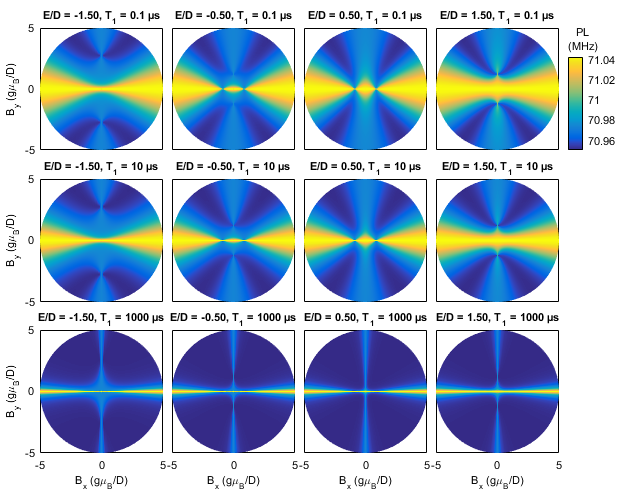}
  \caption{PL as a function of in-plane magnetic field for triplet-GS level diagram (b), assuming coupling coefficients $\bm{m}^{\prime}= (1,0,0)$  and $\bm{m} = (1,0,0)$. }
  \label{fig:triplet_b}
\end{figure}

\begin{figure}[tbp]
  \includegraphics[width=5in]{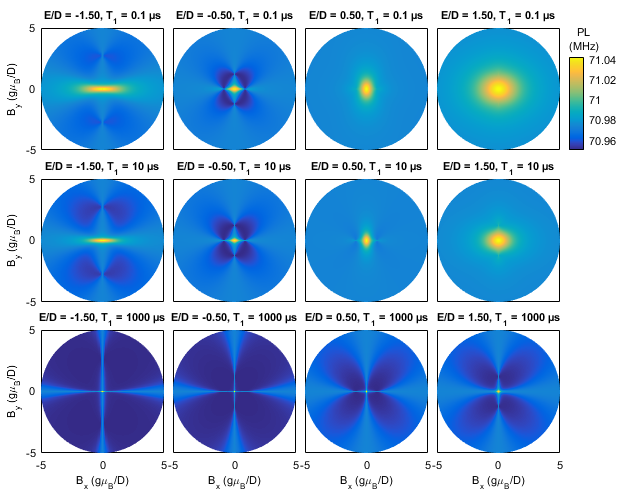}
  \caption{PL as a function of in-plane magnetic field for triplet-GS level diagram (c) or (f), assuming coupling coefficients $\bm{m}^{\prime}= (0,0,1)$  and $\bm{m} = (0,0,1)$. }
  \label{fig:triplet_cf}
\end{figure}

\begin{figure}[tbp]
  \includegraphics[width=5in]{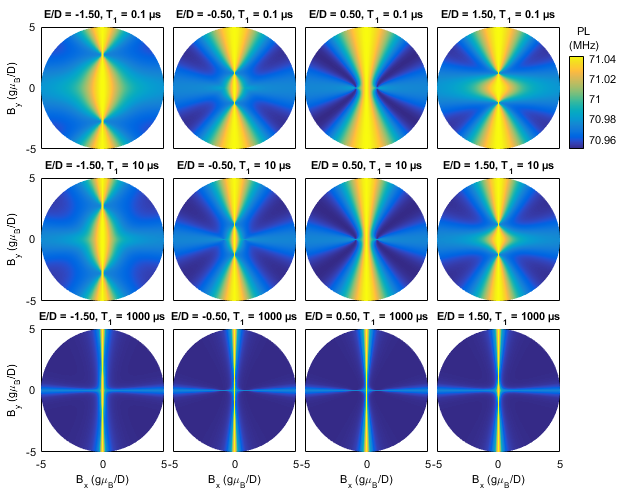}
  \caption{PL as a function of in-plane magnetic field for triplet-GS level diagram (d), assuming coupling coefficients $\bm{m}^{\prime}= (0,1,0)$  and $\bm{m} = (0,1,0)$. }
  \label{fig:triplet_d}
\end{figure}

\begin{figure}[tbp]
  \includegraphics[width=5in]{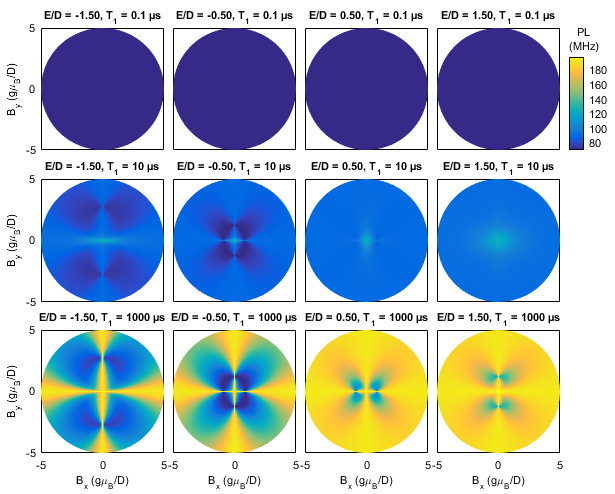}
  \caption{PL as a function of in-plane magnetic field for triplet-GS level diagram (e), assuming coupling coefficients $\bm{m}^{\prime}= (0,0,1)$  and $\bm{m} = (\frac{1}{3},\frac{1}{3},\frac{1}{3})$. }
  \label{fig:triplet_e}
\end{figure}

\begin{figure}[tbp]
  \includegraphics[width=5in]{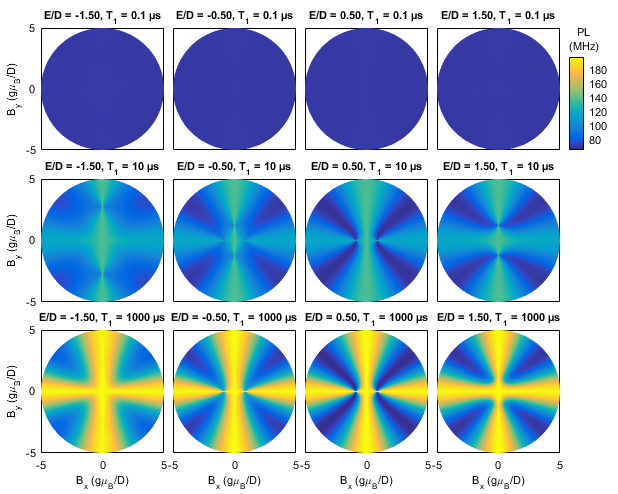}
  \caption{PL as a function of in-plane magnetic field for triplet-GS level diagram (g), assuming coupling coefficients $\bm{m}^{\prime}= (0,1,0)$  and $\bm{m} = (1,0,0)$. }
  \label{fig:triplet_g}
\end{figure}

\begin{figure}[tbp]
  \includegraphics[width=5in]{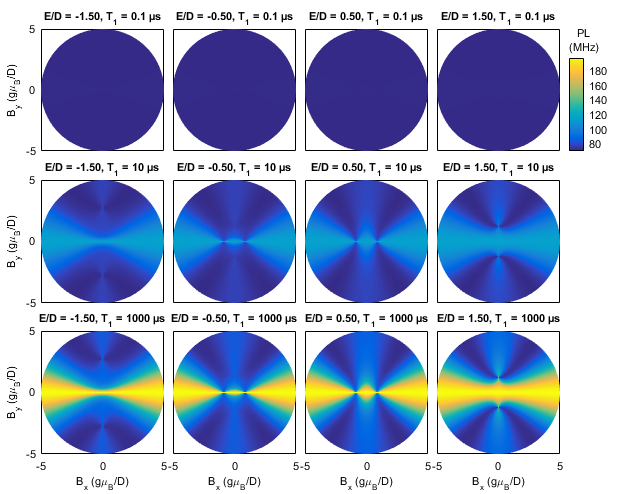}
  \caption{PL as a function of in-plane magnetic field for triplet-GS level diagram (g), assuming coupling coefficients $\bm{m}^{\prime}= (0,\frac{1}{2},\frac{1}{2})$  and $\bm{m} = (1,0,0)$. }
  \label{fig:triplet_g2}
\end{figure}

\begin{figure}[tbp]
  \includegraphics[width=5in]{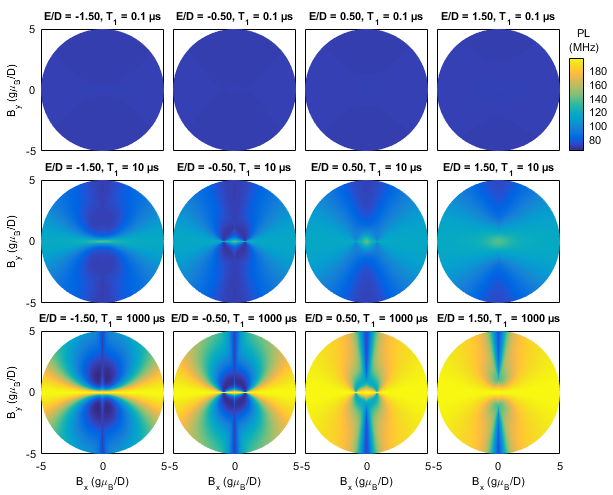}
  \caption{PL as a function of in-plane magnetic field for triplet-GS level diagram (g), assuming coupling coefficients $\bm{m}^{\prime}= (0,0,1)$  and $\bm{m} = (1,0,0)$. }
  \label{fig:triplet_g3}
\end{figure}

\begin{figure}[tbp]
  \includegraphics[width=5in]{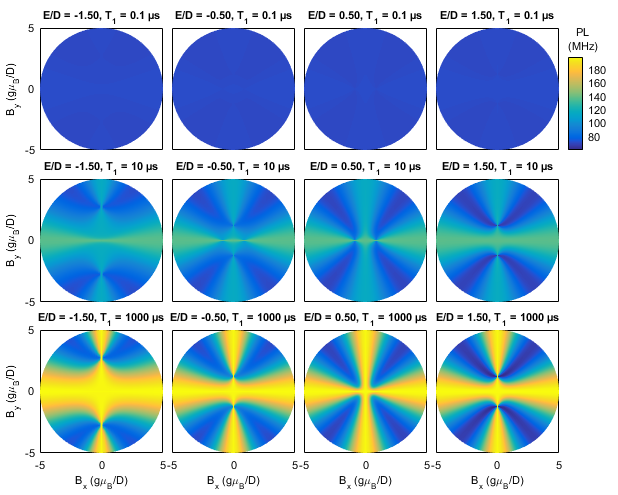}
  \caption{PL as a function of in-plane magnetic field for triplet-GS level diagram (h), assuming coupling coefficients $\bm{m}^{\prime}= (1,0,0)$  and $\bm{m} = (0,1,0)$. }
  \label{fig:triplet_h}
\end{figure}

\begin{figure}[tbp]
  \includegraphics[width=5in]{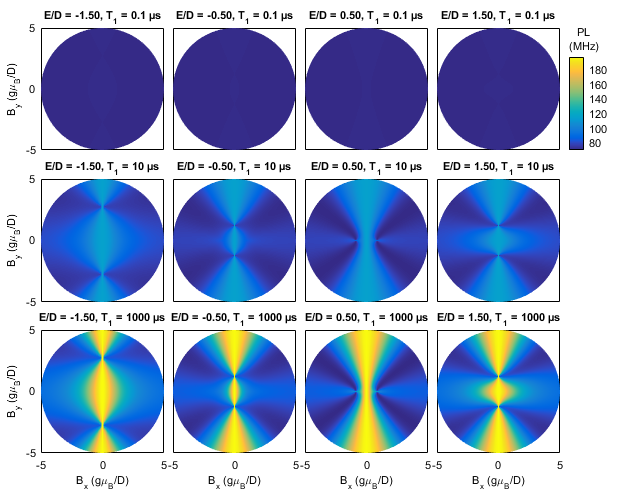}
  \caption{PL as a function of in-plane magnetic field for triplet-GS level diagram (h), assuming coupling coefficients $\bm{m}^{\prime}= (\frac{1}{2},0,\frac{1}{2})$  and $\bm{m} = (0,1,0)$. }
  \label{fig:triplet_h2}
\end{figure}

\begin{figure}[tbp]
  \includegraphics[width=5in]{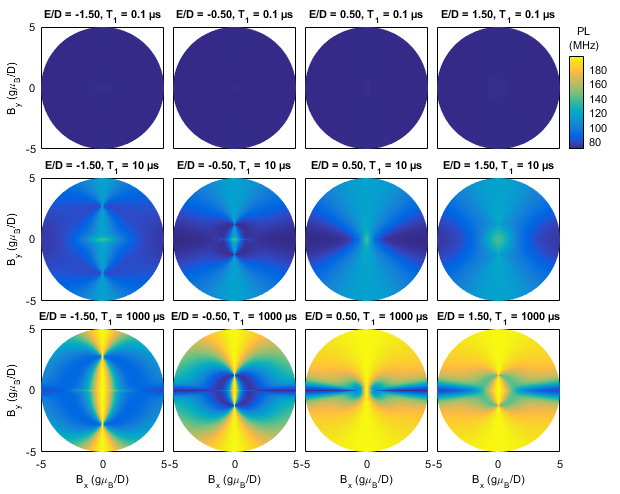}
  \caption{PL as a function of in-plane magnetic field for triplet-GS level diagram (h), assuming coupling coefficients $\bm{m}^{\prime}= (0,0,1)$  and $\bm{m} = (0,1,0)$. }
  \label{fig:triplet_h3}
\end{figure}

In Figures~\ref{fig:singlet_b} to \ref{fig:triplet_h3} below we plot simulations of the PL as a function of in-plane magnetic field vector, for various settings of the spin-triplet parameters $E/D$ and $T_{1}$.  In each case we set the following parameter values based on the analytical three-level 
model: $\Gamma_{s}$ = 600 MHz,  $\Gamma_{e}$ = 300 MHz,  $\Gamma_\mathrm{ISC1}$ = 1.8 MHz,  $\Gamma_\mathrm{ISC2}$ = 0.11 MHz

\vspace{6pt}
\noindent\textit{Singlet Ground-State Models:}

Here only level structure (b) is of interest in Fig.~\ref{fig:SingletDiagrams}, since level structure (a), with no spin-dependent selection rules, does not result in field-dependent PL. Simulations for diagram (b) are shown in Fig.~\ref{fig:singlet_b}

\vspace{6pt}
\noindent\textit{Triplet Ground-State Models:}

Level structure (a), with no spin-dependent selection rules, does not produce field-dependent PL.  Calculations for the other possible level arrangements are shown below.  Unless otherwise noted, we assume that ISC transitions with no spin-orbit-allowed selection rule can proceed via other mechanisms with no spin selectivity.

For level structures (g) and (h), the potential availability of multiple singlet states between the ground- and excited-state triplet levels can be approximated within our model by varying the excited-state coupling coefficients, $\bm{m}^{\prime}$, so we include multiple simulations for these diagrams.  We also assume any singlet-to-singlet relaxation happens quickly compared to the metastable lifetime and therefore the ground-state ISC coupling coefficients, $\bm{m}$, are determined solely by the ground-state singlet.

